\documentclass[fleqn,usenatbib]{mnras}

\usepackage{newtxtext,newtxmath}
\usepackage[T1]{fontenc}

\DeclareRobustCommand{\VAN}[3]{#2}
\let\VANthebibliography\thebibliography
\def\thebibliography{\DeclareRobustCommand{\VAN}[3]{##3}\VANthebibliography}

\usepackage{graphicx}	% Including figure files
\usepackage{amsmath}	% Advanced maths commands
\usepackage{subcaption}
\usepackage{booktabs}
\usepackage{lipsum}
\usepackage[normalem]{ulem}
\usepackage{float}
\newcommand{\uncertainties}[3]{#1{\raisebox{0.5ex}{\tiny$_{-#2}^{+#3}$}}}    %To make the uncertainties smaller

\defcitealias{Rhee_2017_ApJ}{R17}
\defcitealias{Pasquali_2019_MNRAS}{P19}
\defcitealias{Tempel_2017_A&A}{T17}
\defcitealias{Valk_2025_MNRAS}{VR25}

\title[The origin of ionized gas in retired galaxies]{The origin of ionized gas in retired galaxies: dynamical clues}

\author[G. A. Valk et al.]{
Greique A. Valk,$^{1}$\thanks{E-mail: greique.valk@acad.ufsm.br (GAV)}
Sandro B. Rembold$^{1}$
\\
$^{1}$Departamento de F\'isica, Centro de Ci\^encias Naturais e Exatas, Universidade Federal de Santa Maria, 97105-900, Santa Maria, RS, Brazil}

\date{Accepted XXX. Received YYY; in original form ZZZ}

\pubyear{2025}

\begin{document}
\label{firstpage}
\pagerange{\pageref{firstpage}--\pageref{lastpage}}
\maketitle

\begin{abstract}
We investigate the kinematical and dynamical properties of quiescent cluster galaxies with weak emission lines, referred to as retired (R), and those without emission lines, dubbed passive (P), to better understand the origin of the ionized gas in R galaxies and what drives the differences between these populations. We stack 2,907 P and 2,387 R galaxies from 336 relaxed galaxy clusters to build an ensemble cluster and estimate their projected number density and velocity dispersion profiles, $\sigma_P(R)$, as well as their projected phase-space (PPS) distributions. Additionally, we apply the MAMPOSSt code and the Jeans equation inversion technique to constrain the velocity anisotropy profiles, $\beta(r)$. We find that P galaxies tend to reside closer to the cluster centres than R galaxies, and that both populations exhibit similar $\sigma_P(R)$ and $\beta(r)$ profiles, regardless of their stellar mass, stellar age, or morphology. The only exception is elliptical R galaxies, which are marginally more concentrated and display more radial orbits than their P counterparts. PPS analyses indicate that R galaxies were, on average, accreted later than P galaxies, except for those with $D_n4000 > 1.86$ or elliptical morphology. These results suggest that R galaxies, though accreted more recently, have already had enough time to evolve towards a dynamical state more consistent with that of the dynamically relaxed P population. Finally, our findings suggest that the ionized gas in early-type R galaxies likely originates from accretion from their own hot gas haloes, and that its removal triggers the transition toward the P phase over relatively long timescales.
\end{abstract}


%%%%%%%%%%%%%%%%%%%%%%%%%%%%%%%%%%%%%%%%%%%%%%%%%%

%%%%%%%%%%%%%%%%% BODY OF PAPER %%%%%%%%%%%%%%%%%%

%%%%%%%%%%%%%%%%%%%%%%%%
%%%%%%%%%%%%%%%%%%%%%%%%
%%%%%%%%%%%%%%%%%%%%%%%%
%%%%%%%%%%%%%%%%%%%%%%%%
%%%%%%%%%%%%%%%%%%%%%%%%
\section{Introduction}
\label{sec:Introduction}
% %%%%%%%%%%%%%%%%%%%%%%%%%%%%%%%%%%%%%%%%%%%%%%%%%%%%%%%%%%
% %%%%%%%%%%%%%%%%%%%%%%%%%%%%%%%%%%%%%%%%%%%%%%%%%%%%%%%%%%
% %%%%%%%%%%%%%%%%%%%%%%%%%%%%%%%%%%%%%%%%%%%%%%%%%%%%%%%%%%

It is well established that galaxy properties depend on their environment. Early-type galaxies (elliptical/lenticular) dominate the central, high-density regions of groups and clusters, while late-type systems (spiral/irregular) constitute the bulk of the galaxy population in low-density regions. This morphology-density relation was first identified by \citet{Dressler_1980_ApJ} and has since been confirmed by a wide range of studies \citep[e.g.][]{Postman_1984_ApJ, Einasto_1987_MNRAS, Treu_2003_ApJ, Huertas-Company_2009_A&A, Houghton_2015_MNRAS}. In addition, several works have shown that galaxies residing in high-density environments display redder colours, host older stellar populations, and exhibit lower star formation rates compared to similar galaxies in low-density regions \citep[e.g.][]{Kauffmann_2004_MNRAS, Weinmann_2006_MNRAS, Weinmann_2009_MNRAS}. In order to explain the dependence of galaxy properties on their environment, a variety of environmentally dependent mechanisms has been proposed over the years, among which are ram-pressure stripping \citep{Gunn_1972_ApJ}, dynamical friction \citep{Chandrasekhar_1943_ApJ}, starvation \citep{Larson_1980_ApJ}, galaxy mergers \citep[][]{Gerhard_1981_MNRAS, Noguchi_1988_A&A, Hernquist_1995_ApJ, Mihos_1996_ApJ, Gao_2020_A&A}, and harassment \citep[][]{Moore_1996_Natur, Moore_1998_ApJ}. Furthermore, galaxies residing in low-mass halos before entering the cluster environment may also have their properties affected -- a process known as pre-processing \citep{Zabludoff_1998_ApJ, Balogh_2000_ApJ, Fujita_2004_PASJ, Lopes_2024_MNRAS}.

Additionally to environment effects, galaxy properties are also affected by internal processes. For instance, \citet{Peng_2010_ApJ} found that galaxy colours depend on both stellar mass and local environmental density \citep[see also e.g.][]{Pasquali_2010_MNRAS}. These internal mechanisms -- commonly referred to as `mass quenching' since they scale with the galaxy stellar mass -- include feedback from Supernova and Active Galactic Nuclei \citep[AGN, e.g.][]{Springel_2003_MNRAS, Bower_2006_MNRAS, Croton_2006_MNRAS, Agertz_2013_ApJ}, morphological quenching \citep{Martig_2009_ApJ, Martig_2013_MNRAS}, and bar-driven evolution \citep{Spinoso_2017_MNRAS}. \citet{Peng_2010_ApJ} demonstrated that these two modes of quenching (environment and mass) act independently, meaning their effects must be carefully disentangled when studying galaxy evolution.

%%%%%%%%%%%%%%%%%%%%%%%%%%%%%%%%%%%%%%%%%%%%%%%%
%%%%%%%%%%%%%%%%%%%%%%%%%%%%%%%%%%%%%%%%%%%%%%%%
%%%%%%%%%%%%%%%%%%%%%%%%%%%%%%%%%%%%%%%%%%%%%%%%

Galaxies infalling through filaments into the cluster environment are expected to display more radial orbits and exhibit properties that resemble their previous environments, whether they were field regions or low-mass haloes along the filaments themselves \citep[e.g.][]{Martinez_2016_MNRAS, Salerno_2019_MNRAS}. Once inside the cluster, environment mechanisms begin to act on them, simultaneously modifying their morphologies, removing their gas, and gradually isotropising their orbits. As a result, galaxies that entered the cluster at earlier times are expected to be quiescent systems, located deeper within the cluster potential well and exhibiting more isotropic orbits than those currently infalling. This suggests that the properties of the galaxies in clusters are related to their time since infall. Therefore, given that galaxies occupying similar regions in projected phase-space (PPS) diagrams are expected to have entered the cluster at comparable epochs \citep[see e.g.][]{Mahajan_2011_MNRAS, Oman_2013_MNRAS, Rhee_2017_ApJ, Pasquali_2019_MNRAS}, the PPS serves as a powerful tool for probing the environment impact on galaxy properties. Indeed, several authors have employed PPS analyses to better understand the galaxy evolution within clusters \citep[see e.g.][]{Mahajan_2011_MNRAS, Jaffe_2015_MNRAS, Jaffe_2018_MNRAS, deCarvalho_2019_MNRAS, Pasquali_2019_MNRAS, Rhee_2020_ApJS, Roberts_2020_MNRAS, Sampaio_2021_MNRAS, Brambila_2023_MNRAS, Sampaio_2024_MNRAS}. Additionally, the efficiency of some of these environmentally dependent mechanisms also correlates with the orbital properties of the galaxies. For instance, ram-pressure stripping, starvation, and dynamical friction are more effective on galaxies accreted into the cluster environment through more radial orbits \citep[e.g.][]{Vollmer_2001_ApJ, Jaffe_2018_MNRAS, Lotz_2019_MNRAS}. Therefore, orbital analysis can also provide valuable insights into the relative efficiency of these mechanisms in shaping galaxy evolution. Indeed, several works have shown that, at low-redshift ($z\sim0.1$), quiescent/early-type/red galaxies display more isotropic orbits than star-forming/late-type/blue galaxies \citep[e.g.][]{Mahdavi_1999_ApJ, Biviano_2004_A&A, Munari_2014_A&A, Mamon_2019_A&A}. In contrast, at high-redshift ($z>0.4$), both populations appear to share similar orbital profiles, characterized by isotropic orbits in the inner regions that become more radial with increasing cluster-centric distances \citep{Biviano_2009_A&A, Biviano_2013_A&A, Biviano_2016_A&A}. These findings suggest an evolution in the orbital profile of quiescent galaxies towards lower redshifts, while that of star-forming systems remains unchanged. However, there are also works that have found that quiescent/red galaxies move on more radial orbits than star-forming/blue galaxies, in both observations \citep{Aguerri_2017_MNRAS, Mercurio_2021_A&A} and simulations \citep{Iannuzzi_2012_MNRAS, AguirreTagliaferro_2021_A&A}.

%%%%%%%%%%%%%%%%%%%%%%%%%%%%%%%%%%%%%%%%%%%%%%%%%%%%%%%%%%
%%%%%%%%%%%%%%%%%%%%%%%%%%%%%%%%%%%%%%%%%%%%%%%%%%%%%%%%%%
%%%%%%%%%%%%%%%%%%%%%%%%%%%%%%%%%%%%%%%%%%%%%%%%%%%%%%%%%%

Quiescent galaxies can be separated into two classes: those without emission lines, dubbed passive (P) galaxies, and those that exhibit weak emission lines, generally referred to as retired (R) galaxies. The term `retired` was first introduced by \citet{Stasinska_2008_MNRAS} to refer to these galaxies, in the sense that they are retired from star formation. In fact, hot low-mass evolved stars (HOLMES) are the preferred ionization mechanism invoked to explain the gas emission lines observed in R galaxies \citep[see e.g.][]{Stasinska_2008_MNRAS, CidFernandes_2011_MNRAS, Herpich_2018_MNRAS, Stasinska_2022_FrASS}. As shown by \citet{CidFernandes_2011_MNRAS}, P and R galaxies exhibit similar distributions of several physical parameters, such as stellar mass, optical colours, and stellar population mean age and metallicity. Furthermore, the old stellar populations in both classes are expected to contain HOLMES at similar proportions. Therefore, it is surprising that a significant fraction of the quiescent galaxies do not show emission lines at all \citep[see e.g.][]{CidFernandes_2011_MNRAS, Stasinska_2015_MNRAS}. 

In this context, the work by \citet{Herpich_2018_MNRAS} provided significant insights into the origin of the observed differences between P and R galaxies. Using a sample of elliptical galaxies extracted from the Sloan Digital Sky Survey \citep[SDSS,][]{York_2000_AJ} DR7, \citet{Herpich_2018_MNRAS} compared the properties of lineless (P) and liny (R) galaxies after matching them in stellar mass, redshift, and half-light radius in the $r$ band. Through stellar population analysis, they concluded that there is no difference in the number of ionizing photons produced by HOLMES in both populations. Consequently, this result indicates that the difference between P and R galaxies is not due to a lack of ionizing photons in P galaxies, but rather to a difference in their warm gas content. Additionally, \citet{Herpich_2018_MNRAS} found that R galaxies display higher values of stellar extinction compared to P galaxies. Since dust is expected to be mixed with warm gas, this further supports the absence of warm gas in P galaxies. Moreover, independent evidence for this comes from \citet{Ryzhov_2025_ApJS}, who identified that quiescent galaxies with weak emission lines display median cold dust temperatures $\sim 1$ K higher than those observed for galaxies without emission lines. According to the authors, this can be interpreted as the presence of a warm gas reservoir in R galaxies. Finally, \citet{Herpich_2018_MNRAS} also found evidence that R galaxies likely underwent a recent period of star formation ($0.1-5\,\rm{Gyr}$), which did not occur in P galaxies.

The above results point to an absence of a warm gas reservoir in P galaxies compared to R galaxies. The simplest assumption is that such gas is ejected by winds from intermediate-mass stars. Indeed, \citet{Herpich_2018_MNRAS} demonstrated that the mass of gas ejected by intermediate-mass stars is larger by orders of magnitude than that required to explain the observed H$\alpha$ luminosities. However, by analysing the chemical composition of the emitting gas through the [NII]/[OII] emission-line ratio, they concluded that the emitting gas is not nitrogen enriched as it would be if it were originating from mass loss by intermediate-mass stars. In addition, by comparing the [NII]/[OII] ratios of R galaxies with those of star-forming galaxies with the same masses or metallicities, they also discarded merger with a metal-poor galaxy as a possible source of the emitting gas. Consequently, based on their analyses, \citet{Herpich_2018_MNRAS} concluded that the more likely sources for the emitting gas in R galaxies are from accretion from the haloes of the galaxies or from residual streams of metal-rich gas coming from a merger in the recent past. Furthermore, additional support for an external origin for the emitting gas in R galaxies comes from the decoupling between the stellar and gas kinematics in early-type galaxies, which would not be the case if the gas came from stellar winds. For instance, \citet{Belfiore_2017_MNRAS} found that 30 out of their 49 galaxies with extended low-ionization emission-line regions display gas–stellar kinematic misalignments larger than $30^\circ$, suggesting an external origin for the gas \citep[see also e.g.][]{Sarzi_2006_MNRAS, Davis_2016_MNRAS, Gomes_2016_A&A}. Taken together, these results indicate that, in R galaxies, the source of ionization and the origin of the gas producing the emission lines are disconnected. 

In this context, investigating the environments of P and R galaxies can provide valuable insights into the origin of the differences between these two populations. Indeed, environmental and kinematical mapping represents a promising approach for evaluating the various scenarios proposed for the origin of the emitting gas in R galaxies, as environmentally driven mechanisms are expected to affect each proposed mechanism differently. For instance, mass loss by intermediate-mass stars should not exhibit any environmental dependence, while the cooling of the halo gas can be suppressed in dense regions due to processes like strangulation \citep{Larson_1980_ApJ}. Mergers, though generally rare within clusters, may occur in their outskirts, particularly within infalling groups. Thus, examining the environments of P and R galaxies is crucial for constraining the origin of their observed differences.

%%%%%%%%%%%%%%%%%%%%%%%%%%%%%%%%%%%%%%%%%%%%%%%%%%%%%%%%%%
%%%%%%%%%%%%%%%%%%%%%%%%%%%%%%%%%%%%%%%%%%%%%%%%%%%%%%%%%%
%%%%%%%%%%%%%%%%%%%%%%%%%%%%%%%%%%%%%%%%%%%%%%%%%%%%%%%%%%

In \citet[hereafter \citetalias{Valk_2025_MNRAS}]{Valk_2025_MNRAS}, we analysed the kinematical and dynamical properties of cluster galaxy populations classified according to their main source of gas ionization, namely: star-forming (SF) galaxies, optical AGNs, mixed SF plus AGN ionization (transition objects, T) and quiescent (Q) galaxies. In particular, we found that the Q (SF) galaxies exhibit both the lowest (highest) typical cluster-centric distances and velocity dispersion values, and are characterized by inner isotropic orbits, that become increasingly radial with increasing cluster-centric distances. Furthermore, we were able (unable) to recover the observed $\sigma_P(R)$ profile of the Q (SF) population using equilibrium solutions for the mass and velocity anisotropy profiles. These results were interpreted in \citetalias{Valk_2025_MNRAS} as indicating that the Q (SF) galaxies, as a population, are the closest (farthest) to equilibrium within their clusters. In contrast, the results obtained for the AGN and T populations suggest that the galaxies in these populations are in an intermediate dynamical stage between those of the Q and SF populations, likely evolving towards quiescence. In that work, no differentiation was made between P and R galaxies, and they were all included in the Q population. In the present study, we take a step further by providing a detailed analysis of the kinematical and dynamical properties of P and R galaxies within galaxy clusters. We follow a methodology similar to that adopted in \citetalias{Valk_2025_MNRAS}, but we complement our analyses by also exploring the distribution of P and R galaxies across the PPS diagram. In addition to comparing the overall dynamical properties of the full sample of P and R galaxies, we investigate whether these populations exhibit distinct dynamical behaviours when matched in terms of stellar mass, mean stellar population age (as traced by the $4000$-\AA\ break index), and morphology. Through this approach, we aim to determine whether P and R galaxies have experienced different evolutionary pathways within their host clusters, which could account for the presence of a warm gas reservoir in R galaxies, and its absence in P galaxies.

This paper is organized as follows. In Section~\ref{sec:Data}, we describe the data sample. The methods used to derive the kinematical and dynamical properties, as well as the PPS distributions, of the galaxy populations are presented in Section~\ref{sec:Methods}. In Section~\ref{sec:Galaxy_populations}, we present the results obtained for the entire P and R galaxy samples, while the results for subsamples matched in stellar mass, mean stellar population age, and morphology are reported in Section~\ref{sec:Galaxy_parameters}. We discuss our results in Section~\ref{sec:Discussion}, and summarize our main conclusions in Section~\ref{sec:Conclusions}. Throughout this paper we adopt $H_0 = 67.8$ km\,s$^{-1}$\,Mpc$^{-1}$, $\Omega_m = 0.308$ and $\Omega_\Lambda = 0.692$ \citep{PlanckCollaboration_2016_A&A}.

%%%%%%%%%%%%%%%%%%%%%%%%
%%%%%%%%%%%%%%%%%%%%%%%%
%%%%%%%%%%%%%%%%%%%%%%%%
%%%%%%%%%%%%%%%%%%%%%%%%
%%%%%%%%%%%%%%%%%%%%%%%%
%%%%%%%%%%%%%%%%%%%%%%%%

%%%%%%%%%%%%%%%%%%%%%%%%
%%%%%%%%%%%%%%%%%%%%%%%%
%%%%%%%%%%%%%%%%%%%%%%%%
%%%%%%%%%%%%%%%%%%%%%%%%
%%%%%%%%%%%%%%%%%%%%%%%%
%%%%%%%%%%%%%%%%%%%%%%%%

\section{Data}
\label{sec:Data}

In this work, we use the galaxy and groups catalogues presented by \citet[hereafter \citetalias{Tempel_2017_A&A}]{Tempel_2017_A&A}, which are based on data from the SDSS DR12 \citep{Eisenstein_2011_AJ, Alam_2015_ApJS} and complemented with 1349 redshifts from others surveys. The final galaxy sample consists of 584\,449 objects with Petrosian $r$-band magnitudes brighter than $17.77$ and redshifts $z \leq 0.2$. The galaxies were assigned to groups by \citetalias{Tempel_2017_A&A} employing a group finder algorithm based on the Friends-of-Friends (FoF) method \citep[e.g.][]{Turner_1976_ApJS, Beers_1982_ApJ, Zeldovich_1982_Natur}. Furthermore, to improve the reliability of the groups identified by the FoF method, a group membership refinement algorithm was also employed. Several physical parameters for each group were derived by \citetalias{Tempel_2017_A&A}, with the velocity dispersion and virial radius being of particular interest in this work. The velocity dispersion of a system was estimated from the variance in the line-of-sight (LOS) velocities of its confirmed members, while the virial radius is uniquely defined by the system's virial mass, which is derived using the virial theorem and depends on the group's projected extend, velocity dispersion, and the assumed mass density profile. In \citetalias{Tempel_2017_A&A}, the virial masses were derived under the assumption of an NFW profile \citep{Navarro_1996_ApJ} and the mass-concentration relation proposed by \citet{Maccio_2008_MNRAS}. The resulting group catalogue includes 88\,662 systems with at least two members, among which 498 merging systems were identified due to overlapping virial radii among multiple clusters. Finally, the \citetalias{Tempel_2017_A&A} galaxy catalogue also includes morphological information for its galaxies, based on the work of \citet{Huertas-Company_2011_A&A}. These authors performed an automated morphological classification based on support vector machines for $\sim 700\,000$ galaxies from the SDSS DR7 spectroscopic sample. For each object, the classification provides a probability of being in the four morphological classes, namely: elliptical (E), lenticular (S0), early-type spiral (Sab), and late-type spiral (Scd). 

We selected from the \citetalias{Tempel_2017_A&A} catalogues only those galaxies which belong to clusters with 20 or more members, resulting in a final sample of 642 clusters and 23\,977 galaxies. This constraint on cluster richness was adopted to ensure a reliable dynamical analysis, as it ensures that the derived cluster parameters are reliable and that the estimated centre positions are physically meaningful. In addition, as pointed out by \citetalias{Valk_2025_MNRAS}, dynamical analyses can be significantly affected by unrelaxed galaxy clusters, making their exclusion necessary to ensure robust results. Thus, we removed all unrelaxed galaxy clusters from our initial sample of 642 clusters (for details, see section 4 of \citetalias{Valk_2025_MNRAS}), yielding a final sample of 336 relaxed galaxy clusters comprising 10\,898 galaxies.  

The fluxes and equivalent widths ($W$) of the emission lines $\rm{H}\alpha$, $\rm{H}\beta$, $\rm{[NII]}\lambda \, 6584$, and $\rm{[OIII]}\lambda \, 5007$, as well as the stellar masses, $M_\star$, and the $4000-$\AA\, break index, $D_n4000$, for the galaxies in our sample, were obtained from the SDSS Catalogue Archive Server\footnote{\url{https://skyserver.sdss.org/casjobs/}} (CAS). The emission line measurements were performed by \citet{Thomas_2013_MNRAS} using adapted versions of the Absorption Line Fitting \citep[GANDALF,][]{Sarzi_2006_MNRAS} and penalized PiXel Fitting \citep[pPXF,][]{Cappellari_2004_PASP} codes, while the stellar population templates from \citet{Maraston_2011_MNRAS} and \citet {Thomas_2011_MNRAS} were used to model the continuum. Furthermore, an amplitude-over-noise (AoN) ratio larger than two was imposed for each emission line detection. Additionally, the stellar masses were estimated by \citet{Maraston_2013_MNRAS} through broad-band spectral energy distribution fitting of stellar population models to the observed $ugriz$ BOSS \citep[Baryon Oscillation Spectroscopic Survey,][]{Dawson_2013_AJ} magnitudes, using the BOSS spectroscopic redshift. The fit was performed utilising an adaptation of the publicly available \textsc{HYPERZ} code \citep{Bolzonella_2000_A&A}. The stellar masses were derived for both star-forming and passive stellar population models. However, since in this work we will only focus on the study of quiescent galaxies, we adopt the results obtained using the passive templates based on \citet{Maraston_2009_MNRAS}, calculated assuming a \citet{Kroupa_2001_MNRAS} initial mass function. This passive model is composed by two single-burst simple stellar populations (SSP), characterized by identical ages but different metallicities, namely solar and 0.05 solar. The solar metallicity SSP contributes 97 per cent of the template mass, while the metal-poor SSP accounts for the remaining 3 per cent. Moreover, age is the only parameter, with a minimum allowed fitting value of 3 Gyr. Lastly, the break index $D_n4000$, defined as the ratio of the average flux density in the narrow continuum bands \citep[$3850-3950$ and $4000-4100$ \AA,][]{Balogh_1999_ApJ}, was calculated as described in \citet{Brinchmann_2004_MNRAS}.

%%%%%%%%%%%%%%%%%%%%%%%%
%%%%%%%%%%%%%%%%%%%%%%%%
%%%%%%%%%%%%%%%%%%%%%%%%
%%%%%%%%%%%%%%%%%%%%%%%%
%%%%%%%%%%%%%%%%%%%%%%%%
%%%%%%%%%%%%%%%%%%%%%%%%

\section{Methods}
\label{sec:Methods}

%%%%%%%%%%%%%%%%%%%%%%%%
%%%%%%%%%%%%%%%%%%%%%%%%
%%%%%%%%%%%%%%%%%%%%%%%%
%%%%%%%%%%%%%%%%%%%%%%%%
%%%%%%%%%%%%%%%%%%%%%%%%
%%%%%%%%%%%%%%%%%%%%%%%%

%%%%%%%%%%%%%%%%%%%%%%%%
%%%%%%%%%%%%%%%%%%%%%%%%

\subsection{Ensemble Cluster}
\label{subsec:Ensemble_Cluster}

%%%%%%%%%%%%%%%%%%%%%%%%
%%%%%%%%%%%%%%%%%%%%%%%%

The number of spectroscopic members in each cluster of our sample is typically too low to provide reliable dynamical analysis. Therefore, we stacked the individual clusters together to create an \textit{ensemble cluster}, and analyse it under the assumption that it is representative of each cluster in our sample. This is a widely used procedure \citep[e.g][]{Katgert_2004_ApJ, Biviano_2016_A&A, Mamon_2019_A&A} and is supported by both numerical simulations which predict a global mass profile, $M(r)$, for haloes of dark matter \citep[e.g.][]{Navarro_1996_ApJ}, and by the fact that this mass profile is weakly dependent on halo mass and redshift \citep[e.g.][]{DeBoni_2013_MNRAS, Biviano_2016_A&A}. 

We follow the stacking procedure described in detail in \citetalias{Valk_2025_MNRAS}. In summary, the cluster-centric distance and LOS velocity of each galaxy are normalized by the virial radius\footnote{Throughout this paper, we refer to the `virial radius' as the radius of a sphere in which the mean matter density is 200 times the critical density of the Universe, represented by $r_{200}$.} and velocity dispersion of their parent halo, respectively. Subsequently, the virial radius, $\langle r_{200} \rangle$, and the velocity dispersion, $\langle \sigma_v \rangle$, of the ensemble cluster\footnote{The ensemble cluster is characterized by a virial radius, $\langle r_{200} \rangle$, a velocity dispersion, $\langle \sigma_v \rangle$, and a redshift $\langle z \rangle$, defined as the mean values of the respective parameters from all individual clusters comprising the ensemble cluster.} are used as scaling factors to convert these normalized values back into physical quantities with units of distance and velocity, respectively. Finally, it is important to mention that this approach preserves the relative positions and velocities of the galaxies in their parent clusters within the ensemble cluster.

\subsubsection{Clusters centres' locations}

In order to construct the ensemble cluster, the centres of the clusters are required to be well defined. Otherwise, the stacking procedure will spread the galaxies around the centre of the ensemble cluster, thereby affecting the derived profiles of projected number density and velocity dispersion. In \citetalias{Tempel_2017_A&A}, the cluster centre was estimated as the geometric centre of all galaxies, without any weighting by luminosity or mass. However, as shown by \citetalias{Valk_2025_MNRAS}, this centre does not accurately map the density peak of the clusters. For instance, the $I(R)$ profile of the ensemble cluster obtained with the \citetalias{Tempel_2017_A&A} centre is flatter in the inner region and exhibits a slow decline in the outer region (see fig. 4 of \citetalias{Valk_2025_MNRAS}). Therefore, we have decided to correct the centre location of our clusters. To do this, we follow the methodology adopted by \citetalias{Valk_2025_MNRAS} and define the new centre of each cluster as the peak location of a Gaussian kernel applied to its 2D distribution of galaxies. In the remainder of this work, we adopt this centre location, rather than that estimated by \citetalias{Tempel_2017_A&A}, for our clusters. Finally, although we have re-estimated the centre position of each cluster, it is important to mention that we retain the $r_{200}$ value as estimated by \citetalias{Tempel_2017_A&A}.

\subsection{Observed projected profiles}
\label{subsec:Observed_projected_profiles}

The anisotropy profiles of galaxy clusters are inferred from two fundamental observables, namely: the number density, $I(R)$, and velocity dispersion, $\sigma_P(R)$, profiles. Below, we summarize the methods used to estimate and fit these profiles. A detailed explanation is provided in \citetalias{Valk_2025_MNRAS}.

The projected number density, $I$, is estimated in circular rings concentric with the centre of the ensemble cluster, while the LOS velocity dispersion, $\sigma_P$, is calculated using the interquartile range ($\sigma_P = \rm{IQR}/1.349$). The interquartile range (IQR), defined as the difference between the 75th and 25th percentiles of the distribution, was adopted to avoid potential issues at the edges of the velocity distribution generated by both the FoF method and the membership refinement process. Both quantities, $I$ and $\sigma_P$, are estimated in radial bins with a fixed number of galaxies, excepting for the last bin, to create radial profiles of $I(R)$ and $\sigma_P(R)$.

To perform the dynamical analysis continuum versions of the $I(R)$ and $\sigma_P(R)$ profiles are required. We fit the observed $I(R)$ profile using a NFW profile for the number density, $\nu(r)$, given by\footnote{Throughout this work, we use $r$ and $R$ to represent the three-dimensional radius and the projected radius in the sky plane, respectively, both measured with respect to the cluster centre.}
\begin{equation}
    \nu(r) = \dfrac{\nu_0}{r(r+r_\nu)^2},
\end{equation}

\noindent where $\nu_0$ and $r_\nu$ are the normalization factor and the scale radius of the NFW profile, respectively. Additionally, to fit the $\sigma_P(R)$ profile we use the functional form presented by \citetalias{Valk_2025_MNRAS} (see their eq. 4), which, as highlighted by the authors, was found to accurately reproduces the shape of $\sigma_P(R)$ profiles estimated using NFW density/anisotropy profiles pairs with varying parameters.

%%%%%%%%%%%%%%%%%%%%%%%%
%%%%%%%%%%%%%%%%%%%%%%%%

\subsection{MAMPOSSt}
\label{subsec:MAMPOSSt}

%%%%%%%%%%%%%%%%%%%%%%%%
%%%%%%%%%%%%%%%%%%%%%%%%

The dynamics of the galaxies in our ensemble cluster was investigated using MAMPOSSt code \citep[Modelling Anisotropy and Mass Profiles of Observed Spherical Systems,][]{Mamon_2013_MNRAS}. MAMPOSSt assumes parametric forms for the $\nu(r)$, $M(r)$ and velocity anisotropy profiles to perform a maximum likelihood fit of the galaxy distribution in the PPS. Moreover, it has the advantage of requiring only cluster-centric distances and velocities as input to be run. The orbital information of galaxies is encoded in the velocity anisotropy profile, $\beta(r)$, defined as
\begin{equation}
    \beta(r) = 1 - \dfrac{\sigma_\theta^2(r)-\sigma_\phi^2(r)}{2\sigma_r^2(r)},
\end{equation}

\noindent where $\sigma_\theta$, $\sigma_\phi$, and $\sigma_r$ are the two tangential, and the radial component, respectively, of the velocity dispersion tensor. In the case of spherical symmetry, $\sigma_\theta = \sigma_\phi$. Isotropic orbits are characterized by $\beta = 0$, while purely radial and tangential orbits correspond to $\beta = 1$ and $\beta = -\infty$, respectively. Lastly, the code assumes both spherical symmetry for the ensemble cluster and a Gaussian 3D velocity distribution.

Dynamical equilibrium is required from the systems under analysis since MAMPOSSt is based on the Jeans equations. To meet this requirement, we decide to restrict our analysis only to those galaxies with projected cluster-centric distances smaller than the virial radius of the system \citep[$R \leq r_{200}$, see e.g.][\citetalias{Valk_2025_MNRAS}]{Biviano_2016_A&A, Biviano_2021_A&A}. Nevertheless, it is important to mention that this approach does not fully ensure the equilibrium condition due to the presence of interlopers, i.e. galaxies with projected cluster-centric distances lower than $r_{200}$ but real 3D position without the virial sphere. In addition, the dynamics of the satellite galaxies closest to the central regions are mainly driven by the potential of the cluster's central galaxy (or Brightest Cluster Galaxy, BCG) rather than the overall cluster potential. Thus, to ensure a reliable dynamical analysis, we also removed from our sample all objects with $R/r_{200} < 0.1$. 

In order for MAMPOSSt to be run, the user needs to specify the functional forms for the $\nu(r)$, $M(r)$, and $\beta(r)$ profiles. We choose a NFW profile for the $\nu(r)$ profile, which, as shown in Section~\ref{subsec:I_profiles_All_Gal}, accurately fits the observed projected number density profiles estimated using our galaxy sample. Moreover, we execute MAMPOSSt in the \textit{Split} mode, which restrict the maximum likelihood fit to the velocity space. In this mode, the $\nu(r)$ profile is fitted externally to MAMPOSSt and only the best-fitting value of $r_\nu$, along with its respective uncertainty, are provided to the code. In addition, we also utilize a NFW profile for the cluster mass density, $\rho(r)$. Thus, the mass profile is given by
\begin{equation}\label{eq:Mass_profile}
    M(r) = M_{200} \frac{\ln(1+r/r_{-2})-r/r_{-2}(1+r/r_{-2})^{-1}}{\ln(1+c)-c/(1+c)},
\end{equation}

\noindent where $M_{200}$ is the mass contained within $r_{200}$, $r_{-2}$ is the radius where the logarithmic slope of the mass density profile is equal to $-2$, and $c \equiv r_{200}/r_{-2}$ represents the concentration of the mass profile. Lastly, the mass profile estimated by MAMPOSSt corresponds only to the dark matter, as we assume that galaxies are massless tracers of the cluster potential and neglect any contribution from a possible central black hole.

In \citetalias{Valk_2025_MNRAS}, we have shown that the orbits of all cluster galaxy populations analysed are consistent with isotropic orbits in the inner regions that become increasingly radial with increasing cluster-centric distances. In fact, the solution estimated by MAMPOSSt for the Q galaxies, which are of particular interest here, exhibit $\beta = 0.25 \pm 0.13$ at $r_{200}$. In this work, we will provide a further investigation of the Q objects, by splitting them into P and R galaxies. As a consequence, we expect the uncertainties in the MAMPOSSt solutions to increase due to the decrease in the number of objects analysed in each sample. Therefore, in an attempt to minimize this problem, we decided to use a simpler $\beta(r)$ model than that considered in \citetalias{Valk_2025_MNRAS}, and choose the ML model \citep{Mamon_2005_MNRAS} for the $\beta(r)$ profile, given by
\begin{equation}\label{eq:beta_profile}
\beta(r) = \dfrac{1}{2} \dfrac{r}{r+r_\beta},    
\end{equation}

\noindent where $r_\beta$ is the scale radius of the $\beta(r)$ profile. Finally, we constrain the $r_\beta$ parameter to the range $[0.1-4.0]$ in all MAMPOSSt runs, as values outside this range do not provide any improvement in the physical reliability of the results, given the radial coverage of our galaxy sample.

%%%%%%%%%%%%%%%%%%%%%%%%
%%%%%%%%%%%%%%%%%%%%%%%%

\subsection{Inversion of the Jeans equations}
\label{subsec:IJE}

The mass profile of galaxy clusters is generally well fitted by a NFW profile \citep[e.g.][]{Geller_1999_ApJL, Diaferio_1999_MNRAS, Biviano_2003_ApJ, Oguri_2012_MNRAS, Biviano_2013_A&A, Okabe_2013_ApJL}. However, simulations suggests that the $\beta(r)$ profile of cluster-size haloes can vary significantly from system to system \citep[see e.g.][]{Mamon_2013_MNRAS}. Therefore, although we thought that the adopted ML model would be able to satisfactorily reproduce the observed $\beta(r)$ profile, given the results presented in \citetalias{Valk_2025_MNRAS}, we also apply the inversion of the Jeans equations technique (hereafter IJE) to obtain a secondary, non-parametric estimate of the $\beta(r)$ profile. We follow the IJE method of \citet{Solanes_1990_A&A} and execute a bootstrap procedure, as described in detail in \citetalias{Valk_2025_MNRAS}, to estimate a average $\beta(r)$ profile and its respective uncertainties. In short, the IJE uses as input the $\nu(r)$ and $M(r)$ profiles from MAMPOSSt solutions, while the $\sigma_P(R)$ is estimated using random subsamples of the original galaxy sample.

%%%%%%%%%%%%%%%%%%%%%%%%
%%%%%%%%%%%%%%%%%%%%%%%%

\subsection{Projected Phase Space}
\label{subsec:PPS}

In order to complement the results obtained from the dynamical analysis, we will also investigate the distributions of galaxies in the PPS diagram. The PPS diagram consists of the absolute LOS peculiar velocity, $|V_{\rm{LOS}}|$, of a galaxy plotted on the $y$-axis, versus its cluster-centric distance, $R$, on the $x$-axis. In PPS analyses, a common approach involves normalizing the cluster-centric distance and the LOS peculiar velocity of each galaxy by the virial radius and velocity dispersion, respectively, of their parent halo. This method is adopted to allow the stacking procedure of galaxies from different clusters into a single PPS diagram.

Through a galaxy fall into a cluster, it tends to follows a typically path along the PPS \citep[see e.g. fig. 1 of][]{Rhee_2017_ApJ}. In fact, \citet{Mahajan_2011_MNRAS} and \citet{Oman_2013_MNRAS}, utilizing cosmological simulations, demonstrated that galaxies at different orbital positions (virialized, backsplash, and infall) are often found in semi-distinct regions in the PPS. Thus, galaxies that occupy similar regions in the PPS are expected to have entered the cluster at similar early times. This result was confirmed by \citet[hereafter \citetalias{Rhee_2017_ApJ}]{Rhee_2017_ApJ} using the YZiCS set of hydrodynamic zoom-in simulations of galaxy clusters \citep{Smith_2016_ApJ, Choi_2017_ApJ}, which identified that galaxies with similar time since infall, $T_{\rm{inf}}$, defined as the time since a sub-halo first enters the cluster virial radius, occupy nearly distinct regions in the 3D phase space (3D velocities vs 3D cluster-centric distances). Unfortunately, due to projection effects, this relation between $T_{\rm{inf}}$ and phase space location suffers a significant scattering when projected quantities are considered. Nevertheless, \citetalias{Rhee_2017_ApJ} showed that it is still possible to associate different regions in the PPS to galaxies characterized by different times since infall \citep[see also][]{Pasquali_2019_MNRAS}.

Based on their results, \citetalias{Rhee_2017_ApJ} defined five regions in the PPS (labelled A-E). These zones were designed in an attempt to maximize the fraction of galaxies belonging to a particular galaxy population, which are characterized by different $T_{\rm{inf}}$, in each zone (see fig. 6 of \citetalias{Rhee_2017_ApJ}). Galaxies that have not yet fallen into the cluster (first infallers) dominate in the A zone. The B and C zones are dominated by recent infallers ($0 < T_{\rm{inf}} < 3.64$ Gyr), while in the D zone, intermediate infallers ($3.63 < T_{\rm{inf}} < 6.45$\, Gyr) predominate. Finally, the ancient infallers ($6.45 < T_{\rm{inf}} < 13.7$\, Gyr) are the most numerous population in the E zone. We will refer to these five regions as the Rhee zones in the remainder of this work.

To analyse the distribution of galaxies in the PPS, we split it into 2D bins. The normalized density distribution of galaxies in the PPS of a particular galaxy population is obtained by normalizing the number of galaxies in each bin by the total number of galaxies in that population. The matrix resulting from this procedure is typically noising and thus can be difficult identifies trends or differences between populations. Thus, to facilitate the analysis of the results, we smooth the normalized density distribution by applying a Gaussian kernel with standard deviation $\sigma = 1$. This kernel size was carefully chosen to reduce the noise in the results while also preserving their global trends. In this process, bins with no galaxies are recovered through kernel smoothing, as long as at least 50 per cent of their neighbouring bins are occupied; otherwise, they are discarded. 

%%%%%%%%%%%%%%%%%%%%%%%%
%%%%%%%%%%%%%%%%%%%%%%%%

%%%%%%%%%%%%%%%%%%%%%%%%
%%%%%%%%%%%%%%%%%%%%%%%%

\subsection{Characterizing the gas ionization source}
\label{subsec:classification_pops}

%%%%%%%%%%%%%%%%%%%%%%%%
%%%%%%%%%%%%%%%%%%%%%%%%

In \citetalias{Valk_2025_MNRAS}, we classified the galaxies in our sample into populations according to their dominant source of gas ionization, namely: star-forming (SF), optical AGNs, transition objects (T, SF+AGN ionization), and quiescent (Q) galaxies. In short, the classification was performed using both the BPT-NII \citep{Baldwin_1981_PASP} and WHAN \citep{CidFernandes_2011_MNRAS} optical diagnostic diagrams simultaneously. However, a significant fraction of galaxies cannot be classified through these diagnostic diagrams due to the absence of information for at least one of the emission lines required by them. Since most of these objects are truly quiescent galaxies, with intrinsically weak emission lines, a secondary classification based solely on the available emission lines was employed in an attempt to select these quiescent objects instead of discarding them from our sample.

In this work, we use an adapted version of the classification scheme considered in \citetalias{Valk_2025_MNRAS} to account for the separation of the Q objects into P and R galaxies, which are of particular interest here. For those galaxies classified using the BPT-NII and WHAN diagrams, we applied the criteria from \citet{CidFernandes_2011_MNRAS}: P galaxies must have both $W_{\mathrm{H}\alpha}$ and $ W_{\mathrm{[NII]}} < 0.5 $ \AA, while R galaxies must have $W_{\mathrm{H}\alpha} < 3 $ \AA. Additionally, a secondary classification scheme was considered for the remaining objects lacking information on at least one of the emission lines required by both diagrams. Galaxies in which the $\rm{[OIII]}\lambda\,5007$ or H$\beta$ lines are not available, but both H$\alpha$ and $\rm{[NII]}\lambda\,6584$ are detected, are classified as P or R according to the criteria of \citet{CidFernandes_2011_MNRAS}. On the other hand, if the H$\alpha$ line is detected but the $\rm{[NII]}\lambda \, 6584$ line is absent, the classification into P or R is based solely on $W_{\mathrm{H}\alpha}$: P galaxies are defined by \mbox{$W_{\mathrm{H}\alpha} < 0.5$ \AA}, while R galaxies have \mbox{$0.5 \leq W_{\mathrm{H}\alpha} < 3.0$ \AA}. Furthermore, in both cases we applied an additional constraint to ensure that the objects classified in this step are truly quiescent galaxies: whenever the $\rm{[OIII]}\lambda\,5007$ or H$\beta$ lines are detected, their equivalent widths must satisfy $W < 3.0$ \AA.

%%%%%%%%%%%%%%%%%%%%%%%%
%%%%%%%%%%%%%%%%%%%%%%%%

%%%%%%%%%%%%%%%%%%%%%%%%
%%%%%%%%%%%%%%%%%%%%%%%%

\section{Dynamical properties of passive and retired galaxies}
\label{sec:Galaxy_populations}

\begin{figure*}
    \centering
         \includegraphics[width = 0.95\linewidth]{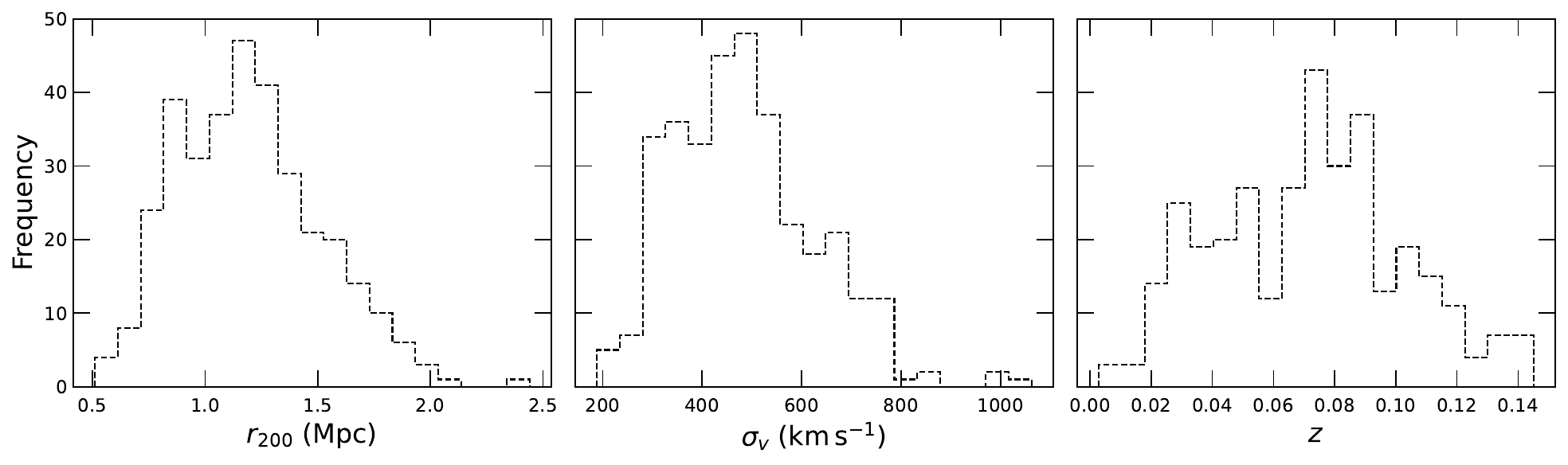}
         \caption{Distributions of virial radius ($r_{200}$, left panel), velocity dispersion ($\sigma_v$, middle panel), and redshift ($z$, right panel) for the 336 relaxed galaxy clusters that compose the ensemble cluster.}
    \label{fig:r200_sigmav_z_dist}
\end{figure*}

We identified 3\,401 P and 2\,833 R galaxies in our sample of 10\,898 galaxies belonging to 336 relaxed galaxy clusters by applying the criteria from Section~\ref{subsec:classification_pops}. These P and R galaxies were stacked together to create a single ensemble cluster, following the procedure outlined in Section~\ref{subsec:Ensemble_Cluster}. To ensure a reliable dynamical analysis, only objects with $0.1 \leq R/r_{200} \leq 1.0$ (see Section~\ref{subsec:MAMPOSSt}) were included, resulting in a final sample of 2\,907 P and 2\,387 R galaxies. The virial radius, velocity dispersion, and redshift of the ensemble cluster are $\langle r_{200} \rangle = 1194.73$\,kpc, $\langle \sigma_v \rangle = 485.74$\, km\,s$^{-1}$ and $\langle z \rangle = 0.071824$, respectively. The distributions of virial radius, velocity dispersion, and redshift for the 336 relaxed galaxy clusters that compose the ensemble cluster are presented in Fig.~\ref{fig:r200_sigmav_z_dist}.

%%%%%%%%%%%%%%%%%%%%%%%%
%%%%%%%%%%%%%%%%%%%%%%%%

%%%%%%%%%%%%%%%%%%%%%%%%
%%%%%%%%%%%%%%%%%%%%%%%%

\subsection{Projected number density profiles}
\label{subsec:I_profiles_All_Gal}

\begin{figure*}
        \centering
        \includegraphics[width = 0.7\linewidth]{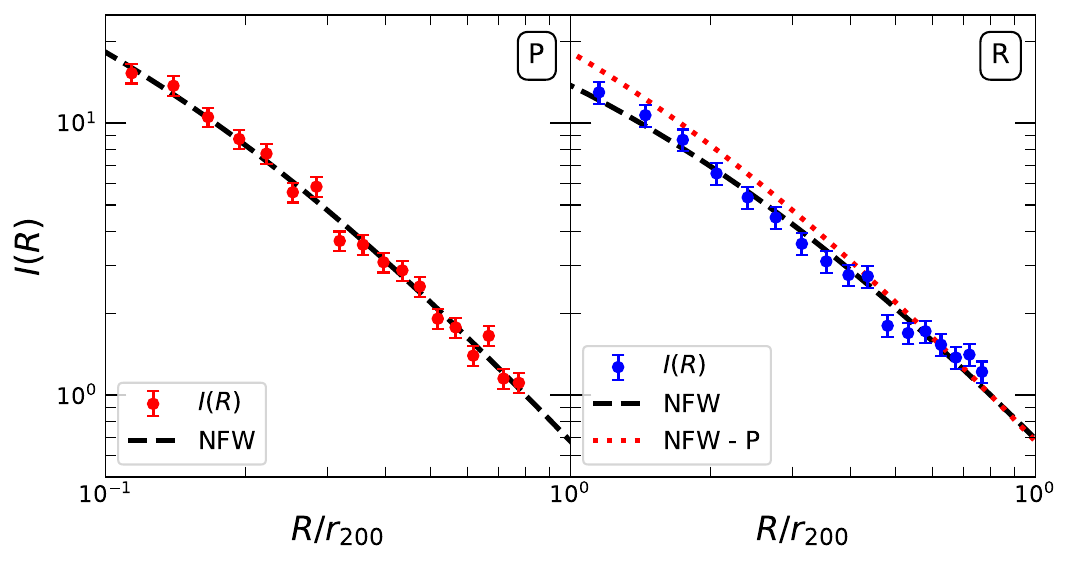}
        \caption{Projected number density profiles $I(R)$ of the P (red dots, left panel) and R (blue dots, right panel) galaxy populations. The fitted NFW profile is represented by the dashed black line in each panel. The $I(R)$ and the NFW profiles are normalized by the value of the fitted NFW profile at $R/r_{200} = 0.8$. The P NFW profile is exhibited in the right panel (dotted red line) to better compare the slope of the profiles. In each panel, the horizontal axis provides the projected radial distance normalized by the $\langle r_{200} \rangle$ value given in Section~\ref{sec:Galaxy_populations}, while the horizontal values indicate the central value of each bin.}
        \label{fig:I_profiles_All_Gal}
    \end{figure*}

The $I(R)$ profiles of the P (red dots, left panel) and R (blue dots, right panel) galaxy populations, estimated using bins containing 5 per cent of the respective population sample (except for the last bin), are shown in Fig.~\ref{fig:I_profiles_All_Gal}. We only show the $I(R)$ profile out to radial distances below $R/r_{200} = 0.8$, where the fraction of individual clusters contributing galaxies to the ensemble cluster remains high ($\sim 95$ per cent). Beyond this radius, the fraction decreases rapidly, producing a artificial drop in the $I(R)$ profile. The NFW profile fitted to the $I(R)$ profile is represented by the dashed black line in each panel of Fig.~\ref{fig:I_profiles_All_Gal}, while the best-fitting values of $r_\nu$ and $\nu_0$, along with their uncertainties, and the $\chi^2$ of the fit are exhibited in Table~\ref{tab:table_I_fit_All_Gal}. It is important to mention that only bins below $R/r_{200} = 0.8$ are considered in the fit. In order to better compare the slope of the profiles, the $I(R)$ and the NFW profile of each population are normalized by the value of the respective fitted NFW profile at $R/r_{200} = 0.8$, and the P NFW profile is shown in the right panel (dotted red line).

We observe in Fig.~\ref{fig:I_profiles_All_Gal} that the P and R galaxy populations exhibit very similar $I(R)$ profiles, indicating that their spatial distributions are alike. Additionally, the results for the $r_\nu$ parameter in Table~\ref{tab:table_I_fit_All_Gal}, which provides an estimate of the typical cluster-centric distances of the galaxies, show that systems in both populations are generally located closer to the centre of the ensemble cluster. Nevertheless, according to these results, P galaxies tend to be significantly more concentrated in the central regions than R galaxies.

    \begin{table}
    	\centering
    	\caption{Best-fitting values of $r_\nu$ and $\nu_0$, along with their uncertainties, and the $\chi^2$ of the fit, for the $I(R)$ profiles of the P and R galaxy populations.}
    	\label{tab:table_I_fit_All_Gal}
    	\begin{tabular}{ccccccc} % four columns, alignment for each
    		\hline
    		  Pop. & $r_\nu$ (kpc) & $\nu_0$ (kpc$^{-3}$) & $\chi^2$\\
    		\hline
    		P & $ 197 \pm 21$ & $208 \pm 12$ & $1.17$ \\
    		R & $ 285 \pm 33$ & $203 \pm 16$ & $1.29$ \\
    		\hline
    	\end{tabular}
    \end{table}

%%%%%%%%%%%%%%%%%%%%%%%%
%%%%%%%%%%%%%%%%%%%%%%%%

%%%%%%%%%%%%%%%%%%%%%%%%
%%%%%%%%%%%%%%%%%%%%%%%%

\subsection{Velocity anisotropy profiles}
\label{subsec:Beta_profiles_All_Gal}

    We run MAMPOSSt for the ensemble cluster considering the P and R galaxy populations as independent tracers of the cluster potential. The execution was performed in the \textit{Split} mode, providing to MAMPOSSt the $r_\nu$ values on Table~\ref{tab:table_I_fit_All_Gal}, and also the redshift of the ensemble cluster given in Section~\ref{sec:Galaxy_populations}.
    
    The results obtained for the $M(r)$ profile of the ensemble cluster are \mbox{$r_{200} = \uncertainties{1047}{24}{25}$\,kpc} and $r_{-2} = \uncertainties{347}{78}{110}$\,kpc, respectively, which imply a concentration factor of $c = \uncertainties{3.02}{0.78}{0.88}$ and a mass enclosed within the virial radius of $M_{200} = 10^{14.11} \,M_\odot$. The central value of each parameter corresponds to the 50th percentile of its marginal distribution, with uncertainties represented by the 5th and 95th percentiles. These results are in very good agreement with those obtained by \citetalias{Valk_2025_MNRAS} (\mbox{$r_{200} = \uncertainties{1047}{24}{25}$\,kpc} and $r_{-2} = \uncertainties{347}{84}{100}$\,kpc) using all galaxy populations comprising the ensemble cluster in their MAMPOSSt analysis. This consistency ensures that the MAMPOSSt execution with only the P and R galaxies does not significantly affect the dynamical results provided by it. In fact, the P and R galaxies in our sample are those that make up the Q population analysed by \citetalias{Valk_2025_MNRAS}, and their results indicated that this population is the most virialized. In addition, the MAMPOSSt solution for $r_{200}$, although slightly lower, is also in accordance with the actual value estimated for the ensemble cluster ($\langle r_{200} \rangle = 1194.73$\,kpc). Finally, the results obtained for the $\beta(r)$ profiles of the P and R populations are \mbox{$\log (r_\beta/\rm{kpc}) = \uncertainties{2.7}{1.8}{1.0}$} and \mbox{$\log (r_\beta/\rm{kpc}) = \uncertainties{2.9}{1.8}{0.9}$}, respectively. 

\begin{figure*}
        \centering
        \includegraphics[width = 0.85\linewidth]{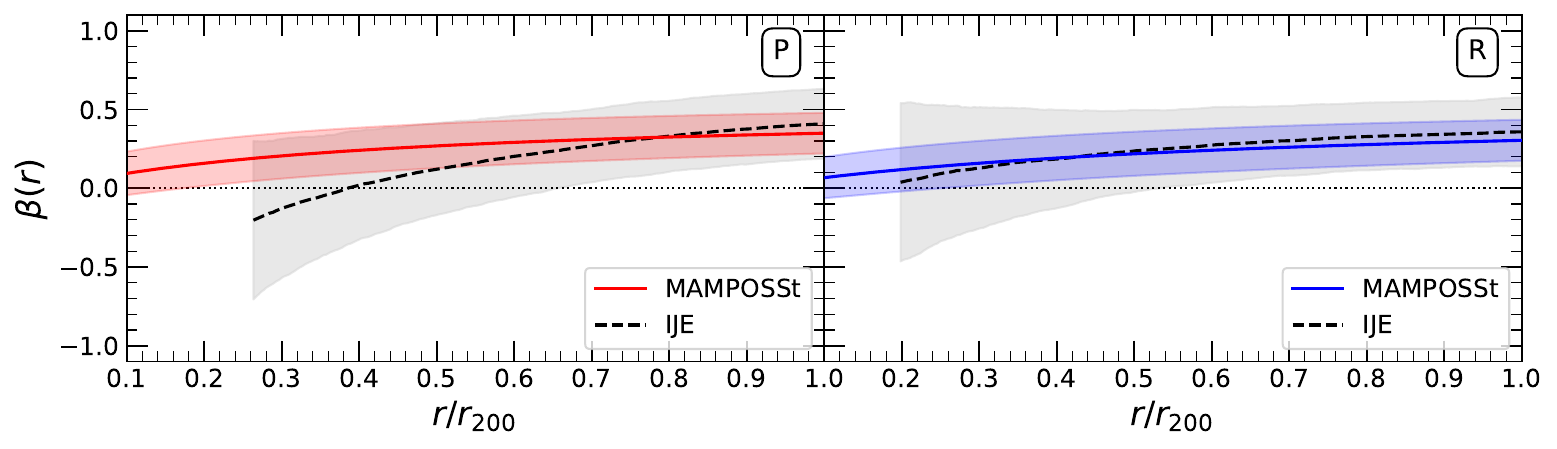}
        \caption{Velocity anisotropy profiles $\beta(r)$ estimated from MAMPOSSt (coloured solid line) and IJE (dashed black line) for the P (left panel) and R (right panel) galaxy populations. The uncertainties on MAMPOSSt and IJE $\beta(r)$ profiles are given by the coloured and grey shaded regions, respectively. The $\beta(r)$ profiles are displayed only in the regions where their uncertainties are lower than 0.5. The horizontal dotted black line indicates $\beta = 0$ (isotropic orbits). Purely radial and tangential orbits correspond to $\beta = 1$ and $\beta = -\infty$, respectively. In each panel, the horizontal axis provides the radial distance normalized by the $\langle r_{200} \rangle$ value given in Section~\ref{sec:Galaxy_populations}.}
        \label{fig:Beta_profiles_All_Gal}
    \end{figure*}

    The $\beta(r)$ profiles of the P (left panel) and R (right panel) galaxy populations are shown in Fig.~\ref{fig:Beta_profiles_All_Gal}. The MAMPOSSt profile is given by the coloured solid line in each panel. The uncertainties on the MAMPOSSt profile, represented by the coloured shaded region, are estimated from bootstraps and calculated using the normalized interquartile range ($\rm{NIQR} = \rm{IQR}/1.349$). The dashed black line in each panel indicates the median $\beta(r)$ profile of the IJE (see Section~\ref{subsec:IJE}), while the grey shaded region depicts its uncertainties, which also are estimated from bootstraps and calculated using the NIQR. The uncertainties on the $\beta(r)$ profiles estimated tend to be higher the lower the number of tracers. In our case, this effect is more pronounced on the IJE $\beta(r)$ profiles at the very internal regions, where the information available is poorer making our results less reliable. Thus, we only show the $\beta(r)$ profiles in Fig.~\ref{fig:Beta_profiles_All_Gal} for the regions where the uncertainties are lower than 0.5. As pointed out by \citetalias{Valk_2025_MNRAS}, this cut-off limit was arbitrarily selected to exclude poorly constrained regions while still ensuring a sufficiently large spatial coverage of the $\beta(r)$ profiles.

    The MAMPOSSt solution for the $\beta(r)$ profile indicates that the P population is characterized by more isotropic orbits in the inner regions ($\beta = 0.09 \pm 0.14$ at $r/r_{200} = 0.1$), that become more radial at larger cluster-centric distances, reaching $\beta = 0.35 \pm 0.13$ at $r/r_{200} = 1.0$. The IJE $\beta(r)$ profile of the P population suggests slightly tangential orbits at the inner regions ($\beta = -0.20 \pm 0.50$ at $r/r_{200} = 0.26$), which become increasingly radial towards the cluster outskirts ($\beta = 0.41 \pm 0.22$ at $r/r_{200} = 1.0$). The R galaxy population is, according to the MAMPOSSt solution, characterized by inner isotropic orbits ($\beta = 0.07 \pm 0.13$ at $r/r_{200} = 0.1$) that become more radial at the cluster outskirts ($\beta = 0.30 \pm 0.13$ at $r/r_{200} = 1.0$). The same orbital behaviour is also suggested by the IJE solution. In fact, the IJE $\beta(r)$ profile of the R population is nearly a copy of the one obtained with MAMPOSSt. In addition, it is important to mention that the MAMPOSSt and IJE solutions, for both populations, are compatible within the uncertainties and indicate inner isotropic orbits that become more radial with increasing cluster-centric distance. Therefore, there are no significant differences between the P and R galaxies in terms of their orbital profiles. Finally, as no significant improvement was provided by the IJE solution and since its requires considerable computational time to compute the median $\beta(r)$ profile along with its associated uncertainties, we will restrict all subsequent orbital analysis in this work to the MAMPOSSt $\beta(r)$ profile only.  
    
\subsection{Projected Velocity Dispersion Profiles}
\label{subsec:SigmaP_profiles_All_Gal}

\begin{figure*}
        \centering
        \includegraphics[width = 0.8\linewidth]{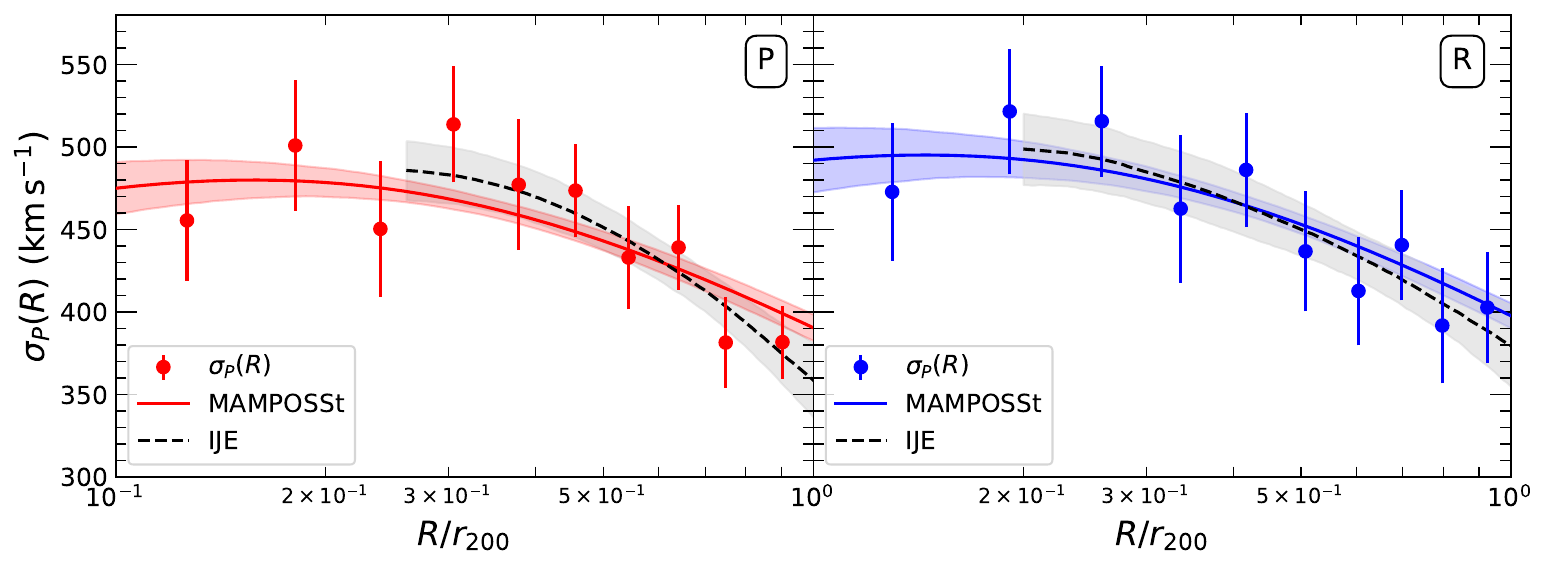}
        \caption{Observed LOS velocity dispersion profiles $\sigma_P(R)$ of the P (red dots, left panel) and R (blue dots, right panel) galaxy populations. The MAMPOSSt (coloured solid line) and IJE (dashed black line) $\sigma_P(R)$ profiles are shown (see text for details), with their uncertainties represented by coloured and grey shaded regions, respectively. In each panel, the horizontal axis provides the projected radial distance normalized by the $\langle r_{200} \rangle$ value given in Section~\ref{sec:Galaxy_populations}, while the horizontal values indicate the central value of each bin.} 
        \label{fig:SigmaP_profiles_All_Gal}
    \end{figure*}

    The observed $\sigma_P(R)$ profiles of the P (red dots, left panel) and R (blue dots, right panel) galaxy populations, estimated using bins containing 10 per cent of the respective population sample (except for the last bin), are shown in Fig.~\ref{fig:SigmaP_profiles_All_Gal}. We observe that P and R galaxies exhibit similar velocity dispersions at all radii, and, given the uncertainties, no significant difference is observed between the $\sigma_P(R)$ profiles of the two populations.
    
    In order to assess the reliability of the MAMPOSSt solution, we estimate the $\sigma_P(R)$ of each galaxy population using the $M(r)$ and $\beta(r)$ profiles obtained by MAMPOSSt. This profile is represented by the coloured solid line in each panel of Fig.~\ref{fig:SigmaP_profiles_All_Gal}, while its uncertainties, estimated from bootstraps, are given by the coloured shaded regions. We also show in Fig.\ref{fig:SigmaP_profiles_All_Gal}, the median $\sigma_P(R)$ profile from the IJE (dashed black line) for each population, estimated from the $\sigma_P(R)$ profiles used to derive the respective IJE $\beta(r)$ profile, along with its uncertainties, represented by the grey shaded regions. The MAMPOSSt and IJE $\sigma_P(R)$ profiles are shown only in the regions where the uncertainties in their respective $\beta(r)$ profiles are lower than 0.5 (see Section~\ref{subsec:Beta_profiles_All_Gal}).

    Analysis of Fig.~\ref{fig:SigmaP_profiles_All_Gal} reveals that the MAMPOSSt and IJE $\sigma_P(R)$ profiles accurately reproduce the observed $\sigma_P(R)$ profiles of both the P and R galaxy populations. These results are indicative that the galaxies in both populations are close to equilibrium within their clusters (see e.g. \citetalias{Valk_2025_MNRAS}). Otherwise, MAMPOSSt would not be able to reproduce their observed $\sigma_P(R)$ profiles using equilibrium solutions for the $M(r)$ and $\beta(r)$ profiles.

\subsection{Projected Phase-Space Distributions}
\label{subsec:PPS_Dist_All_Gal}

\begin{figure*}
        \centering
        \includegraphics[width = \linewidth]{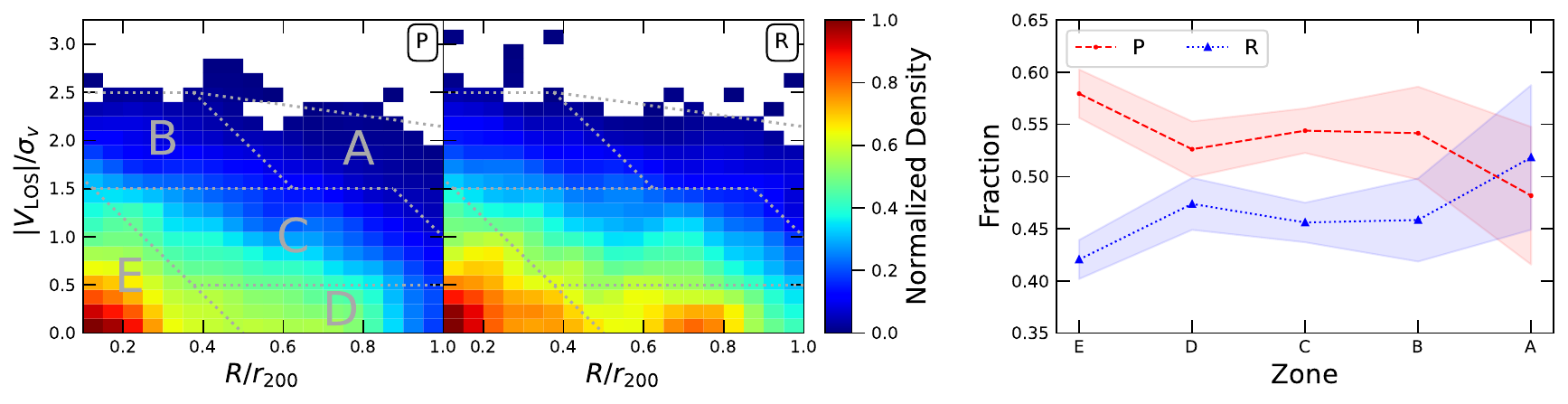}
        \caption{Normalized density distributions of P (left panel) and R (middle panel) galaxies in the PPS. The Rhee zones are represented by the dotted grey lines (see Section~\ref{subsec:PPS}). The right panel exhibits the fractions of P (red dots) and R (blue triangles) galaxies relative to the total number of galaxies in each of the Rhee zones, along with their respective uncertainties represented by the coloured shaded regions.} 
        \label{fig:PPS_Dist_All_Gal}
    \end{figure*}

    The normalized density distributions of P (left panel) and R (middle panel) galaxies in the PPS, estimated following the procedure outlined in Section~\ref{subsec:PPS}, are shown in Fig.~\ref{fig:PPS_Dist_All_Gal}. The PPS was divided into 2D bins of $0.15|V_{\rm{LOS}}|/\sigma_v \times 0.05R/r_{200}$ to ensure a good sampling of all Rhee zones (see Section~\ref{subsec:PPS}), which are represented by the dotted grey lines in Fig.~\ref{fig:PPS_Dist_All_Gal}. The percentages of galaxies in each population located in each of the five zones, along with their respective uncertainties, are presented in Table~\ref{tab:Frac_Rhee_All_Gal}. Galaxies located outside the Rhee zones are not considered in these estimates. It should be noted that the five Rhee regions cover very different areas of the PPS, and therefore the same density in the PPS would lead to different percentages of galaxies in each region. In addition, to complement the analysis of the PPS distributions, we also show in Fig.~\ref{fig:PPS_Dist_All_Gal} the fractions of P (red dots) and R (blue triangles) galaxies relative to the total number of galaxies in each zone (right panel), along with their respective uncertainties, given by the coloured shaded regions. 

    The highest densities (red/orange colours) of P galaxies occur in the E zone (dominated by galaxies with ancient infall times). Nevertheless, the C and D zones (dominated by galaxies with recent/intermediate infall times, respectively) also exhibit densities significantly higher than those observed in the A and B zones. In fact, according to the results on Table~\ref{tab:Frac_Rhee_All_Gal}, the overwhelming majority of P galaxies are located in the C, D, and E zones. The B zone contains only $7.9\pm0.5$ per cent of the P population, while $2.7\pm0.3$ per cent are located in the A zone. On the other hand, we observe that the R population displays two high-density envelopes in the PPS. The first is located in the same zone as that of the P population, but extends to larger radial distances ($R/r_{200} \sim 0.4$) than in the P case. The second density peak, which is not observed for the P population, occurs in the D zone (dominated by galaxies with intermediate infall time). Furthermore, although this peak is most pronounced at small radial distances and low velocities, the high densities (cyan/green colours) also extend to larger radial distances ($R/r_{200} \sim 1.0$) and higher velocities ($|V_{\rm{LOS}}|/\sigma_v \sim 1.0$). Despite the differences in their PPS distributions, R galaxies are distributed similarly to P galaxies along the Rhee zones (see Table~\ref{tab:Frac_Rhee_All_Gal}): the vast majority of R galaxies are located in the C-E zones, with only $11.7\pm1.0$ per cent populating the A-B zones. However, it is worth noting that while the fraction of P galaxies tends to decrease as we move from the E zone (dominated by galaxies with ancient infall times) towards the A zone (dominated by galaxies that have not yet fallen into the cluster), the fraction of R galaxies increases, as shown in the right panel of Fig.~\ref{fig:PPS_Dist_All_Gal}. Moreover, the otherwise smooth decrease/increase in the P/R fractions is interrupted in the D zone, likely due to the higher densities observed for the R population in this region.
    
    In this section, we compared the kinematical and dynamical properties of P and R galaxies, as well as their distributions in the PPS. We found that both populations exhibit similar $I(R)$ profiles and that, according to the $r_\nu$ parameter, they are predominantly located near the central regions of their clusters, with P galaxies being even more centrally concentrated than R galaxies. This difference becomes more apparent when analysing their PPS distributions. The highest density of P galaxies is located in the bottom-left region of the PPS diagram, which corresponds to galaxies with small cluster-centric distances and low peculiar velocities. According to \citetalias{Rhee_2017_ApJ}, this region is dominated by galaxies with ancient infall times ($6.45 < T_{\rm{inf}} < 13.7$\, Gyr). Similarly, the R population also exhibits a high concentration of galaxies in the E zone, but in this case, it extends to larger radial distances compared to the P population. Moreover, the R population displays an additional concentration of galaxies at large cluster-centric distances ($R/r_{200} \sim 0.6-0.8$) and low peculiar velocities ($|V_{\rm{LOS}}|/\sigma_v < 0.5$), located in the region dominated by galaxies with intermediate infall times ($3.63 < T_{\rm{inf}} < 6.45$\, Gyr). This second peak is not observed in the distribution of P galaxies and indicates a distinct dynamical component within the R population. When analysing the global distributions of P and R galaxies across the PPS, both populations appear similarly distributed: the vast majority of galaxies are located in the C–E zones, with only a small fraction falling into the A–B zones. Taken together, these results indicate that both P and R populations are predominantly composed of galaxies that entered the cluster environment at early times. However, only the R population shows a significant concentration in the region associated with intermediate infall times, a feature not observed in the distribution of P galaxies. Finally, it is also worth noting that, despite the differences in their spatial distributions in the PPS, P and R galaxies show comparable velocity dispersions and exhibit similar orbital profiles. 

    To better understand the origin of these differences between P and R galaxies, it is essential to examine the physical properties of the galaxies in each population. This will allow us to investigate the properties of the R galaxies associated with the high concentration observed in the region of intermediate infall times, and why such galaxies are absent from the P population. Additionally, it will help determine whether P and R galaxies with comparable physical properties also share similar kinematic and orbital characteristics, as observed for their respective populations as a whole. We will address these issues in the next section.

    \begin{table}
    	\centering
    	\caption{Percentage of P and R galaxies located in each Rhee zone, estimated relative to the total number of galaxies in each respective population. Galaxies located outside the Rhee zones are excluded from the computations.}
    	\label{tab:Frac_Rhee_All_Gal}
    	\begin{tabular}{ccccccc} % four columns, alignment for each
    		\hline
    		  Zone & P & R\\
    		\hline
                A & $\phantom{1}2.7 \pm 0.3$ & $\phantom{1}3.6  \pm 0.4$ \\	
                B & $\phantom{1}7.9  \pm 0.5$ & $\phantom{1}8.1 \pm 0.6$ \\ 
                C & $34.6 \pm 1.3$ & $35.4 \pm 1.4$ \\
                D & $20.5 \pm 0.9$ & $22.5 \pm 1.1$ \\
                E & $34.3 \pm 1.3$ & $30.4 \pm 1.3$ \\        
                \hline
    	\end{tabular}
    \end{table}
    
%%%%%%%%%%%%%%%%%%%%%%%%
%%%%%%%%%%%%%%%%%%%%%%%%

%%%%%%%%%%%%%%%%%%%%%%%%
%%%%%%%%%%%%%%%%%%%%%%%%

\section{Trends with stellar mass, stellar population age and morphology}
\label{sec:Galaxy_parameters}

In this section, we investigate the kinematical and dynamical properties, as well as the PPS distributions, of P and R galaxies characterized by similar stellar masses, $D_n4000$ values, and morphologies, with the aim of better understanding the differences observed between the P and R galaxy populations in Section~\ref{sec:Galaxy_populations}. To this end, we selected from our galaxy sample only those objects with available stellar mass, $D_n4000$ index, and morphological information from the databases considered in this work (Section~\ref{sec:Data}). As a result, we lost 86 galaxies (9 lacking $D_n4000$ measurements and 77 lacking morphological classifications), leading to a final sample of 2\,860 P and 2\,348 R galaxies with $0.1 \leq R/r_{200} \leq 1.0$. 

%%%%%%%%%%%%%%%%%%%%%%%%%%%%%%%%%%%%%%%%%%%%%%%%%%%%%%%%%%%%%%%%%%%%%%%%%%%%%%
%%%%%%%%%%%%%%%%%%%%%%%%%%%%%%%%%%%%%%%%%%%%%%%%%%%%%%%%%%%%%%%%%%%%%%%%%%%%%%
%%%%%%%%%%%%%%%%%%%%%%%%%%%%%%%%%%%%%%%%%%%%%%%%%%%%%%%%%%%%%%%%%%%%%%%%%%%%%%

\subsection{Stellar mass}
\label{subsection_stellar_mass}
%%%%%%%%%%%%%%%%%%%%%%%%%%%%%%%%%%%%%%%%%%%%%%%%%%%%%%%%%%%%%%%%%%%%%%%%%%%%%%
%%%%%%%%%%%%%%%%%%%%%%%%%%%%%%%%%%%%%%%%%%%%%%%%%%%%%%%%%%%%%%%%%%%%%%%%%%%%%%
%%%%%%%%%%%%%%%%%%%%%%%%%%%%%%%%%%%%%%%%%%%%%%%%%%%%%%%%%%%%%%%%%%%%%%%%%%%%%%

\begin{figure}
    \centering
    \includegraphics[width = \linewidth]{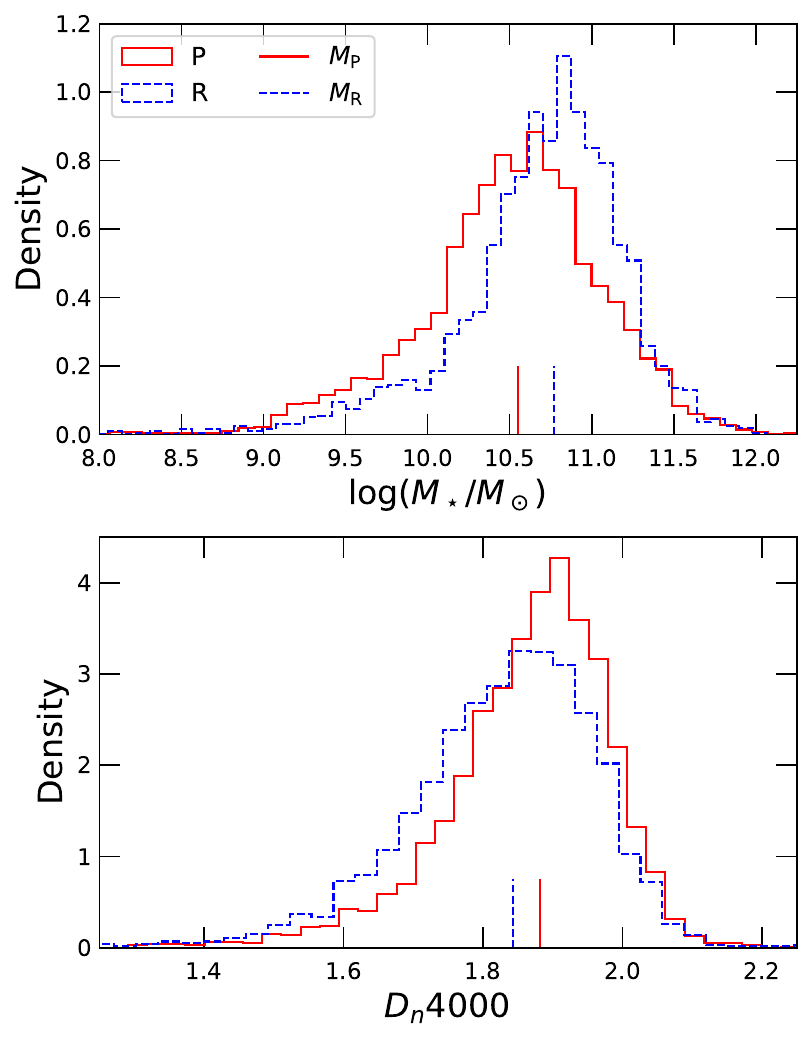}
    \caption{Distributions of stellar mass (upper panel) and $D_n4000$ (bottom panel) for the P (solid red line) and R (dashed blue line) galaxy populations. The median values of each parameter are indicated by the solid red (P) and dashed blue (R) vertical lines.}
    \label{fig:Hist_StellarMass_Dn4000_Dist}
\end{figure}

The stellar mass distributions of the P (solid red line) and R (dashed blue line) galaxy populations are shown in the upper panel of Fig.~\ref{fig:Hist_StellarMass_Dn4000_Dist}. Galaxies in the P population are less massive than those in the R population, with a Kolmogorov-Smirnov (KS) test confirming that the two distributions do not originate from the same parent distribution at a confidence level of $\sim 13\sigma$. The median stellar masses of the P and R galaxies, along with their uncertainties, estimated using the 5th and 95th percentiles, are $\log M_\star = \uncertainties{10.6}{1.1}{0.8}$ and $\log M_\star = \uncertainties{10.8}{1.1}{0.6}$, respectively. 

To disentangle the effect of stellar mass on the differences observed between the P and R populations in Section~\ref{sec:Galaxy_populations}, we separate the galaxies into two bins of stellar mass, namely: low-mass galaxies (LM$_\star$, $\log M_\star \leq 10.5$) and high-mass galaxies (HM$_\star$, $\log M_\star > 10.5$). This threshold roughly corresponds to the central value of the stellar mass distribution. The number of P and R galaxies in each sub-sample is: $1\,320 \,\, \rm{P}_{\rm{LM}_\star}$, $1\,540 \,\, \rm{P}_{\rm{HM}_\star}$, $663 \,\, \rm{R}_{\rm{LM}_\star}$, and $1\,685 \,\, \rm{R}_{\rm{HM}_\star}$, respectively.

We estimate and fit the $I(R)$ profiles for each of the P and R stellar mass subsamples. The best-fitting values of $r_\nu$ and $\nu_0$, along with their respective uncertainties, and the $\chi^2$ of the fit are presented in Table~\ref{tab:table_I_fit_Subsamples}. We note that the general trend of P galaxies being more concentrated than R galaxies (Section~\ref{subsec:I_profiles_All_Gal}) is also recovered when the populations are split into bins of similar stellar mass. However, we observe that although high-mass P galaxies appear to be more concentrated than their low-mass counterparts, this difference is not observed for the R population, given the uncertainties.

\begin{table}
    \centering
     \caption{Best-fitting values of $r_\nu$ and $\nu_0$, along with their uncertainties, and the $\chi^2$ of the fit, for the $I(R)$ profiles of the low-mass (LM$_\star$), high-mass (HM$_\star$), low-$D_n4000$ (LD$_n$), high-$D_n4000$ (HD$_n$), elliptical (E), lenticular (S0), and spiral (S) galaxy subsamples from both the P and R populations. The last column shows the MAMPOSSt results for the $\log r_\beta$ parameter of the $\beta(r)$ profile for each subsample. The central value corresponds to the 50th percentile of its marginal distribution, and the uncertainties are given by the 5th and 95th percentiles.}
    \label{tab:table_I_fit_Subsamples}
    \begin{tabular}{lccccccc} % four columns, alignment for each
        \hline
          Sub. & $r_\nu$ (kpc) & $\nu_0$ (kpc$^{-3}$) & $\chi^2$ & $\log r_\beta$ (kpc) \\
        \hline
        $\rm{P}_{\rm{LM}_\star}$ & $ 220 \pm 23$ & $ 101\pm 6\phantom{0}$ & $0.57$ & $\uncertainties{1.6}{1.3}{1.6}$  \\
        $\rm{P}_{\rm{HM}_\star}$ & $ 177 \pm 27 $ & $104 \pm 8\phantom{0}$ & $1.24$ & $\uncertainties{3.1}{1.3}{0.7}$ \\
        $\rm{R}_{\rm{LM}_\star}$ & $ 295 \pm 69$ & $\phantom{0} 58 \pm 9\phantom{0}$ & $1.44$ & $\uncertainties{1.9}{1.5}{1.4}$ \\
        $\rm{R}_{\rm{HM}_\star}$ & $ 285 \pm 39 $ & $144 \pm 13$ & $1.36$ & $\uncertainties{3.2}{2.0}{0.7}$ \\
        \hline
        $\rm{P}_{\rm{LD}_n}$ & $  261\pm 25$ & $100 \pm 6\phantom{0}$ & $0.42$ & $\uncertainties{1.8}{1.4}{1.5}$ \\
        $\rm{P}_{\rm{HD}_n}$ & $ 152 \pm 16$ & $105 \pm 6\phantom{0}$ & $0.60$ & $\uncertainties{3.0}{1.6}{0.8}$ \\
        $\rm{R}_{\rm{LD}_n}$ & $ 363 \pm 48 $ & $131 \pm 13$ & $0.95$ & $\uncertainties{1.9}{1.6}{1.5}$ \\
        $\rm{R}_{\rm{HD}_n}$ & $  203\pm 33$ & $\phantom{0} 73 \pm 7\phantom{0}$ & $0.98$ & $\uncertainties{3.1}{2.0}{0.8}$ \\
        \midrule
        $\rm{P}_{\rm{E}}$ & $ 212 \pm 38$ & $\phantom{0} 58 \pm 6 \phantom{0}$ & $0.80$ & $\uncertainties{3.3}{2.0}{0.6}$ \\
        $\rm{P}_{\rm{S0}}$ & $ 135 \pm 32$ & $\phantom{0} 46 \pm 5 \phantom{0}$ & $1.20$ & $\uncertainties{2.3}{1.9}{1.4}$ \\
        $\rm{P}_{\rm{S}}$ & $ 302 \pm 51$ & $\phantom{0} 64 \pm 8 \phantom{0}$ & $0.82$ &  $\uncertainties{1.5}{1.2}{1.7}$ \\
        $\rm{R}_{\rm{E}}$ & $ 164 \pm36 $ & $\phantom{0} 34 \pm 4 \phantom{0}$ & $0.89$ & $\uncertainties{2.2}{1.8}{1.4}$ \\
        $\rm{R}_{\rm{S0}}$ & $ 279 \pm49 $ & $\phantom{0} 43 \pm 5 \phantom{0}$ & $0.68$ & $\uncertainties{2.8}{2.2}{1.1}$ \\
        $\rm{R}_{\rm{S}}$ & $ 458 \pm 73$ & $\phantom{0} 98 \pm 13$ & $0.88$ & $\uncertainties{1.5}{1.3}{1.7}$ \\
        \bottomrule
    \end{tabular}
\end{table}

Next, we execute MAMPOSSt in the \textit{Split} mode considering each one of the P and R stellar mass subsamples as independent tracers of the cluster potential. We provided to MAMPOSSt the $r_\nu$ values from Table~\ref{tab:table_I_fit_Subsamples}, and also the redshift of the ensemble cluster given in Section~\ref{sec:Galaxy_populations}. MAMPOSSt results for the $\beta(r)$ profile of each subsample are presented in Table~\ref{tab:table_I_fit_Subsamples}. The central value of each parameter corresponds to the 50th percentile of its marginal distribution, with uncertainties represented by the 5th and 95th percentiles.

\begin{figure*}
    \centering
    \includegraphics[width = 0.8\linewidth]{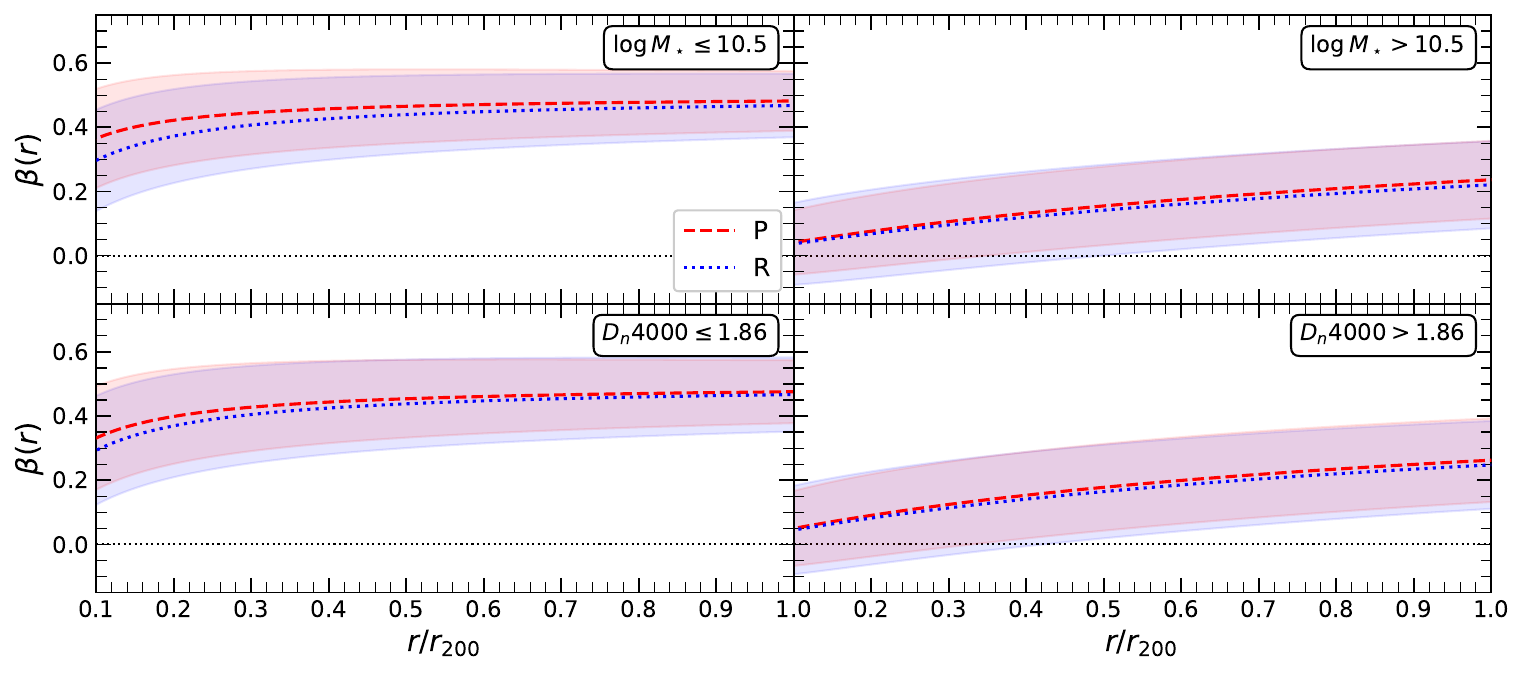}
    \caption{Velocity anisotropy profiles $\beta(r)$ estimated from MAMPOSSt for the low-mass (upper-left panel), high-mass (upper-right panel), low-$D_n4000$ (bottom-left panel), and high-$D_n4000$ (bottom-right panel) galaxy subsamples from both the P (dashed red line) and R (dotted blue line) populations, along with their uncertainties, represented by the respective coloured shaded regions. The horizontal dotted black line indicates $\beta = 0$ (isotropic orbits). Purely radial and tangential orbits correspond to $\beta = 1$ and $\beta = -\infty$, respectively. In each panel, the horizontal axis provides the radial distance normalized by the $\langle r_{200} \rangle$ value given in Section~\ref{sec:Galaxy_populations}.} 
    \label{fig:Beta_profiles_StellarMass_Dn4000}
\end{figure*}

The velocity anisotropy profiles for the low-mass (upper-left panel) and high-mass (upper-right panel) galaxy subsamples of both P (dashed red line) and R (dotted blue line) populations are displayed in Fig.~\ref{fig:Beta_profiles_StellarMass_Dn4000}. The uncertainties in the $\beta(r)$ profiles are represented by the coloured shaded regions. Analysing Fig.~\ref{fig:Beta_profiles_StellarMass_Dn4000}, we realize that P and R galaxies of comparable stellar mass display very similar $\beta(r)$ profiles, implying that there is no significant difference between the orbits of the galaxies in both populations. Nevertheless, we observe that low-mass galaxies of both populations exhibit significantly more radial orbits at all radii than their high-mass counterparts. In fact, while high-mass galaxies show nearly isotropic orbits in the inner regions ($\beta \sim 0.04 $ at $r/r_{200} = 0.1$) that become modestly radial at larger distances ($\beta \sim 0.2 $ at $r/r_{200} = 1.0$), low-mass galaxies display consistently radial orbits, with $\beta$ increasing from $\sim 0.3$ at $r/r_{200} = 0.1$ to $\sim 0.5 $ at $r/r_{200} = 1.0$.

\begin{figure*}
    \centering
    \includegraphics[width = 0.85\linewidth]{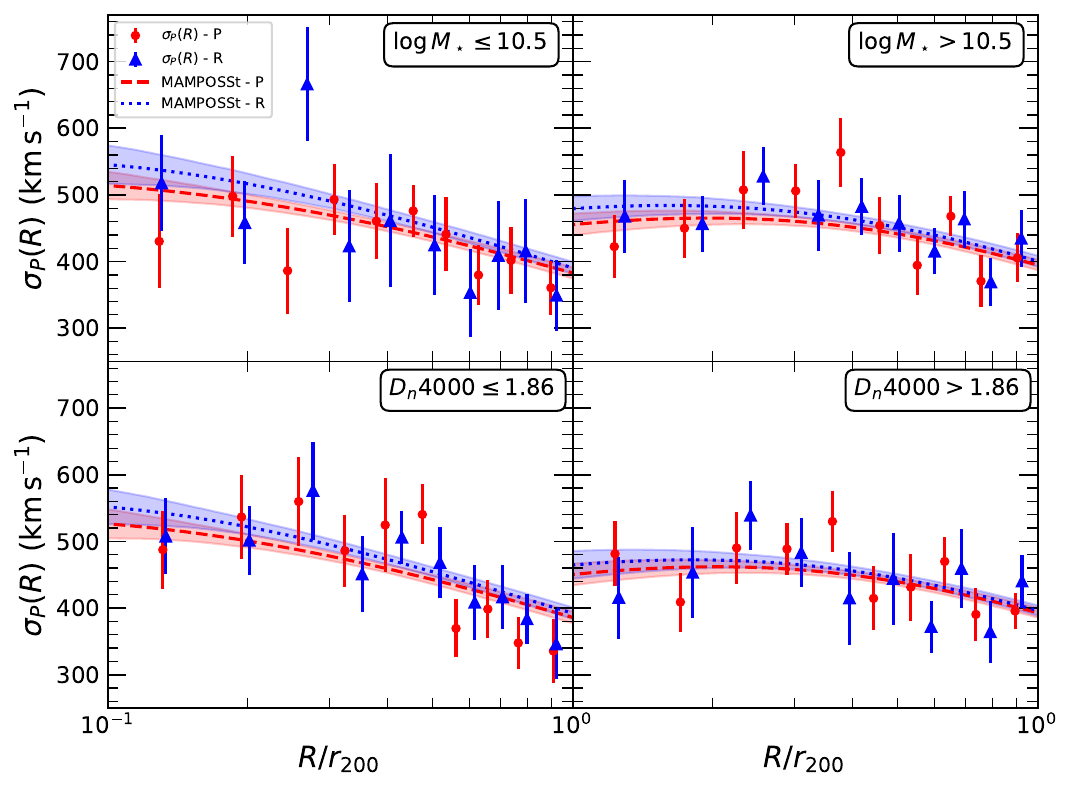}
    \caption{Observed LOS velocity dispersion profiles $\sigma_P(R)$ of the low-mass (upper-left panel), high-mass (upper-right panel), low-$D_n4000$ (bottom-left panel), and high-$D_n4000$ (bottom-right panel) galaxy subsamples from both the P (red dots) and R (blue triangles) populations. The MAMPOSSt $\sigma_P(R)$ profile of each P (dashed red lines) and R (dotted blue lines) subsamples are displayed, along with their uncertainties, represented by the respective coloured shaded regions. In each panel, the horizontal axis provides the projected radial distance normalized by the $\langle r_{200} \rangle$ value given in Section~\ref{sec:Galaxy_populations}, while the horizontal values indicate the central value of each bin.} 
    \label{fig:SigmaP_profiles_StellarMass_Dn4000}
\end{figure*}

In addition, we show in Fig.~\ref{fig:SigmaP_profiles_StellarMass_Dn4000} the $\sigma_P(R)$ profiles of the low-mass (upper-left panel) and high-mass (upper-right panel) galaxy subsamples of both P (red dots) and R (blue triangles) populations. Note that, considering the uncertainties, there is no significant difference between the $\sigma_P(R)$ profiles of the P and R subsamples of similar stellar mass, nor between the low- and high-mass subsamples within each population. To complement the discussion, the $\sigma_P(R)$ profiles obtained through the MAMPOSSt solutions for the P (dashed red line) and R (dotted blue line) stellar mass subsamples are displayed in Fig.~\ref{fig:SigmaP_profiles_StellarMass_Dn4000}, along with their respective uncertainties, given by the coloured shaded regions. In all cases, the MAMPOSSt $\sigma_P(R)$ profile reproduces the observed $\sigma_P(R)$ profile accurately, within the uncertainties. Additionally, although no significant difference was found between the observed $\sigma_P(R)$ profiles of the P and R stellar mass subsamples, the MAMPOSSt solution indicates slightly higher velocity dispersions at all radii for R galaxies compared to P galaxies. Furthermore, the MAMPOSSt $\sigma_P(R)$ profile of the low-mass subsample of both populations shows a steeper inner slope than that of the high-mass subsample, likely due to the more radial orbits observed for these systems (Fig.~\ref{fig:Beta_profiles_StellarMass_Dn4000}).

\begin{figure*}
    \centering
    \includegraphics[width = 0.9\linewidth]{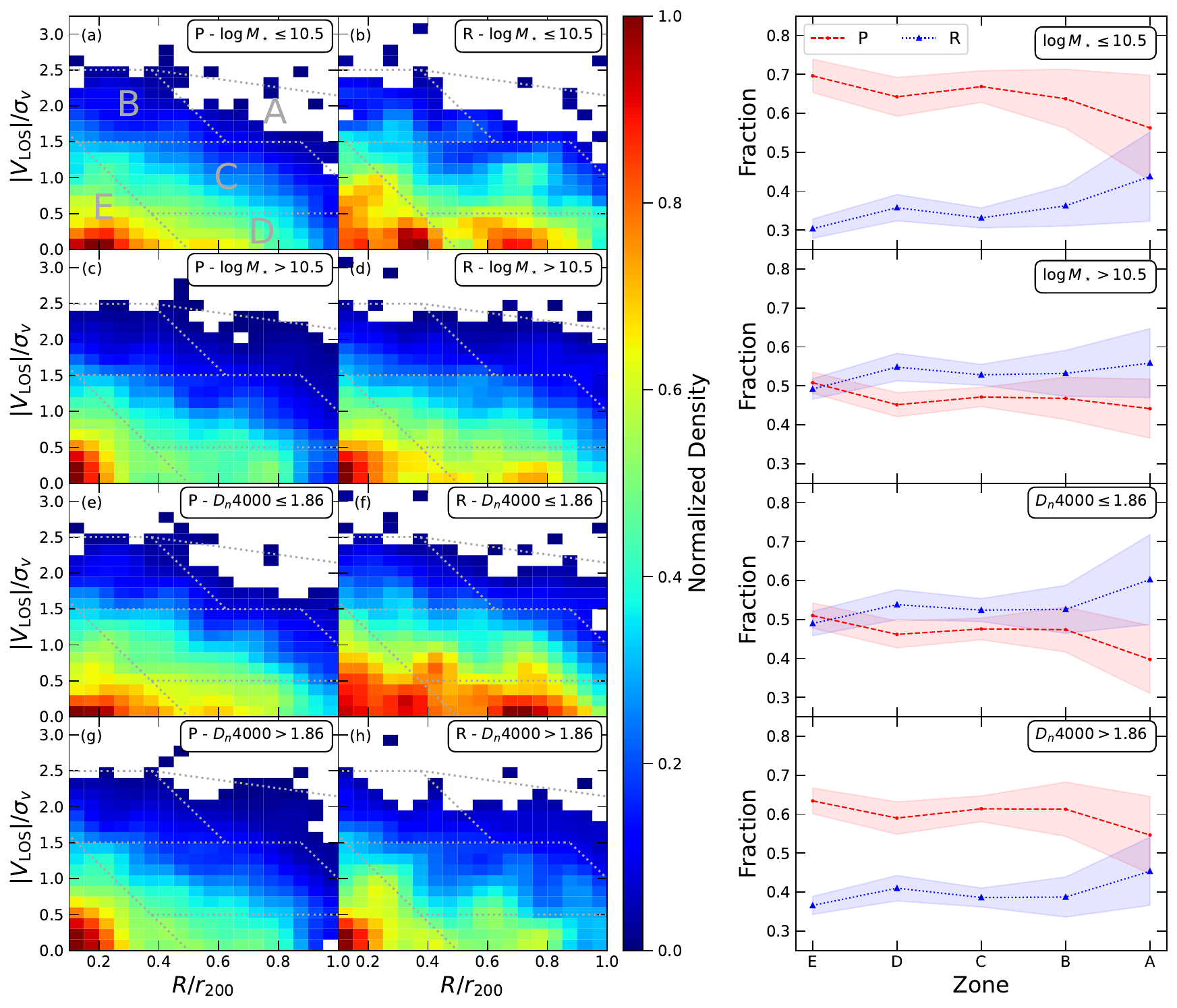}
    \caption{Normalized density distributions of low-mass (panels a and b), high-mass (panels c and d), low-$D_n4000$ (panels e and f), and high-$D_n4000$ (panels g and h) galaxy subsamples from both the P (left panels) and R (middle panels) populations. The Rhee zones are represented by the dotted grey lines (see Section~\ref{subsec:PPS}). The right panels show the fractions of P (red dots) and R (blue triangles) galaxies in each subsample, relative to the total number of galaxies in the corresponding subsample within each Rhee zone, along with their respective uncertainties, represented by the coloured shaded regions.} 
    \label{fig:PPS_Dist_StellarMass_Dn4000}
\end{figure*}

Finally, the normalized density distributions for the low-mass (panels a and b) and high-mass (panels c and d) galaxy subsamples of both P (left panels) and R (middle panels) populations are shown in Fig.~\ref{fig:PPS_Dist_StellarMass_Dn4000}. The percentage of galaxies located in each of the Rhee zones, with their respective uncertainties, for each subsample are displayed in Table~\ref{tab:Frac_Rhee_Subsamples}. Furthermore, we also present in Fig.~\ref{fig:PPS_Dist_StellarMass_Dn4000} the fractions of P (red dots) and R (blue triangles) galaxies of each subsample, relative to the total number of galaxies in that subsample, within each Rhee zone (right panels), along with their respective uncertainties, represented by the coloured shaded regions. Passive galaxies of both low and high stellar mass exhibit similar distributions in the PPS and show the highest densities in the E zone, although slightly elevated densities are also observed in the D zone for the low-mass subsample. In addition, when comparing the PPS distributions of P and R galaxies with similar stellar mass, we find that their differences are essentially the same as those reported in Section~\ref{subsec:PPS_Dist_All_Gal} for their complete populations. Nevertheless, Fig.~\ref{fig:PPS_Dist_StellarMass_Dn4000} clearly shows that, while high-mass R galaxies show a density peak in the bottom-left corner of the PPS diagram, the highest densities for low-mass R galaxies occur in the E zone at $R/r_{200} \sim 0.35$ and in the D zone. Consequently, these low-mass R galaxies appear to be the primary drivers of the differences observed between the overall P and R galaxy populations. Lastly, analysing the right panels on Fig.~\ref{fig:PPS_Dist_StellarMass_Dn4000}, we note that the overall trend of decreasing/increasing fractions of P/R galaxies across the Rhee zones is preserved when the populations are split into bins of comparable stellar mass, although this tendency is more evident for the low-mass subsample. 

\begin{table}
    \centering
    \caption{Percentage of low-mass ($\rm{LM}_\star$), high-mass ($\rm{HM}_\star$), low-$D_n4000$ ($\rm{LD}_n$), high-$D_n4000$ ($\rm{HD}_n$), elliptical (E), lenticular (S0), and spiral (S) galaxies of each population (P and R) located in each Rhee zone, estimated relative to the total number of galaxies in each respective subsample. Galaxies located outside the Rhee zones are excluded from the computations.}
    \label{tab:Frac_Rhee_Subsamples}
    \begin{tabular}{lccccc} % four columns, alignment for each
        \hline
          Sub. & E & D & C & B & A \\
        \hline
            $\rm{P}_{\rm{LM}_\star}$ & $ 34.0 \pm 1.9$ & $20.8 \pm 1.4$ & $34.3 \pm 1.9$ & $\phantom{0}8.8 \pm 0.9$ & $2.1 \pm 0.4$ \\  
            $\rm{P}_{\rm{HM}_\star}$ & $ 34.6\pm 1.7$ & $20.1 \pm 1.3$ & $35.1 \pm 1.8$ & $\phantom{0}7.1 \pm 0.7$ & $3.2 \pm 0.5$ \\   
            $\rm{R}_{\rm{LM}_\star}$ & $ 29.7 \pm 2.4$ & $23.1 \pm 2.1$ & $ 33.9 \pm 2.6$ & $10.0 \pm 1.3$ & $3.2 \pm 0.7$ \\     
            $\rm{R}_{\rm{HM}_\star}$ & $30.6 \pm 1.5$ & $22.3 \pm 1.3$ & $36.0 \pm 1.7$ & $\phantom{0}7.4 \pm 0.7$ & $3.7 \pm 0.5$ \\   
            \hline
            $\rm{P}_{\rm{LD}_n}$ & $31.4 \pm 1.9$ & $21.7 \pm 1.5$ & $36.1 \pm 2.0$ & $8.3 \pm 0.9$ & $2.4 \pm 0.5$ \\  
            $\rm{P}_{\rm{HD}_n}$ & $36.4 \pm 1.7$ & $19.5 \pm 1.2$ & $33.7 \pm 1.7$ & $7.6 \pm 0.7$ & $2.9 \pm 0.4$ \\     
            $\rm{R}_{\rm{LD}_n}$ & $27.9 \pm 1.7$ & $23.4 \pm 1.5$ & $36.8 \pm 2.0$ & $8.6 \pm 0.8$ & $3.4 \pm 0.5$ \\     
            $\rm{R}_{\rm{HD}_n}$ & $33.4 \pm 2.1$ & $21.5 \pm 1.6$ & $33.7 \pm 2.1$ & $7.6 \pm 0.9$ & $3.8 \pm 0.6$ \\                 
            \hline
            $\rm{P}_{\rm{E}}$ & $35.7 \pm 2.5$ & $21.4 \pm 1.8$ & $32.6 \pm 2.4$ & $7.3 \pm 1.0$ & $3.1 \pm 0.6$ \\                 
            $\rm{P}_{\rm{S0}}$ & $38.0 \pm 2.6$ & $19.4 \pm 1.7$ & $33.2 \pm 2.4$ & $6.7 \pm 1.0$ & $2.7 \pm 0.6$ \\                 
            $\rm{P}_{\rm{S}}$ & $28.8 \pm 2.3$ & $22.0 \pm 2.0$ & $38.6 \pm 2.8$ & $8.4 \pm 1.1$ & $2.2 \pm 0.6$ \\                 
            $\rm{R}_{\rm{E}}$ & $35.3 \pm 3.0$ & $18.4 \pm 2.0$ & $34.9 \pm 3.0$ & $8.4 \pm 1.3$ & $2.9 \pm 0.8$ \\                 
            $\rm{R}_{\rm{S0}}$ & $30.6 \pm 2.8$ & $24.0 \pm 2.4$ & $32.3 \pm 2.9$ & $7.7 \pm 1.3$ & $5.4 \pm 1.0$ \\                 
            $\rm{R}_{\rm{S}}$ & $24.8 \pm 1.9$ & $25.0 \pm 2.0$ & $38.9 \pm 2.6$ & $7.8 \pm 1.0$ & $3.5 \pm 0.7$ \\                 
            \hline
    \end{tabular}
\end{table}

\subsection{Mean stellar population ages}
\label{subsection:dn4000}

%%%%%%%%%%%%%%%%%%%%%%%%%%%%%%%%%%%%%%%
%   Results from Dn4000 Subsamples
%%%%%%%%%%%%%%%%%%%%%%%%%%%%%%%%%%%%%%%

In addition to stellar mass, we also compare the properties of P and R systems with comparable $D_n4000$ break indices. The $D_n4000$ index depends on the mean age of the stellar populations, and -- to a lesser extent -- on their metallicities, especially for ages greater than 1 Gyr \citep[e.g.][]{Poggianti_1997_A&A,Kauffmann_2003_MNRAS_Stellar}. At fixed metallicity, older stellar populations display higher $D_n4000$ values compared to younger ones, while metal-poor populations exhibit lower $D_n4000$ values than more metal-rich populations of similar mass. Most of the galaxies in our sample have stellar masses in the range $\sim 10^{9.6} - 10^{11.4} \, M_\odot$, with a median value of $\log M_\star = \uncertainties{10.6}{1.1}{0.7}$. Thus, if we assume that our sample follows the stellar mass-metallicity relation that is found locally \citep[e.g.][]{Gallazzi_2005_MNRAS, Peng_2015_Natur}, we do not expect to observe significant variation in the metallicity of the stellar populations for the galaxies in our sample. We therefore use the $D_n4000$ index as a proxy for the mean stellar age.

The $D_n4000$ distributions of the P (solid red line) and R (dashed blue line) galaxy populations are shown in the bottom panel of Fig.~\ref{fig:Hist_StellarMass_Dn4000_Dist}. As can be seen, P galaxies are characterized by higher $D_n4000$ values compared to R galaxies, a difference confirmed by the KS test at a confidence level of $\sim 10\sigma$. The median $D_n4000$ values for the P and R populations are $\uncertainties{1.88}{0.24}{0.14}$ and $\uncertainties{1.84}{0.26}{0.17}$, respectively. In order to assess what distinguishes P galaxies from R galaxies of comparable mean stellar population ages, we divide the galaxies into two $D_n4000$ bins. The median value of the full population of galaxies (P+R, $\uncertainties{1.86}{0.25}{0.16}$) was adopted as the threshold to separate low-$D_n4000$ galaxies (LD$_n$, $D_n4000 \leq 1.86$) from the high-$D_n4000$ galaxies (HD$_n$, $D_n4000 > 1.86$). The number of galaxies in each sub-sample is: $1206 \,\, \rm{P}_{\rm{LD}_n}$, $1654 \,\, \rm{P}_{\rm{HD}_n}$, $1308 \,\, \rm{R}_{\rm{LD}_n}$, and $1040 \,\, \rm{R}_{\rm{HD}_n}$.

We estimate and fit the $I(R)$ profiles for each of the P and R $D_n4000$ subsamples, and the results obtained are presented in Table~\ref{tab:table_I_fit_Subsamples}. We note that, regardless of the population, galaxies with high $D_n4000$ values -- i.e. those hosting older stellar populations -- are typically located closer to the central regions of their parent clusters compared to galaxies with younger stellar populations. On the other hand, we also observe that, at fixed $D_n4000$, P galaxies are more centrally concentrated than R galaxies.

The MAMPOSSt analysis was performed for the $D_n4000$ P and R subsamples in the same way as for the stellar mass subsamples, and the results obtained for the $\beta(r)$ profile are presented in Table~\ref{tab:table_I_fit_Subsamples}. In Fig.~\ref{fig:Beta_profiles_StellarMass_Dn4000}, we show the $\beta(r)$ profiles obtained through MAMPOSSt for the LD$_n$ (bottom-left panel) and HD$_n$ (bottom-right panel) galaxy subsamples from both the P (dashed red line) and R (dotted blue line) populations, along with their respective uncertainties, represented by the coloured shaded regions. We find that P and R galaxies with comparable $D_n4000$ values display very similar $\beta(r)$ profiles. Consequently, there is no significant difference between the orbits of P and R systems that host stellar populations of comparable mean ages. In addition, we also identify that, regardless of population, galaxies with high-$D_n4000$ values exhibit isotropic inner orbits ($\beta(r)\sim 0 $ at $R/r_{200} = 0.1$) that become increasingly radial towards larger cluster-centric distances ($\beta(r)\sim 0.25 $ at $R/r_{200} = 1.0$), while galaxies with low-$D_n4000$ values display more radial orbits at all radii. Interestingly, this same behaviour is also observed in the high- and low-mass subsamples, respectively.

Additionally, Fig.~\ref{fig:SigmaP_profiles_StellarMass_Dn4000} presents the observed $\sigma_P(R)$ profiles of the LD$_n$ (bottom-left panel) and HD$_n$ (bottom-right panel) galaxy subsamples from both the P (red dots) and R (blue triangles) populations. Taking the uncertainties into account, no significant difference is observed between the $\sigma_P(R)$ profiles of P and R galaxies with similar $D_n4000$ values. Moreover, regardless of population, galaxies with low and high $D_n4000$ values exhibit similar velocity dispersions. We also show in Fig.~\ref{fig:SigmaP_profiles_StellarMass_Dn4000} the $\sigma_P(R)$ profiles derived from MAMPOSSt solution for both the P (dashed red line) and R (dotted blue line) $D_n4000$ subsamples, along with their respective uncertainties, represented by the coloured shaded regions. The observed $\sigma_P(R)$ profiles of all subsamples are well reproduced by MAMPOSSt within the uncertainties. Furthermore, the MAMPOSSt solution suggests that LD$_n$ R galaxies exhibit slightly higher velocity dispersions than their P counterparts, while no significant difference is found for the HD$_n$ galaxies. Lastly, according to the MAMPOSSt results, LD$_n$ galaxies are characterized by higher velocity dispersions in the inner cluster regions ($R/r_{200} \lesssim 0.35$) than their HD$_n$ counterparts.

We present in Fig.~\ref{fig:PPS_Dist_StellarMass_Dn4000} the normalized density distributions for the LD$_n$ (panels e and f) and HD$_n$ (panels g and h) galaxy subsamples from both the P (left panels) and R (middle panels) populations. The fractions of P (red dots) and R (blue triangles) galaxies in each subsample, relative to the total number of galaxies in that subsample, within each Rhee zone, are displayed in Fig.~\ref{fig:PPS_Dist_StellarMass_Dn4000} (right panels), while the percentages of galaxies located in each Rhee zone for each subsample are listed in Table~\ref{tab:Frac_Rhee_Subsamples}. We find that P and R galaxies characterized by high-$D_n4000$ values appear to be similarly distributed across the PPS zones. In fact, HD$_n$ R galaxies exhibit densities that are only marginally higher in the E zone at $R/r_{200} \sim 0.4$, as well as in the D zone, compared to their P counterparts. Nevertheless, significant differences are observed between the distributions of the LD$_n$ subsamples. LD$_n$ P galaxies show the highest densities in the E zone, with only slightly elevated densities observed in the C and D zones. In contrast, LD$_n$ R galaxies display high concentrations in C, D, and E zones, with the highest densities occurring in the E zone and in the D zone at $R/r_{200} \sim 0.7$. Thus, while P and R galaxies with older stellar populations are distributed similarly across the PPS, R galaxies with younger stellar populations exhibit a significantly different distribution from their P counterparts, likely driving the difference found between the full P and R populations in Section~\ref{subsec:PPS_Dist_All_Gal}. Finally, the right panels on Fig.~\ref{fig:PPS_Dist_StellarMass_Dn4000} confirms that the general trend of decreasing/increasing fractions of P/R galaxies from the E zone towards the A zone is preserved when the sample is split into bins of $D_n4000$.

We have found that HM$_\star$/HD$_n$ galaxies of both populations exhibit systematically lower cluster-centric distances and display more isotropic orbits at all radii then their LM$_\star$/LD$_n$ counterparts. In addition, according to the $\sigma_P(R)$ profiles recovered using the MAMPOSSt solution, LM$_\star$/LD$_n$ galaxies display slightly higher velocity dispersions in the inner regions than HM$_\star$/HD$_n$ galaxies. It is worth noting that, for both populations, the stellar mass and $D_n4000$ subsamples yield comparable results. This similarity is expected given that low-mass galaxies tend do exhibit lower $D_n4000$ values -- indicative of younger stellar populations -- whereas high-mass galaxies typically show higher $D_n4000$ values, associated with older stellar populations \citep[see e.g.][]{Kauffmann_2003_MNRAS_The}. In fact, in our sample, 72 (66) per cent of the systems in the low-mass (high-mass) subsample exhibit low-$D_n4000$ (high-$D_n4000$) values. Therefore, although there is a considerable overlap between these samples, they are not equivalent. Furthermore, we highlight that, although one can argue that this similarity arises due to a well-established stellar mass-metallicity relation, implying that metallicity rather than age is the main driver of the observed results, we do not expect to observe a significant variation in the metallicity of the stellar populations between the galaxies in our sample, as previously discussed. Consequently, this implies that the mean stellar age, rather than metallicity, is likely the primary driver of the observed similarities. Additionally, the results obtained for HM$_\star$/HD$_n$ galaxies and their LM$_\star$/LD$_n$ counterparts resemble the trends reported for the well-known effects of mass and colour segregation in galaxy clusters. High-mass, early-type, quiescent galaxies typically reside closer to the cluster centre \citep[e.g.][]{Aguerri_2007_A&A, Biviano_2004_A&A, Munari_2014_A&A}, display lower velocity dispersions \citep[e.g.][]{Adami_1998_A&A, Aguerri_2007_A&A, Cava_2017_A&A}, and are characterized by more isotropic orbits \citep{Biviano_2004_A&A,Munari_2014_A&A,Mamon_2019_A&A,Biviano_2021_A&A} than their low-mass, late-type, star-forming counterparts. These findings have generally been interpreted as the consequence of different accretion times into the cluster for the two populations, with the high-mass, quiescent galaxies being older infallers than the low-mass, star-forming population \citep[see e.g.][]{Lotz_2019_MNRAS}. This interpretation is in good agreement with our results. HM$_\star$/HD$_n$ galaxies of both populations show the highest densities in the zone E, while their LM$_\star$/LD$_n$ counterparts display significantly high densities also in the D zone, indicative that their were typically accreted later into their clusters. Nevertheless, some results deviate from this overall trend \citep[see e.g.][]{Annunziatella_2016_A&A, Aguerri_2017_MNRAS, Mercurio_2021_A&A}.

%%%%%%%%%%%%%%%%%%%%%%%%%%%%%%%%%%%%%%%%%%%%%%%%%%%%
%%%%%%%%%%%%%%%%%%%%%%%%%%%%%%%%%%%%%%%%%%%%%%%%%%%%
%%%%%%%%%%%%%%%%%%%%%%%%%%%%%%%%%%%%%%%%%%%%%%%%%%%%

\subsection{Morphological Type}
\label{subsection:morphology}

%%%%%%%%%%%%%%%%%%%%%%%%%%%%%%%%%%%%%%%%%%%%%%%%%%%%
%%%%%%%%%%%%%%%%%%%%%%%%%%%%%%%%%%%%%%%%%%%%%%%%%%%%
%%%%%%%%%%%%%%%%%%%%%%%%%%%%%%%%%%%%%%%%%%%%%%%%%%%%

The criteria used to select the P and R galaxies analysed in this work are based solely on the primary source of gas ionization in the galaxies, with no constrains imposed on their morphology. As a result, both P and R populations include galaxies from different morphological classes. In order to gain a complete understanding of the similarities and differences between these two populations, it is also important to account for the effect of morphology. In the following analysis, we investigate the kinematical and dynamical behaviours, as well as the PPS distributions, of P and R galaxies as a function of their morphological class. In this way, we aim to examine what differentiates P galaxies from R ones of the same morphological type.

\begin{figure}
        \centering
        \includegraphics[width = 0.95\linewidth]{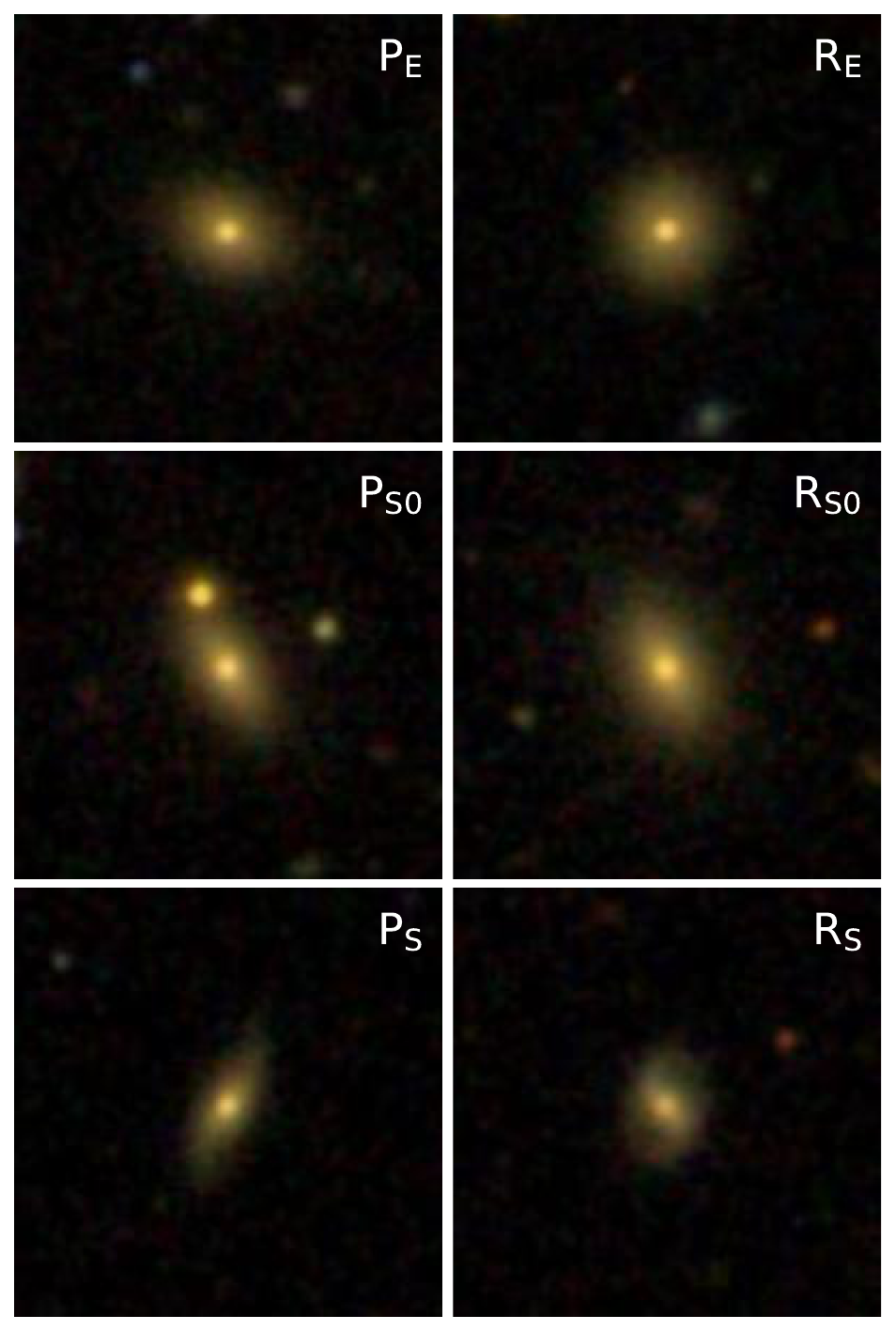}
        \caption{SDSS composite $gri$ images of a subset of galaxies from our sample representative of each morphological class -- elliptical (top row), lenticular (middle row), and spiral (bottom row) -- for the P (left column) and R (right column) populations. Each image has a field of view of $40\times40$ arcsec$^2$, with north up and east to the left.}
        \label{fig:Morpho_Classes}
    \end{figure}

For this task, we use the morphological classification scheme provided by \citet{Huertas-Company_2011_A&A}. In their work, \citet{Huertas-Company_2011_A&A} assigned to each galaxy a probability value of being in the four morphological classes, namely: elliptical (E), lenticular (S0), early-type spiral (Sab), and late-type spiral (Scd). We use these probability values to assign each galaxy in our sample to the morphological class with the highest probability. However, only galaxies with a probability value greater than or equal to $0.5$ of belonging to their assigned class were kept in the sample. We conducted a careful visual inspection of a significant number of randomly selected galaxies from each morphological category and confirmed that this $0.5$ threshold value is needed to ensure that the galaxies classified into each type resemble the expected morphology according to canonical morphological classes. In contrast, when no threshold is imposed and galaxies are assigned solely based on the highest probability, all classes suffer from contamination by objects that deviate from their expected morphology, with the Sab and Scd classes being particularly affected. Due to the imposed threshold, a considerable fraction of galaxies ($\sim21$ per cent) could not be classified into any of the defined morphological classes. The Scd galaxies are the most affected by this constraint, and the remaining number of P and R galaxies classified as Scd prevents any reliable analysis for this class. Thus, to overcome this limitation, we combined galaxies classified as Sab and Scd into a single Spiral (S) class. Finally, the number of P and R galaxies classified into each morphological class is as follows: $775 \, \rm{P}_{\rm{E}}$, $771 \, \rm{P}_{\rm{S0}}$, $696 \, \rm{P}_{\rm{S}}$, $524 \, \rm{R}_{\rm{E}}$, $523 \, \rm{R}_{\rm{S0}}$, and $825 \, \rm{R}_{\rm{S}}$. A representative example of a galaxy from each morphological class is shown in Fig.~\ref{fig:Morpho_Classes} for both the P and R populations.

We estimate and fit the $I(R)$ profiles for each of the morphological classes. The best-fitting values of $r_\nu$ and $\nu_0$, along with their respective uncertainties, and the $\chi^2$ of the fit are listed in Table~\ref{tab:table_I_fit_Subsamples}. Surprisingly, $\rm{P}_{\rm{S0}}$ galaxies are typically found closer to the cluster central regions than $\rm{P}_{\rm{E}}$ galaxies, whereas $\rm{P}_{\rm{S}}$ galaxies tend to inhabit the outermost regions among the P population. With respect to the R population, $\rm{R}_{\rm{E}}$ galaxies appear to be the most centrally concentrated, followed by $\rm{R}_{\rm{S0}}$ and $\rm{R}_{\rm{S}}$ galaxies, respectively. In addition, we find that $\rm{P}_{\rm{S0}}$ and $\rm{P}_{\rm{S}}$ are more centrally located than their R counterparts, while $\rm{P}_{\rm{E}}$ galaxies are as concentrated as $\rm{R}_{\rm{E}}$ galaxies, given the uncertainties. 

\begin{figure*}
        \centering
        \includegraphics[width = 1\linewidth]{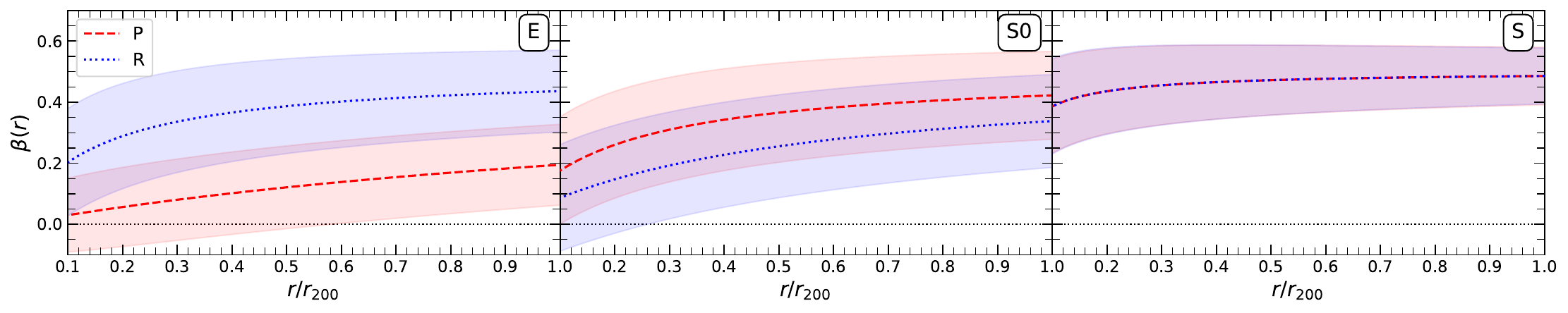}
        \caption{Velocity anisotropy profiles $\beta(r)$ estimated from MAMPOSSt for the E (left panel), S0 (middle panel), and S (right panel) galaxy subsamples from both the P (dashed red line) and R (dotted blue line) populations, along with their uncertainties, represented by the respective coloured shaded regions. The horizontal dotted black line indicates $\beta = 0$ (isotropic orbits). Purely radial and tangential orbits correspond to $\beta = 1$ and $\beta = -\infty$, respectively. In each panel, the horizontal axis provides the radial distance normalized by the $\langle r_{200} \rangle$ value given in Section~\ref{sec:Galaxy_populations}.} 
        \label{fig:Beta_profiles_Morpho}
    \end{figure*}

We run MAMPOSSt in the \textit{Split} mode, considering each morphological class as an independent tracer of the cluster potential, and the results obtained for the $\beta(r)$ profile are listed in Table~\ref{tab:table_I_fit_Subsamples}. The corresponding velocity anisotropy profiles $\beta(r)$ for the E (left panel), S0 (middle panel), and S (right panel) galaxy subsamples from both the P (dashed red line) and R (dotted blue line) populations are displayed in Fig.~\ref{fig:Beta_profiles_Morpho}, with uncertainties indicated by shaded regions. As shown in Fig.~\ref{fig:Beta_profiles_Morpho}, $\rm{P}_{\rm{E}}$ galaxies are characterized by more isotropic orbits at all radii compared to $\rm{R}_{\rm{E}}$ galaxies. In fact, $\rm{P}_{\rm{E}}$ galaxies exhibit inner isotropic orbits ($\beta = 0.03 \pm 0.12$ at $r/r_{200} = 0.1$) that become modestly radial towards the cluster outskirts ($\beta \sim 0.20 \pm 0.14$ at $r/r_{200} = 1.0$). In contrast, $\rm{R}_{\rm{E}}$ galaxies already display slightly radial orbits at small radii ($\beta = 0.20 \pm 0.14$ at $r/r_{200} = 0.1$), which become even more radial at larger radial distances ($\beta \sim 0.44 \pm 0.13$ at $r/r_{200} = 1.0$). For the S0 subsample, P galaxies appear to have slightly more radial orbits than R galaxies, although this difference is only marginally significant given the uncertainties. Finally, no significant difference is observed between the $\beta(r)$ profiles of $\rm{P}_{\rm{S}}$ and $\rm{R}_{\rm{S}}$ galaxies.

\begin{figure*}
        \centering
        \includegraphics[width = 0.95\linewidth]{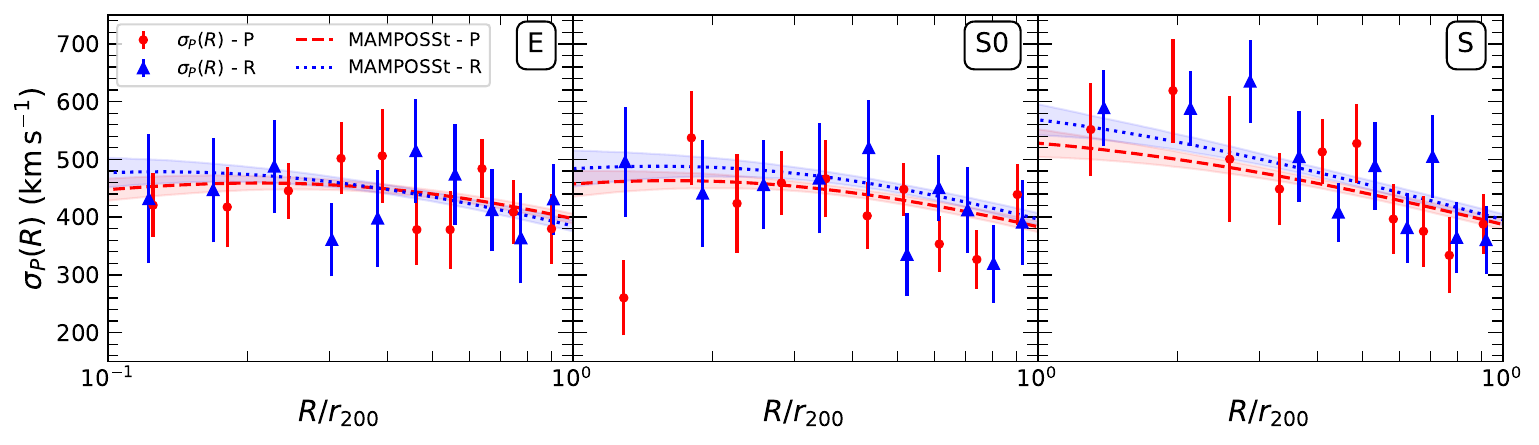}
        \caption{Observed LOS velocity dispersion profiles $\sigma_P(R)$ of the E (left panel), S0 (middle panel), and S (right panel) galaxy subsamples from both the P (red dots) and R (blue triangles) populations. The MAMPOSSt $\sigma_P(R)$ profile of each P (dashed red lines) and R (dotted blue lines) subsamples are displayed, along with their uncertainties, represented by the respective coloured shaded regions. In each panel, the horizontal axis provides the projected radial distance normalized by the $\langle r_{200} \rangle$ value given in Section~\ref{sec:Galaxy_populations}, while the horizontal values indicate the central value of each bin.} 
        \label{fig:SigmaP_profiles_Morpho}
    \end{figure*}

Furthermore, Fig.~\ref{fig:SigmaP_profiles_Morpho} shows the observed $\sigma_P(R)$ profiles for the E (left panel), S0 (middle panel), and S (right panel) galaxy subsamples from both the P (red dots) and R (blue triangles) populations. Note that, taking the uncertainties into account, no significant difference is observed between the $\sigma_P(R)$ profiles of P and R galaxies of the same morphological class. Nevertheless, $\rm{P}_{\rm{S0}}$ galaxies exhibit a notably lower velocity dispersion in the first bin compared to $\rm{R}_{\rm{S0}}$ galaxies. In addition, we find that, for both populations, E and S0 galaxies display similar $\sigma_P(R)$ profiles within the uncertainties, while S galaxies tend to show slightly higher velocity dispersions in the very inner regions compared to their E and S0 counterparts. The $\sigma_P(R)$ profiles estimated from the MAMPOSSt solution for the P (dashed red line) and R (dotted blue line) galaxy subsamples are also shown in Fig.~\ref{fig:SigmaP_profiles_Morpho}, with uncertainties represented by the respective coloured shaded regions. The MAMPOSSt $\sigma_P(R)$ profile accurately reproduces the observed $\sigma_P(R)$ profile for all morphological classes, from both populations, within the uncertainties. It is worth mentioning that the MAMPOSSt solution indicates slightly higher velocity dispersions for the $\rm{R}_{\rm{S0}}$ and $\rm{R}_{\rm{S}}$ galaxies compared to their P counterparts, while no significant difference is observed between the $\rm{R}_{\rm{E}}$ and $\rm{P}_{\rm{E}}$ galaxies. Finally, according to the MAMPOSSt results, S galaxies from both populations display slightly higher velocity dispersions in the inner regions compared to their E and S0 counterparts, in agreement with the observed trend previously noted.
 
\begin{figure*}
        \centering
        \includegraphics[width = 0.9\linewidth]{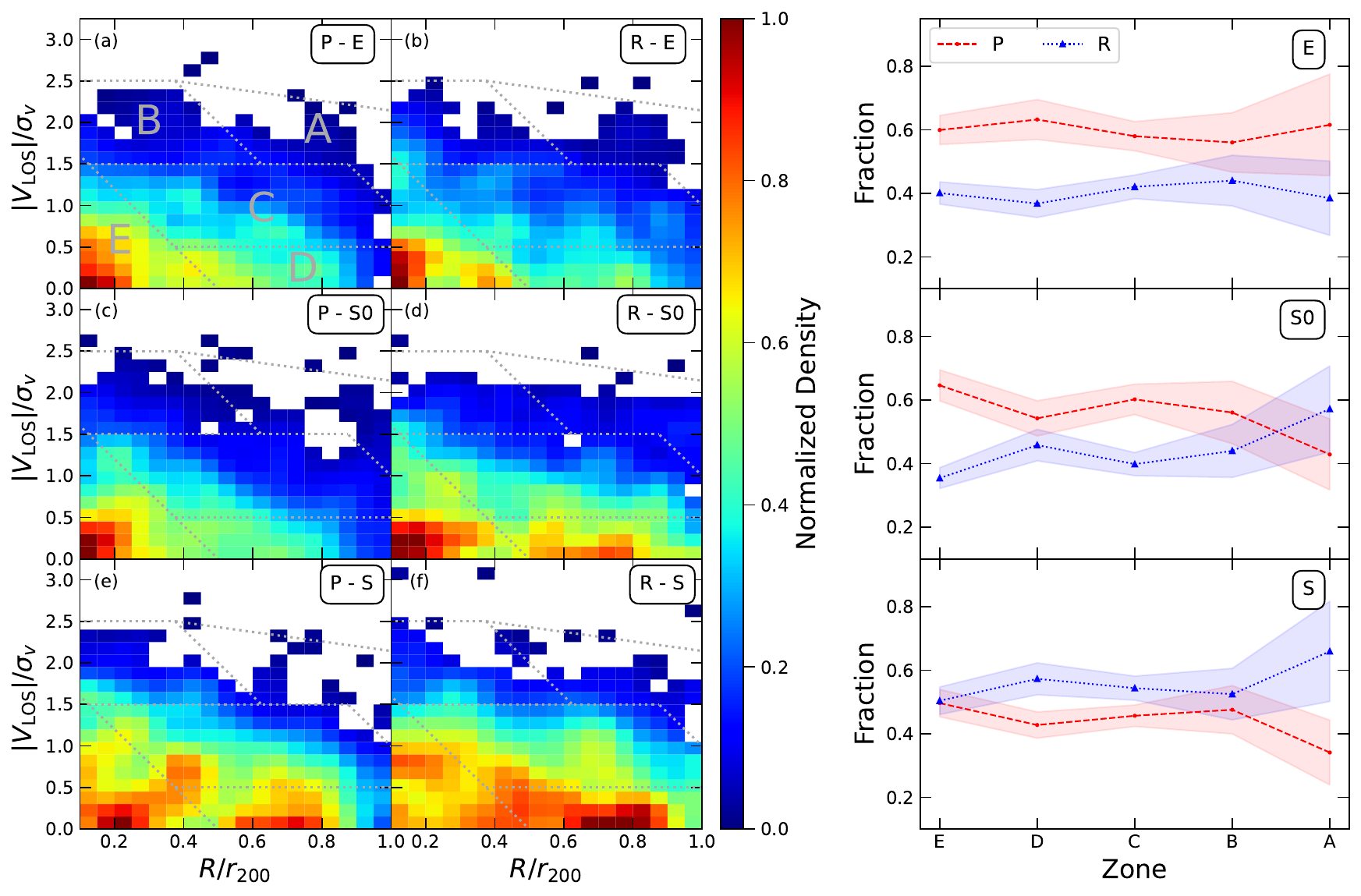}
        \caption{Normalized density distributions of E (panel a and b), S0 (panels c and d), and S (panels e and f) galaxy subsamples from both the P (left panels) and R (middle panels) populations. The Rhee zones are represented by the dotted grey lines (see Section~\ref{subsec:PPS}). The right panels exhibits the fractions of P (red dots) and R (blue triangles) galaxies in each subsample, relative to the total number of galaxies in the corresponding subsample, within each Rhee zone, along with their respective uncertainties, represented by the coloured shaded regions.} 
        \label{fig:PPS_Dist_Morpho}
    \end{figure*}

    Finally, the normalized density distributions for the E (panels a and b), S0 (panels c and d), and S (panels e and f) galaxy subsamples from both the P (left panels) and R (middle panels) populations are shown in Fig.~\ref{fig:PPS_Dist_Morpho}. The fractions of P (red dots) and R (blue triangles) galaxies in each morphological subsample, relative to the total number of galaxies in that subsample, within each Rhee zone, are displayed in the right panels of Fig.~\ref{fig:PPS_Dist_Morpho}, while the percentages of galaxies located in each Rhee zone for each morphological subsample are listed in Table~\ref{tab:Frac_Rhee_Subsamples}. $\rm{P}_{\rm{E}}$ and $\rm{R}_{\rm{E}}$ galaxies are similarly distributed across the PPS. Regarding the S0 subsample, we find that the highest densities of $\rm{P}_{\rm{S0}}$ galaxies are located in the E zone, characterized by small cluster-centric distances and low peculiar velocities. In contrast, although the highest concentration of $\rm{R}_{\rm{S0}}$ galaxies also occurs in the E zone, their high densities extend towards larger radial distances ($R/r_{200}\sim 0.4$) compared to those of the P galaxies. Moreover, $\rm{R}_{\rm{S0}}$ galaxies show a notable high concentration in the D zone, which is not observed for their P counterparts. The distribution of $\rm{P}_{\rm{S}}$ galaxies across the PPS is in some respects similar to that of the $\rm{R}_{\rm{S}}$ galaxies. However, while most of the $\rm{P}_{\rm{S}}$ galaxies populating the E zone are located in the bottom left corner of the PPS diagram (i.e. exhibiting both small cluster-centric distances and low peculiar velocities), $\rm{R}_{\rm{S}}$ galaxies tends to avoid this region. Instead, they are typically found in regions within the E zone characterized by either higher peculiar velocities ($|V_{\rm{LOS}}|/\sigma_v \sim 0.8-0.9$) or larger cluster-centric distances ($R/r_{200} \sim 0.4$). Furthermore, $\rm{R}_{\rm{S}}$ galaxies appears to be more homogeneously distributed across the D zone than $\rm{P}_{\rm{S}}$ galaxies. It is worth noting that, although a high-density envelope is observed in the D zone for the $\rm{P}_{\rm{S}}$ galaxies, the full P population does not exhibit this feature in the same zone, but instead displays a nearly smooth trend of decreasing galaxy density with increasing radial distance. This is likely because both $\rm{P}_{\rm{E}}$ and $\rm{P}_{\rm{S0}}$ galaxies do not display a comparable high-density peak in the D zone, and the number of galaxies in these two subsamples is almost twice that of the $\rm{P}_{\rm{S}}$ galaxies. As a result, the concentration of $\rm{P}_{\rm{S}}$ galaxies in the D zone is smoothed out in the full P population, and no high-density envelope is observed in this zone for the full P population in Fig.~\ref{fig:PPS_Dist_All_Gal}. In contrast, both $\rm{R}_{\rm{S0}}$ and $\rm{R}_{\rm{S}}$ galaxies exhibit high-density envelopes in the D zone, whereas only the $\rm{R}_{\rm{E}}$ galaxies do not display this feature. Consequently, since the combined number of $\rm{R}_{\rm{S0}}$ and $\rm{R}_{\rm{S}}$ galaxies is significantly higher than that of the $\rm{R}_{\rm{E}}$ galaxies, the high-density structures in the D zone remain visible in the full R population (see Fig~\ref{fig:PPS_Dist_All_Gal}). Lastly, analysing the right panels of Fig.~\ref{fig:PPS_Dist_Morpho}, we note that the fractions of P and R galaxies exhibit a generally smooth decrease/increase from the E zone towards the A zone for both the S0 (middle panel) and S (bottom panel) subsamples. However, this otherwise smooth trend is interrupted in the D zone, as previously observed for the stellar mass and $D_n4000$ subsamples, as well as for the full P and R galaxy populations. In contrast, a different behaviour is observed for the E subsample (top right panel), where the fraction of $\rm{P}_{\rm{E}}$/$\rm{R}_{\rm{E}}$ galaxies increases/decreases from the E zone towards the D zone and also from the B zone towards the A zone. We highlight that this is the only case where such behaviour is observed among all the subsamples analysed in this work.

    %%%%%%%%%%%%%%%%%%%%%%%%%%%%%%%%%%%%%%%%%%
    %%%%%%%%%%%%%%%%%%%%%%%%%%%%%%%%%%%%%%%%%%

    The results obtained for the morphological subsamples clearly reflect the morphology-density relation \citep{Dressler_1980_ApJ}, according to which early-type galaxies are typically found deeper in cluster environments than late-type ones. Despite that, $\rm{P}_{\rm{S0}}$ galaxies appear more centrally concentrated than the $\rm{P}_{\rm{E}}$ galaxies. However, this result is significant only at the $1\sigma$ level. In addition, both $\rm{P}_{\rm{E}}$ and $\rm{P}_{\rm{S0}}$ galaxies exhibit peak concentrations of density exclusively in the E zone of the PPS, which indicates that they were, on average, accreted at similarly early times into their clusters. By contrast, the broader distribution of $\rm{P}_{\rm{S}}$ galaxies across zones C, D, and E implies that they were typically accreted at later times. On the other hand, the PPS distributions of R galaxies indicate that $\rm{R}_{\rm{E}}$ galaxies are generally the oldest within their clusters, followed in age by $\rm{R}_{\rm{S0}}$ and $\rm{R}_{\rm{S}}$ galaxies, respectively. In addition, for both populations, we find that E and S0 galaxies tend to exhibit similar $\sigma_P(R)$ profiles, which are less steep than those observed for S galaxies. These results are consistent with previous studies showing that E and S0 galaxies have similar velocity dispersions, both lower than those observed for S galaxies \citep[e.g.][]{Adami_1998_A&A_Segregation, Cava_2017_A&A}.
        
    As for the orbital profiles, E and S0 galaxies from both populations exhibit more isotropic orbits in the inner regions, which become increasingly radial at larger cluster-centric distances. In contrast, S galaxies show nearly constant orbital anisotropy, indicating radial orbits at all radii. Additionally, $\rm{P}_{\rm{E}}$ galaxies display more isotropic orbits than $\rm{P}_{\rm{S0}}$ galaxies, which in turn appear to have slightly less radial orbits than $\rm{P}_{\rm{S}}$ galaxies. On the other hand, and somewhat surprisingly, both $\rm{R}_{\rm{E}}$ and $\rm{R}_{\rm{S}}$ galaxies exhibit more radial orbits than $\rm{R}_{\rm{S0}}$ galaxies. To the best of our knowledge, \citet{Mamon_2019_A&A} is the only work to date that has individually estimated the orbital profiles of E, S0, and S galaxies. Their results show that the orbits of E and S0 galaxies are consistent with isotropy at all radii, although mildly radial orbits at $r_{200}$ are also acceptable for both populations. In this case, the S0 population ($\beta_{\rm{S0}} \simeq 0.31 \pm 0.17$) displays more radial orbits at $r_{200}$ than the E population ($\beta_{\rm{E}} \simeq 0.19 \pm 0.25$). In contrast, S galaxies exhibit isotropic orbits in the inner regions that become more radially elongated ($\beta \simeq 0.45 \pm 0.08$) at $r_{200}$. These findings are in overall agreement with our results for the P population, as well as for both $\rm{R}_{\rm{S}}$ and $\rm{R}_{\rm{S0}}$ galaxies. However, it is particularly surprising that $\rm{R}_{\rm{E}}$ galaxies exhibit orbital properties similar to those of $\rm{R}_{\rm{S}}$ galaxies, and more radial than those of $\rm{R}_{\rm{S0}}$ galaxies. Finally, we emphasise that our comparison with previous work aimed only to assess whether our findings follow expected trends. Notably, two results deviate from the literature: $\rm{P}_{\rm{E}}$ galaxies galaxies are more centrally concentrated than $\rm{P}_{\rm{S0}}$ galaxies, and $\rm{R}_{\rm{E}}$ galaxies exhibit radial orbits similar to those of $\rm{R}_{\rm{S}}$ galaxies. These differences may arise because our sample includes only quiescent (P or R) galaxies, making direct comparison with morphology-based studies that ignore star formation content inappropriate. 

%%%%%%%%%%%%%%%%%%%%%%%%
%%%%%%%%%%%%%%%%%%%%%%%%
%%%%%%%%%%%%%%%%%%%%%%%%
%%%%%%%%%%%%%%%%%%%%%%%%
%%%%%%%%%%%%%%%%%%%%%%%%
%%%%%%%%%%%%%%%%%%%%%%%%

\section{Discussion}
\label{sec:Discussion}

The positions of the galaxies on the PPS diagram are connected to their orbital evolution within clusters (see e.g. Fig. 1 of \citetalias{Rhee_2017_ApJ}). Indeed, galaxies characterized by distinct times since infall tend to occupy semi-distinct regions in the PPS \citep[e.g.][]{Mahajan_2011_MNRAS, Oman_2013_MNRAS, Haines_2015_ApJ, Rhee_2017_ApJ, Pasquali_2019_MNRAS}. Therefore, we can interpret the difference between the PPS distributions of the P and R populations, in particular the high galaxy concentration observed in the D zone for the latter, as an indication that R galaxies were, on average, accreted later than the P ones in their clusters. This scenario is corroborated by the apparent increase in the fractions of P galaxies from the A zone towards the E zone, while the opposite trend is observed for R galaxies (Fig.~\ref{fig:PPS_Dist_All_Gal}). Moreover, the otherwise smooth increase/decrease in the P/R fractions is interrupted in the D zone, likely due to the high densities observed for the R population in this region.

In this scenario, one would expect R galaxies to still retain dynamical information from their infall epoch and, consequently, to exhibit kinematical and dynamical properties that differ from those of the more dynamically evolved P population. However, this is not the case: P and R galaxies exhibit comparable velocity dispersions (Fig.~\ref{fig:SigmaP_profiles_All_Gal}) and display similar orbital profiles (Fig.~\ref{fig:Beta_profiles_All_Gal}). In fact, MAMPOSSt and IJE $\beta(r)$ profiles, for both populations, are compatible within the uncertainties and indicate inner isotropic orbits that become increasingly radial with increasing cluster-centric distances. These profiles are in agreement with the profile obtained for the Q population in \citetalias{Valk_2025_MNRAS}, as expected. Moreover, this same orbital behaviour was also found by \citet{Biviano_2013_A&A, Biviano_2016_A&A, Biviano_2021_A&A} for their colour-selected passive galaxies and by \citet{Capasso_2019_MNRAS} for their sample of non-emission-line passive galaxies. Indeed, such shape for the $\beta(r)$ profile is expected to be the result of the growth of structures occurring in two phases: an initial phase of fast collapse, which produces the inner isotropic orbits, followed by a slower, smoother accretion phase, responsible for the outer radial anisotropy \citep{Lapi_2011_ApJ}. Nevertheless, fully isotropic orbits inside the virial region were found by \citet{Biviano_2009_A&A} for their sample of non-emission line galaxies \citep[see also e.g.][]{Mahdavi_1999_ApJ}, while \citet{Munari_2014_A&A} found inner isotropic orbits that become slightly tangential beyond the virial radius for their red galaxy population. Lastly, MAMPOSSt was able to successfully reproduce the observed $\sigma_P(R)$ profile of both populations, given the uncertainties (Fig.~\ref{fig:SigmaP_profiles_All_Gal}), thus indicating that galaxies on both populations are close to dynamical equilibrium within their clusters. 

These results support the idea that R galaxies are not recent infallers in their clusters. They must have been accreted at earlier epochs, in order to have had sufficient time for the dynamical information from their infall epoch to be partially erased, which explains why their dynamical properties resemble those of a more equilibrated P population. Indeed, the almost complete absence of P and R galaxies in the A zone (Fig.~\ref{fig:PPS_Dist_All_Gal}), which are populated by the most recent arrivals to the cluster environment, confirms that galaxies from both populations are not recent infallers within their host haloes and, consequently, have already been subjected to the action of environmental mechanisms. In addition, the dynamical results obtained for the entire R population indicate that the processes responsible for shutting down the R signature, whatever their nature, must operate on longer time-scales than that of the orbital isotropization. Otherwise, we would expect to observe differences in the velocity dispersion and orbital properties between the P and R galaxies. There are several mechanisms capable of isotropising the orbits of galaxies in clusters, including violent relaxation, dynamical friction, radial orbit instability \citep[ROI, e.g.][]{Bellovary_2008_ApJ}, and galaxy interactions with the intra-cluster medium \citep[e.g.][]{Dolag_2009_MNRAS}. However, their relative efficiency on shaping galaxy orbits remains under debate. Numerical simulations can offer valuable insights into this topic by estimating the relative time-scales of these processes across different stages of the cluster evolution. Yet, although simulations have so far successfully reproduced the general shape of the $\beta(r)$ profile for galaxies in clusters, they still disagree on its evolution with redshift. \citep[see e.g.][]{Wetzel_2011_MNRAS,Lemze_2012_ApJ,Iannuzzi_2012_MNRAS,Munari_2013_MNRAS,Lotz_2019_MNRAS}. Nevertheless, based on our results, one may argue that the time-scale required for a galaxy to erase the dynamical signatures of its infall epoch and evolve towards a state more consistent with that of a more dynamically evolved P population should be shorter than the typical time R galaxies have resided within their clusters. In this context, the high concentration of R galaxies in zones 3 and 4 of \citet[Fig.~\ref{fig:PPS_Dist_All_Gal_Pasquali}]{Pasquali_2019_MNRAS} indicates that a significant fraction of them were, on average, accreted more recently, with infall times in the range $3.89- 4.5 \, \rm{Gyr}$. Furthermore, this upper limit for orbital isotropization ($\sim 4 \, \rm{Gyr}$) sets a lower bound on the time-scale required for a galaxy to lose its R signature and transition to the P class.

The results obtained in this work provide further insights into the origin of the warm gas reservoir in R galaxies. In the following discussion, we focus particularly on systems with early-type morphologies, which are the most numerous within the Q population (comprising $\sim 69$ per cent and $\sim 56$ per cent of the P and R populations, respectively), and for which explaining the origin and presence of a warm gas reservoir is especially challenging. A more detailed discussion of the results for P and R galaxies as a function of morphological class is provided later in this section. Based on the chemical composition of the emitting gas in elliptical R galaxies, \citet{Herpich_2018_MNRAS} proposed that it likely originates from accretion from the haloes of the galaxies or from residual streams of metal-rich gas coming from a merger in the recent past. Galaxy mergers are more common in low-mass groups or in the outskirts of galaxy clusters. Consequently, if the warm gas reservoir in R galaxies originates from streams of metal-rich gas supplied by a recent merger, we would not expect to find these systems in the denser central regions of clusters. Instead, they would likely be located in the outskirts of clusters, where they still retain dynamical information from their infall epoch. Therefore, in this scenario, one would expect R galaxies to be more concentrated in the outer zones of the PPS -- indicative of recent infall -- as well as to exhibit higher velocity dispersions and more radial orbits compared to the P population. Since none of these trends are observed, this likely suggests that metal-rich gas streams from mergers are not the main contributors to the warm gas reservoir in R galaxies. In contrast, galaxies falling into the cluster environment for the first time are expected to retain their own hot gas haloes -- unless they have undergone significant pre-processing \citep[e.g.][]{Zabludoff_1998_ApJ, Balogh_2000_ApJ, Fujita_2004_PASJ, Lopes_2024_MNRAS} -- which serve as reservoirs for the replenishment of cold gas. Once within the cluster, however, environmental mechanisms such as ram-pressure stripping \citep{Gunn_1972_ApJ} and starvation \citep{Larson_1980_ApJ} are expected to strip the galaxies of their hot gas haloes, ultimately leading to the quenching of star formation. Ram-pressure stripping, being a more violent process, is thought to quench star-formation on much shorter time-scales \citep[$\lesssim 1\,\rm{Gyr}$, e.g.][]{Lotz_2019_MNRAS, Pallero_2022_MNRAS} than those typically associated with starvation \citep[$\sim 3-4\,\rm{Gyr}$, e.g.][]{Peng_2015_Natur, Trussler_2020_MNRAS}. In this context, if a galaxy initially sustains its R signature due to the presence of a hot gas halo -- which provides the minimal amount of ionised gas required for emission -- the removal of this halo triggers a gradual transition towards the P phase. This process, however, is not instantaneous. If the timescale for the transition is short (as expected for ram-pressure stripping), R galaxies would remain in this phase only briefly after infall. Conversely, if the transition occurs over longer timescales (as in the case of starvation), there would be sufficient time for R galaxies to dynamically relax within the cluster potential before fully quenching. As a result, R galaxies may still be observed with orbital properties similar to those of P galaxies. 

In order to better understand the differences reported above between the P and R populations, we compared the properties of P and R systems characterized by similar stellar mass, mean stellar population age (parametrized by the $D_n4000$ index), and morphology. The results obtained for the stellar mass and $D_n4000$ subsamples support a scenario in which low-mass, younger galaxies have undergone significant gas removal during their infall phase, which likely led to the progressive suppression of their star formation. These galaxies now appear as a more quiescent population in the D zone. In fact, ram-pressure stripping is more effective at removing gas from both low-mass galaxies and those on more radial orbits \citep[e.g.][]{Vollmer_2001_ApJ, Jaffe_2018_MNRAS, Lotz_2019_MNRAS}. Therefore, as these low-mass, younger galaxies spend more time within the cluster and gradually migrate towards the central regions, they continue to lose gas and may eventually transition from the R to the P population. In this context, the $D_n4000$ threshold value adopted to separate galaxy populations broadly corresponds to mean stellar population ages of $\sim 4 - 6\,\rm{Gyr}$\footnote{These estimates were obtained using the SSP models of \citet[2016 version]{Bruzual_2003_MNRAS}, adopting the MILES spectral libraries \citep{Falcon-Barroso_2011_A&A} and the \citet{Kroupa_2001_MNRAS} IMF, and assuming reasonable metallicity values given the median galaxy stellar mass of our sample.}. Conversely, backsplash galaxies typically have time-since-infall values in the range $1-4\,\rm{Gyr}$ \citep[e.g.][]{Haines_2015_ApJ}. Therefore, our $D_n4000$ cut approximately separates galaxies that are likely still in the backsplash phase (accreted $\lesssim 4\,\rm{Gyr}$ ago) from those that have already passed through this phase and become part of the older, virialized cluster population. Notably, the most pronounced differences between P and R galaxies are found among those with low $D_n4000$ values -- that is, galaxies likely still in the backsplash phase. Such result supports the interpretation that, after leaving the backsplash phase, galaxies are likely to lose their R signature and transition to the P population, which is also consistent with the lack of significant differences between the PPS distributions of HD$_n$ P and R galaxies. In contrast, high-mass, older galaxies are likely less affected by ram-pressure stripping as they orbit through the cluster, which could explain why R galaxies in this regime are less commonly found in the D zone compared to their younger counterparts. Nevertheless, they are still subject to environmentally driven mechanisms that may gradually deplete their remaining gas reservoirs. As these galaxies continue to move towards denser central regions, they are expected to eventually lose their remaining gas and transition from the R phase to the P one.

The comparative morphological analysis of P and R galaxies is a particularly important tool to understand the intrinsic differences between these populations, since galaxies with different morphologies are likely to exhibit distinct origins for the warm gas reservoir observed in R galaxies. We find that $\rm{R}_{\rm{E}}$ galaxies appear to be marginally more centrally concentrated than $\rm{P}_{\rm{E}}$ galaxies. However, both types exhibit high-density envelopes solely in the E zone, suggesting they were accreted at similar early times. In contrast, $\rm{P}_{\rm{S0}}$ and $\rm{P}_{\rm{S}}$ galaxies tend to lie closer to the cluster centres than their R counterparts. Furthermore, the higher concentration of $\rm{R}_{\rm{S0}}$ and $\rm{R}_{\rm{S}}$ galaxies in the D zone, relative to their P counterparts, suggests that they were typically accreted into their clusters at later times. These findings indicate that galaxies with S0 and S morphology are the primary contributors to the differences observed between the P and R populations. Additionally, we do not observe any significant difference in the velocity dispersions of P and R galaxies with the same morphology, although the MAMPOSSt $\sigma_P(R)$ profiles suggest slightly higher values for $\rm{R}_{\rm{S0}}$ and $\rm{R}_{\rm{S}}$ galaxies compared to their P counterparts. As for the orbital profiles, $\rm{P}_{\rm{S}}$ and $\rm{R}_{\rm{S}}$ galaxies exhibit indistinguishable $\beta(r)$ profiles, while $\rm{P}_{\rm{S0}}$ galaxies appear to have slightly more radial orbits than $\rm{R}_{\rm{S0}}$ galaxies -- though this difference is only marginally significant given the uncertainties. Finally, although $\rm{P}_{\rm{E}}$ and $\rm{R}_{\rm{E}}$ galaxies were likely accreted at similar times and exhibit comparable velocity dispersions, the latter are characterized by more radial orbits all radii. This result shows that the $\beta(r)$ profile can provide kinematical information beyond that provided by the PPS, which we interpret here as signatures of distinct accretion processes. Furthermore, this also highlights the complementary nature of these techniques, since neither alone can fully extract the dynamical information of a galaxy population.

Spiral galaxies falling into clusters are expected to follow nearly radial orbits and to be rapidly transformed into S0 and E galaxies as they orbit through the cluster \citep[e.g.][]{Cava_2017_A&A, Mamon_2019_A&A, Sampaio_2024_MNRAS}. Several environmentally driven mechanisms can account for such morphological transformations, including galaxy mergers \citep{Gerhard_1981_MNRAS}, and galaxy harassment \citep{Moore_1996_Natur}, as well as the stripping of a galaxy's cold gas supply, either through ram-pressure stripping \citep{Gunn_1972_ApJ} or tidal stripping \citep{Larson_1980_ApJ}. The spiral galaxies analysed in this work cannot yet have undergone significant morphological transformation, as they still exhibit clear spiral features. However, their star formation has already been quenched, at least in the inner regions probed by the SDSS fibre. Furthermore, among all quiescent spirals galaxies, those with R signature -- which we interpreted here as still retaining some gas -- display a strong excess in the D zone relative to their P counterparts. Galaxies located at this zone are characterized by large cluster-centric distances and low peculiar velocities, consistent with our expectations from backsplash systems \citep[e.g.][]{Gill_2005_MNRAS, Mahajan_2011_MNRAS, Oman_2013_MNRAS}. Therefore, this excess is consistent with the interpretation that these galaxies are star-forming spirals that gradually lost their gas during their first infall and now appear as a more quenched backsplash population in the D zone -- while still retaining some gas, enough to appears as R galaxies. In contrast, $\rm{P}_{\rm{S}}$ galaxies likely underwent a much more intense gas removal event that efficiently quenched their star formation. Such an event -- possibly driven by environmental mechanisms such as ram-pressure stripping \citep{Gunn_1972_ApJ} -- caused these galaxies to transition directly from the star-forming phase to the passive one, thus justifying why, although already passive, they still appear in excess in the D zone. %This evolutionary path, driven by cluster environmental processes, is expected to ultimately transform infalling star-forming spirals into dynamically relaxed, passive S0 galaxies residing in the innermost cluster regions \citep[e.g.][]{Cava_2017_A&A, Sampaio_2024_MNRAS}. 
In addition, spiral galaxies that have already been transformed into S0s -- but that still retain some gas, allowing them to be detected as R galaxies -- exhibit a peak in density in the D zone. This finding can be interpreted as evidence that the time-scale required for morphological transformation from S to S0 is shorter than the time-scale needed for a galaxy to migrate to the innermost regions. Indeed, galaxies located in the zones 3 and 4 of \citet[Fig.~\ref{fig:PPS_Dist_Morpho_Pasquali}]{Pasquali_2019_MNRAS} were accreted $3.89 - 4.50\,\rm{Gyr}$ ago, which is in agreement with previous work that have pointed out a time-scale $\sim 3\,\rm{Gyr}$ for morphological evolution \citep[e.g.][]{Cava_2017_A&A}. This points to a scenario where S0 galaxies likely transition from the R population to the P one as they approach virialization and settle into the central regions of the cluster.

Finally, $\rm{P}_{\rm{E}}$ galaxies are the best representatives of a virialized cluster galaxy population. They show no evidence of recent star formation, are predominantly located in the innermost regions of clusters, exhibit lower velocity dispersions, and follow nearly isotropic orbits that become only slightly radial near $r_{200}$ ($\beta \sim 0.20 \pm 0.14$). In contrast, the interpretation of the results obtained for the $\rm{R}_{\rm{E}}$ galaxies is less straightforward. Although marginally more concentrated than their P counterparts, their PPS distributions indicate similarly early infall times, and their velocity dispersions are likewise comparable. Taken together, these results suggest that $\rm{R}_{\rm{E}}$ galaxies are as dynamically relaxed as $\rm{P}_{\rm{E}}$ galaxies. However, they exhibit systematically more radial orbits than $\rm{P}_{\rm{E}}$ galaxies, which may indicate a less evolved dynamical state. To investigate this further, we examined the stellar mass and $D_n4000$ distributions of both populations. We find that $\rm{R}_{\rm{E}}$ galaxies are more massive ($\log M_\star = \uncertainties{10.9}{0.7}{0.6}$ vs $\log M_\star = \uncertainties{10.8}{0.9}{0.7}$) and slightly younger ($D_n4000 = \uncertainties{1.89}{0.19}{0.14}$ vs $D_n4000 = \uncertainties{1.92}{0.17}{0.12}$) than $\rm{P}_{\rm{E}}$ galaxies. Moreover, the median value of $D_n4000$ indicates that the mean stellar population age of $\rm{R}_{\rm{E}}$ galaxies is on the order of $\sim 5 \, \rm{Gyr}$. Interestingly, \citet{Herpich_2018_MNRAS} found that R elliptical galaxies have an excess of intermediate-age stellar populations ($0.1 - 5 \, \rm{Gyr}$) compared to their P counterparts. Assuming that their mean stellar population ages cannot be significantly younger than their infall times -- otherwise, they would have undergone a rejuvenation event within clusters, which is very unlikely to occur -- these galaxies are expected to have had sufficient time to virialize. Nevertheless, they still retain residual gas, which allows them to maintain their R signature, and exhibit more radial orbits than their P counterparts, a feature that can be interpreted as indicating that they were, on average, accreted more recently than the $\rm{P}_{\rm{E}}$ galaxies. Based on these results, we propose that $\rm{R}_{\rm{E}}$ galaxies, being the most massive within the R population ($\log M_{\star,\rm{S0}} = \uncertainties{10.8}{0.9}{0.6}$ and $\log M_{\star,\rm{S}} = \uncertainties{10.7}{1.2}{0.6}$) and likely having been accreted at later times, have been able to preserve their R signature. This supports the view that gas removal processes in clusters do not act on a universal timescale \citep[e.g.][]{Lotz_2019_MNRAS,Boselli_2022_A&ARv,Wright_2022_MNRAS}; rather, they depend on intrinsic galaxy properties, allowing some galaxies to retain gas -- and hence their R signature -- over extended periods. Indeed, the stellar population-based estimate of the infall time for $\rm{R}_{\rm{E}}$ galaxies is in good agreement with dynamical inferences: their high densities in zones 1 and 2 of \citet[Fig.~\ref{fig:PPS_Dist_Morpho_Pasquali}]{Pasquali_2019_MNRAS} suggest that a significant fraction of $\rm{R}_{\rm{E}}$ galaxies were accreted, on average, between $5.18$ and $5.42\,\rm{Gyr}$ ago. Therefore, it is plausible that this population was actively forming stars in the distant past, and, over time, has been evolving towards the P class -- while still retaining an R signature due to the presence of residual warm gas from a hot gas halo that has survived longer than in other populations. This is consistent with previous studies that reported relatively long time-scales for starvation to significantly reduce the star formation activity on galaxies \citep[e.g.][]{Wetzel_2013_MNRAS}.

An important caveat of our analysis is the fact that the galaxy classification, as well as the $D_n4000$ measurements, are based on SDSS single fibre spectra \citep[covering 2 arcsec;][]{Eisenstein_2011_AJ}, and are therefore biased to the properties of the innermost region of the galaxies in the sample. Besides, given the known relation between stellar mass and galaxy size \citep[e.g.][]{Shen_2003_MNRAS}, the spectra of more massive objects sample proportionally smaller regions of the galaxy. For late-type objects in particular, this relation could cause the fibre spectra to be more dominated by the galaxy bulge, rendering such galaxies more similar to old systems in general. However, we find that only 66 per cent of the HM$_\star$ galaxies also have HD$_n$ values, and only 22 per cent of the HM$_\star$ and HD$_n$ galaxies display spiral morphologies. Also, the fact that $r_\nu$ of the S population is consistently larger than that of the HD$_n$ population shows that this effect is not dominant. In fact, such bias would tend to reduce the differences between e.g. the PPS distributions and the $\beta(r)$ profiles of the S and HD$_n$ subsamples. Therefore the differences between such populations are expected to be intrinsically even larger than the observed ones. Finally, it is important to note that the SDSS fibre encompasses proportionally different regions of a galaxy as a function of redshift. However, in the short redshift span of our sample, this effect is not expected to be large. Indeed, we find that the observed differences between populations in the PPS are only slightly reduced at high redshifts, mostly due to the fact that the high-$z$ subsample is dominated by more massive galaxies and the contrasts between the P and R populations are less pronounced for the HM$_\star$ galaxies (see Fig.~\ref{fig:PPS_Dist_StellarMass_Dn4000}). Our conclusions are therefore expected to be robust against biases due to the limited fibre size.

%%%%%%%%%%%%%%%%%%%%%%%%
%%%%%%%%%%%%%%%%%%%%%%%%
%%%%%%%%%%%%%%%%%%%%%%%%
%%%%%%%%%%%%%%%%%%%%%%%%
%%%%%%%%%%%%%%%%%%%%%%%%
%%%%%%%%%%%%%%%%%%%%%%%%

\section{Conclusions}
\label{sec:Conclusions}

In this work, we analysed the kinematical and dynamical properties, as well as the PPS distributions, of quiescent galaxies with weak emission-lines, referred to as retired (R), and those without emission-lines, referred to as passive (P). Our sample comprises 2907 P and 2387 R galaxies, belonging to 336 relaxed galaxy clusters ($z<0.2$). In addition to comparing the properties of the full P and R galaxy populations, we also investigate what distinguishes P and R galaxies with similar stellar masses, mean stellar population ages (parametrized by the $D_n4000$ index), and morphology. 

Our main conclusions are summarized below:

\begin{itemize}

    \item [(i)] P galaxies are more centrally concentrated than R galaxies and, based on their PPS distributions, were typically accreted at earlier epochs into their clusters than their R counterparts. Nevertheless, both populations exhibit similar $\sigma_P(R)$ and $\beta(r)$ profiles. Additionally, the observed $\sigma_P(R)$ profiles of both populations were successfully reproduced by MAMPOSSt, suggesting that they are close to dynamical equilibrium within their clusters.
    
    \item [(ii)] P galaxies tend to reside closer to the cluster centre than R galaxies, regardless of their stellar mass, mean stellar population age, or morphology. The only exception involves the $\rm{P}_{\rm{E}}$ galaxies, which are marginally less concentrated than their R counterparts, although this difference is significant only at the $1\sigma$ level.

    \item [(iii)] Low-mass ($\log M_\star \leq 10.5$) and low-$D_n4000$ ($D_n4000 \leq 1.86$) R galaxies exhibit PPS distributions that differ significantly from those of their P counterparts, indicating that they were typically accreted at later times than similar P galaxies. A similar, though slightly less pronounced, trend is observed for high-mass systems. However, no significant differences are found between the PPS distributions of P and R galaxies with high-$D_n4000$ values. Nevertheless, P and R galaxies show similar $\sigma_P(R)$ and $\beta(r)$ profiles when matched in stellar mass and mean stellar population age.
    
    \item [(iv)] The PPS distributions of $\rm{P}_{\rm{S0}}$ and $\rm{P}_{\rm{S}}$ galaxies indicate that they were accreted, on average, at earlier times than their R counterparts. Despite that, no significant differences are observed between their $\sigma_P(R)$ and $\beta(r)$ profiles.
    
    \item [(v)] $\rm{P}_{\rm{E}}$ and $\rm{R}_{\rm{E}}$ galaxies exhibit similar distributions across the PPS, suggesting that they were typically accreted at similarly early times. In contrast, although their $\sigma_P(R)$ profiles are also similar, $\rm{R}_{\rm{E}}$ galaxies display more radial orbits at all radii than $\rm{P}_{\rm{E}}$ galaxies.
    
\end{itemize}

    We interpret these results within a scenario in which, although R galaxies were typically accreted later than P galaxies into their clusters, they already had sufficient time to partially erase the dynamical information from their infall epoch and evolve towards a state more consistent with that of a dynamically evolved P population. In addition, our findings suggest that the emitting gas in early-type R galaxies likely originates from accretion of their own hot gas haloes, and that the removal of this halo triggers the transition towards the P phase. Moreover, this process must occur over relatively long timescales, allowing R galaxies sufficient time to dynamically relax within the cluster potential before fully quenching. On the other hand, the gas in late-type R galaxies is probably a remnant of the cold phase that was already present in these systems before their infall into the cluster environment.
    
    Additionally, the results obtained for the stellar mass and $D_n4000$ subsamples are interpreted as the consequence of significant gas removal experienced by low-mass, younger galaxies on more radial orbits during their infall phase, which are also likely still in the backsplash phase. In contrast, high-mass, older galaxies on more isotropic orbits likely underwent a more gradual gas depletion during their orbits through the cluster. Such galaxies, having likely already left the backsplash phase and become part of the older, virialized cluster population, complete their transition from the R phase to the P one as they move towards the denser central regions.

    Finally, $\rm{R}_{\rm{S}}$ galaxies likely lost their gas gradually during their infall phase, in contrast to their P counterparts, which must have undergone a much more intense gas removal to efficiently quench their star formation before morphological transformations became evident. In this context, spirals galaxies that have been transformed into S0s but still retain some gas will appear as $\rm{R}_{\rm{S0}}$ galaxies, which then transition from the R population to the P one as they approach virialization and settle into the central regions of the cluster. Lastly, we propose that $\rm{R}_{\rm{E}}$ galaxies have been able to retain their gas for longer periods, thereby preserving their R signature, likely because they constitute the most massive subset within the R population and, as suggested by their orbits, were probably, on average, accreted later than their P counterparts.

%%%%%%%%%%%%%%%%%%%%%%%%
%%%%%%%%%%%%%%%%%%%%%%%%
%%%%%%%%%%%%%%%%%%%%%%%%

%%%%%%%%%%%%%%%%%%%%%%%%
%%%%%%%%%%%%%%%%%%%%%%%%
%%%%%%%%%%%%%%%%%%%%%%%%

\section*{Acknowledgements}

We thank an anonymous referee for constructive comments and suggestions that greatly improved the readability and structure of the manuscript. The authors also acknowledge support from the Brazilian research agencies Conselho Nacional de Desenvolvimento Científico e Tecnológico (CNPq) and Fundação de Amparo à pesquisa do Estado do Rio Grande do Sul (FAPERGS). GAV thanks for the financial support from Coordenação de Aperfeiçoamento de Pessoal de Nível Superior -- Brasil (CAPES) -- Finance Code 001.

Funding for SDSS-III has been provided by the Alfred P. Sloan Foundation, the Participating Institutions, the National Science Foundation, and the U.S. Department of Energy Office of Science. The SDSS-III web site is \url{http://www.sdss3.org/}.

SDSS-III is managed by the Astrophysical Research Consortium for the Participating Institutions of the SDSS-III Collaboration including the University of Arizona, the Brazilian Participation Group, Brookhaven National Laboratory, Carnegie Mellon University, University of Florida, the French Participation Group, the German Participation Group, Harvard University, the Instituto de Astrofisica de Canarias, the Michigan State/Notre Dame/JINA Participation Group, Johns Hopkins University, Lawrence Berkeley National Laboratory, Max Planck Institute for Astrophysics, Max Planck Institute for Extraterrestrial Physics, New Mexico State University, New York University, Ohio State University, Pennsylvania State University, University of Portsmouth, Princeton University, the Spanish Participation Group, University of Tokyo, University of Utah, Vanderbilt University, University of Virginia, University of Washington, and Yale University.

%%%%%%%%%%%%%%%%%%%%%%%%%%%%%%%%%%%%%%%%%%%%%%%%%%
\section*{Data Availability}

No new data have been produced in this work.



\bibliographystyle{mnras}
\bibliography{bibliography_valk} % if your bibtex file is called example.bib

\begin{thebibliography}{}
\makeatletter
\relax
\def\mn@urlcharsother{\let\do\@makeother \do\$\do\&\do\#\do\^\do\_\do\%\do\~}
\def\mn@doi{\begingroup\mn@urlcharsother \@ifnextchar [ {\mn@doi@}
  {\mn@doi@[]}}
\def\mn@doi@[#1]#2{\def\@tempa{#1}\ifx\@tempa\@empty \href
  {http://dx.doi.org/#2} {doi:#2}\else \href {http://dx.doi.org/#2} {#1}\fi
  \endgroup}
\def\mn@eprint#1#2{\mn@eprint@#1:#2::\@nil}
\def\mn@eprint@arXiv#1{\href {http://arxiv.org/abs/#1} {{\tt arXiv:#1}}}
\def\mn@eprint@dblp#1{\href {http://dblp.uni-trier.de/rec/bibtex/#1.xml}
  {dblp:#1}}
\def\mn@eprint@#1:#2:#3:#4\@nil{\def\@tempa {#1}\def\@tempb {#2}\def\@tempc
  {#3}\ifx \@tempc \@empty \let \@tempc \@tempb \let \@tempb \@tempa \fi \ifx
  \@tempb \@empty \def\@tempb {arXiv}\fi \@ifundefined
  {mn@eprint@\@tempb}{\@tempb:\@tempc}{\expandafter \expandafter \csname
  mn@eprint@\@tempb\endcsname \expandafter{\@tempc}}}

\bibitem[\protect\citeauthoryear{{Adami}, {Biviano}  \& {Mazure}}{{Adami}
  et~al.}{1998a}]{Adami_1998_A&A_Segregation}
{Adami} C.,  {Biviano} A.,   {Mazure} A.,  1998a, \mn@doi [\aap]
  {10.48550/arXiv.astro-ph/9709268}, \href
  {https://ui.adsabs.harvard.edu/abs/1998A&A...331..439A} {331, 439}

\bibitem[\protect\citeauthoryear{{Adami}, {Mazure}, {Biviano}, {Katgert}  \&
  {Rhee}}{{Adami} et~al.}{1998b}]{Adami_1998_A&A}
{Adami} C.,  {Mazure} A.,  {Biviano} A.,  {Katgert} P.,   {Rhee} G.,  1998b,
  \aap, \href {https://ui.adsabs.harvard.edu/abs/1998A&A...331..493A} {331,
  493}

\bibitem[\protect\citeauthoryear{{Agertz}, {Kravtsov}, {Leitner}  \&
  {Gnedin}}{{Agertz} et~al.}{2013}]{Agertz_2013_ApJ}
{Agertz} O.,  {Kravtsov} A.~V.,  {Leitner} S.~N.,   {Gnedin} N.~Y.,  2013,
  \mn@doi [\apj] {10.1088/0004-637X/770/1/25}, \href
  {https://ui.adsabs.harvard.edu/abs/2013ApJ...770...25A} {770, 25}

\bibitem[\protect\citeauthoryear{{Aguerri}, {S{\'a}nchez-Janssen}  \&
  {Mu{\~n}oz-Tu{\~n}{\'o}n}}{{Aguerri} et~al.}{2007}]{Aguerri_2007_A&A}
{Aguerri} J.~A.~L.,  {S{\'a}nchez-Janssen} R.,   {Mu{\~n}oz-Tu{\~n}{\'o}n} C.,
  2007, \mn@doi [\aap] {10.1051/0004-6361:20066478}, \href
  {https://ui.adsabs.harvard.edu/abs/2007A&A...471...17A} {471, 17}

\bibitem[\protect\citeauthoryear{{Aguerri}, {Agulli}, {Diaferio}  \& {Dalla
  Vecchia}}{{Aguerri} et~al.}{2017}]{Aguerri_2017_MNRAS}
{Aguerri} J.~A.~L.,  {Agulli} I.,  {Diaferio} A.,   {Dalla Vecchia} C.,  2017,
  \mn@doi [\mnras] {10.1093/mnras/stx457}, \href
  {https://ui.adsabs.harvard.edu/abs/2017MNRAS.468..364A} {468, 364}

\bibitem[\protect\citeauthoryear{{Aguirre Tagliaferro}, {Biviano}, {De Lucia},
  {Munari}  \& {Garcia Lambas}}{{Aguirre Tagliaferro}
  et~al.}{2021}]{AguirreTagliaferro_2021_A&A}
{Aguirre Tagliaferro} T.,  {Biviano} A.,  {De Lucia} G.,  {Munari} E.,
  {Garcia Lambas} D.,  2021, \mn@doi [\aap] {10.1051/0004-6361/202140326},
  \href {https://ui.adsabs.harvard.edu/abs/2021A&A...652A..90A} {652, A90}

\bibitem[\protect\citeauthoryear{{Alam} et~al.,}{{Alam}
  et~al.}{2015}]{Alam_2015_ApJS}
{Alam} S.,  et~al., 2015, \mn@doi [\apjs] {10.1088/0067-0049/219/1/12}, \href
  {https://ui.adsabs.harvard.edu/abs/2015ApJS..219...12A} {219, 12}

\bibitem[\protect\citeauthoryear{{Annunziatella} et~al.,}{{Annunziatella}
  et~al.}{2016}]{Annunziatella_2016_A&A}
{Annunziatella} M.,  et~al., 2016, \mn@doi [\aap]
  {10.1051/0004-6361/201527399}, \href
  {https://ui.adsabs.harvard.edu/abs/2016A&A...585A.160A} {585, A160}

\bibitem[\protect\citeauthoryear{{Baldwin}, {Phillips}  \&
  {Terlevich}}{{Baldwin} et~al.}{1981}]{Baldwin_1981_PASP}
{Baldwin} J.~A.,  {Phillips} M.~M.,   {Terlevich} R.,  1981, \mn@doi [\pasp]
  {10.1086/130766}, \href
  {https://ui.adsabs.harvard.edu/abs/1981PASP...93....5B} {93, 5}

\bibitem[\protect\citeauthoryear{{Balogh}, {Morris}, {Yee}, {Carlberg}  \&
  {Ellingson}}{{Balogh} et~al.}{1999}]{Balogh_1999_ApJ}
{Balogh} M.~L.,  {Morris} S.~L.,  {Yee} H.~K.~C.,  {Carlberg} R.~G.,
  {Ellingson} E.,  1999, \mn@doi [\apj] {10.1086/308056}, \href
  {https://ui.adsabs.harvard.edu/abs/1999ApJ...527...54B} {527, 54}

\bibitem[\protect\citeauthoryear{{Balogh}, {Navarro}  \& {Morris}}{{Balogh}
  et~al.}{2000}]{Balogh_2000_ApJ}
{Balogh} M.~L.,  {Navarro} J.~F.,   {Morris} S.~L.,  2000, \mn@doi [\apj]
  {10.1086/309323}, \href
  {https://ui.adsabs.harvard.edu/abs/2000ApJ...540..113B} {540, 113}

\bibitem[\protect\citeauthoryear{{Beers}, {Geller}  \& {Huchra}}{{Beers}
  et~al.}{1982}]{Beers_1982_ApJ}
{Beers} T.~C.,  {Geller} M.~J.,   {Huchra} J.~P.,  1982, \mn@doi [\apj]
  {10.1086/159958}, \href
  {https://ui.adsabs.harvard.edu/abs/1982ApJ...257...23B} {257, 23}

\bibitem[\protect\citeauthoryear{{Belfiore} et~al.,}{{Belfiore}
  et~al.}{2017}]{Belfiore_2017_MNRAS}
{Belfiore} F.,  et~al., 2017, \mn@doi [\mnras] {10.1093/mnras/stw3211}, \href
  {https://ui.adsabs.harvard.edu/abs/2017MNRAS.466.2570B} {466, 2570}

\bibitem[\protect\citeauthoryear{{Bellovary}, {Dalcanton}, {Babul}, {Quinn},
  {Maas}, {Austin}, {Williams}  \& {Barnes}}{{Bellovary}
  et~al.}{2008}]{Bellovary_2008_ApJ}
{Bellovary} J.~M.,  {Dalcanton} J.~J.,  {Babul} A.,  {Quinn} T.~R.,  {Maas}
  R.~W.,  {Austin} C.~G.,  {Williams} L. L.~R.,   {Barnes} E.~I.,  2008,
  \mn@doi [\apj] {10.1086/591120}, \href
  {https://ui.adsabs.harvard.edu/abs/2008ApJ...685..739B} {685, 739}

\bibitem[\protect\citeauthoryear{{Biviano} \& {Girardi}}{{Biviano} \&
  {Girardi}}{2003}]{Biviano_2003_ApJ}
{Biviano} A.,  {Girardi} M.,  2003, \mn@doi [\apj] {10.1086/345893}, \href
  {https://ui.adsabs.harvard.edu/abs/2003ApJ...585..205B} {585, 205}

\bibitem[\protect\citeauthoryear{{Biviano} \& {Katgert}}{{Biviano} \&
  {Katgert}}{2004}]{Biviano_2004_A&A}
{Biviano} A.,  {Katgert} P.,  2004, \mn@doi [\aap]
  {10.1051/0004-6361:20041306}, \href
  {https://ui.adsabs.harvard.edu/abs/2004A&A...424..779B} {424, 779}

\bibitem[\protect\citeauthoryear{{Biviano} \& {Poggianti}}{{Biviano} \&
  {Poggianti}}{2009}]{Biviano_2009_A&A}
{Biviano} A.,  {Poggianti} B.~M.,  2009, \mn@doi [\aap]
  {10.1051/0004-6361/200911757}, \href
  {https://ui.adsabs.harvard.edu/abs/2009A&A...501..419B} {501, 419}

\bibitem[\protect\citeauthoryear{{Biviano} et~al.,}{{Biviano}
  et~al.}{2013}]{Biviano_2013_A&A}
{Biviano} A.,  et~al., 2013, \mn@doi [\aap] {10.1051/0004-6361/201321955},
  \href {https://ui.adsabs.harvard.edu/abs/2013A&A...558A...1B} {558, A1}

\bibitem[\protect\citeauthoryear{{Biviano}, {van der Burg}, {Muzzin},
  {Sartoris}, {Wilson}  \& {Yee}}{{Biviano} et~al.}{2016}]{Biviano_2016_A&A}
{Biviano} A.,  {van der Burg} R.~F.~J.,  {Muzzin} A.,  {Sartoris} B.,  {Wilson}
  G.,   {Yee} H.~K.~C.,  2016, \mn@doi [\aap] {10.1051/0004-6361/201628697},
  \href {https://ui.adsabs.harvard.edu/abs/2016A&A...594A..51B} {594, A51}

\bibitem[\protect\citeauthoryear{{Biviano} et~al.,}{{Biviano}
  et~al.}{2021}]{Biviano_2021_A&A}
{Biviano} A.,  et~al., 2021, \mn@doi [\aap] {10.1051/0004-6361/202140564},
  \href {https://ui.adsabs.harvard.edu/abs/2021A&A...650A.105B} {650, A105}

\bibitem[\protect\citeauthoryear{{Bolzonella}, {Miralles}  \&
  {Pell{\'o}}}{{Bolzonella} et~al.}{2000}]{Bolzonella_2000_A&A}
{Bolzonella} M.,  {Miralles} J.~M.,   {Pell{\'o}} R.,  2000, \mn@doi [\aap]
  {10.48550/arXiv.astro-ph/0003380}, \href
  {https://ui.adsabs.harvard.edu/abs/2000A&A...363..476B} {363, 476}

\bibitem[\protect\citeauthoryear{{Boselli}, {Fossati}  \& {Sun}}{{Boselli}
  et~al.}{2022}]{Boselli_2022_A&ARv}
{Boselli} A.,  {Fossati} M.,   {Sun} M.,  2022, \mn@doi [\aapr]
  {10.1007/s00159-022-00140-3}, \href
  {https://ui.adsabs.harvard.edu/abs/2022A&ARv..30....3B} {30, 3}

\bibitem[\protect\citeauthoryear{{Bower}, {Benson}, {Malbon}, {Helly}, {Frenk},
  {Baugh}, {Cole}  \& {Lacey}}{{Bower} et~al.}{2006}]{Bower_2006_MNRAS}
{Bower} R.~G.,  {Benson} A.~J.,  {Malbon} R.,  {Helly} J.~C.,  {Frenk} C.~S.,
  {Baugh} C.~M.,  {Cole} S.,   {Lacey} C.~G.,  2006, \mn@doi [\mnras]
  {10.1111/j.1365-2966.2006.10519.x}, \href
  {https://ui.adsabs.harvard.edu/abs/2006MNRAS.370..645B} {370, 645}

\bibitem[\protect\citeauthoryear{{Brambila}, {Lopes}, {Ribeiro}  \&
  {Cortesi}}{{Brambila} et~al.}{2023}]{Brambila_2023_MNRAS}
{Brambila} D.,  {Lopes} P. A.~A.,  {Ribeiro} A. L.~B.,   {Cortesi} A.,  2023,
  \mn@doi [\mnras] {10.1093/mnras/stad1233}, \href
  {https://ui.adsabs.harvard.edu/abs/2023MNRAS.523..785B} {523, 785}

\bibitem[\protect\citeauthoryear{{Brinchmann}, {Charlot}, {White}, {Tremonti},
  {Kauffmann}, {Heckman}  \& {Brinkmann}}{{Brinchmann}
  et~al.}{2004}]{Brinchmann_2004_MNRAS}
{Brinchmann} J.,  {Charlot} S.,  {White} S.~D.~M.,  {Tremonti} C.,  {Kauffmann}
  G.,  {Heckman} T.,   {Brinkmann} J.,  2004, \mn@doi [\mnras]
  {10.1111/j.1365-2966.2004.07881.x}, \href
  {https://ui.adsabs.harvard.edu/abs/2004MNRAS.351.1151B} {351, 1151}

\bibitem[\protect\citeauthoryear{{Bruzual} \& {Charlot}}{{Bruzual} \&
  {Charlot}}{2003}]{Bruzual_2003_MNRAS}
{Bruzual} G.,  {Charlot} S.,  2003, \mn@doi [\mnras]
  {10.1046/j.1365-8711.2003.06897.x}, \href
  {https://ui.adsabs.harvard.edu/abs/2003MNRAS.344.1000B} {344, 1000}

\bibitem[\protect\citeauthoryear{{Capasso} et~al.,}{{Capasso}
  et~al.}{2019}]{Capasso_2019_MNRAS}
{Capasso} R.,  et~al., 2019, \mn@doi [\mnras] {10.1093/mnras/sty2645}, \href
  {https://ui.adsabs.harvard.edu/abs/2019MNRAS.482.1043C} {482, 1043}

\bibitem[\protect\citeauthoryear{{Cappellari} \& {Emsellem}}{{Cappellari} \&
  {Emsellem}}{2004}]{Cappellari_2004_PASP}
{Cappellari} M.,  {Emsellem} E.,  2004, \mn@doi [\pasp] {10.1086/381875}, \href
  {https://ui.adsabs.harvard.edu/abs/2004PASP..116..138C} {116, 138}

\bibitem[\protect\citeauthoryear{{Cava} et~al.,}{{Cava}
  et~al.}{2017}]{Cava_2017_A&A}
{Cava} A.,  et~al., 2017, \mn@doi [\aap] {10.1051/0004-6361/201730785}, \href
  {https://ui.adsabs.harvard.edu/abs/2017A&A...606A.108C} {606, A108}

\bibitem[\protect\citeauthoryear{{Chandrasekhar}}{{Chandrasekhar}}{1943}]{Chandrasekhar_1943_ApJ}
{Chandrasekhar} S.,  1943, \mn@doi [\apj] {10.1086/144517}, \href
  {https://ui.adsabs.harvard.edu/abs/1943ApJ....97..255C} {97, 255}

\bibitem[\protect\citeauthoryear{{Choi} \& {Yi}}{{Choi} \&
  {Yi}}{2017}]{Choi_2017_ApJ}
{Choi} H.,  {Yi} S.~K.,  2017, \mn@doi [\apj] {10.3847/1538-4357/aa5e4b}, \href
  {https://ui.adsabs.harvard.edu/abs/2017ApJ...837...68C} {837, 68}

\bibitem[\protect\citeauthoryear{{Cid Fernandes}, {Stasi{\'n}ska}, {Mateus}  \&
  {Vale Asari}}{{Cid Fernandes} et~al.}{2011}]{CidFernandes_2011_MNRAS}
{Cid Fernandes} R.,  {Stasi{\'n}ska} G.,  {Mateus} A.,   {Vale Asari} N.,
  2011, \mn@doi [\mnras] {10.1111/j.1365-2966.2011.18244.x}, \href
  {https://ui.adsabs.harvard.edu/abs/2011MNRAS.413.1687C} {413, 1687}

\bibitem[\protect\citeauthoryear{{Croton} et~al.,}{{Croton}
  et~al.}{2006}]{Croton_2006_MNRAS}
{Croton} D.~J.,  et~al., 2006, \mn@doi [\mnras]
  {10.1111/j.1365-2966.2005.09675.x}, \href
  {https://ui.adsabs.harvard.edu/abs/2006MNRAS.365...11C} {365, 11}

\bibitem[\protect\citeauthoryear{{Davis} \& {Bureau}}{{Davis} \&
  {Bureau}}{2016}]{Davis_2016_MNRAS}
{Davis} T.~A.,  {Bureau} M.,  2016, \mn@doi [\mnras] {10.1093/mnras/stv2998},
  \href {https://ui.adsabs.harvard.edu/abs/2016MNRAS.457..272D} {457, 272}

\bibitem[\protect\citeauthoryear{{Dawson} et~al.,}{{Dawson}
  et~al.}{2013}]{Dawson_2013_AJ}
{Dawson} K.~S.,  et~al., 2013, \mn@doi [\aj] {10.1088/0004-6256/145/1/10},
  \href {https://ui.adsabs.harvard.edu/abs/2013AJ....145...10D} {145, 10}

\bibitem[\protect\citeauthoryear{{De Boni}, {Ettori}, {Dolag}  \&
  {Moscardini}}{{De Boni} et~al.}{2013}]{DeBoni_2013_MNRAS}
{De Boni} C.,  {Ettori} S.,  {Dolag} K.,   {Moscardini} L.,  2013, \mn@doi
  [\mnras] {10.1093/mnras/sts235}, \href
  {https://ui.adsabs.harvard.edu/abs/2013MNRAS.428.2921D} {428, 2921}

\bibitem[\protect\citeauthoryear{{Diaferio}}{{Diaferio}}{1999}]{Diaferio_1999_MNRAS}
{Diaferio} A.,  1999, \mn@doi [\mnras] {10.1046/j.1365-8711.1999.02864.x},
  \href {https://ui.adsabs.harvard.edu/abs/1999MNRAS.309..610D} {309, 610}

\bibitem[\protect\citeauthoryear{{Dolag}, {Borgani}, {Murante}  \&
  {Springel}}{{Dolag} et~al.}{2009}]{Dolag_2009_MNRAS}
{Dolag} K.,  {Borgani} S.,  {Murante} G.,   {Springel} V.,  2009, \mn@doi
  [\mnras] {10.1111/j.1365-2966.2009.15034.x}, \href
  {https://ui.adsabs.harvard.edu/abs/2009MNRAS.399..497D} {399, 497}

\bibitem[\protect\citeauthoryear{{Dressler}}{{Dressler}}{1980}]{Dressler_1980_ApJ}
{Dressler} A.,  1980, \mn@doi [\apj] {10.1086/157753}, \href
  {https://ui.adsabs.harvard.edu/abs/1980ApJ...236..351D} {236, 351}

\bibitem[\protect\citeauthoryear{{Einasto} \& {Einasto}}{{Einasto} \&
  {Einasto}}{1987}]{Einasto_1987_MNRAS}
{Einasto} M.,  {Einasto} J.,  1987, \mn@doi [\mnras] {10.1093/mnras/226.3.543},
  \href {https://ui.adsabs.harvard.edu/abs/1987MNRAS.226..543E} {226, 543}

\bibitem[\protect\citeauthoryear{{Eisenstein} et~al.,}{{Eisenstein}
  et~al.}{2011}]{Eisenstein_2011_AJ}
{Eisenstein} D.~J.,  et~al., 2011, \mn@doi [\aj] {10.1088/0004-6256/142/3/72},
  \href {https://ui.adsabs.harvard.edu/abs/2011AJ....142...72E} {142, 72}

\bibitem[\protect\citeauthoryear{{Falc{\'o}n-Barroso},
  {S{\'a}nchez-Bl{\'a}zquez}, {Vazdekis}, {Ricciardelli}, {Cardiel}, {Cenarro},
  {Gorgas}  \& {Peletier}}{{Falc{\'o}n-Barroso}
  et~al.}{2011}]{Falcon-Barroso_2011_A&A}
{Falc{\'o}n-Barroso} J.,  {S{\'a}nchez-Bl{\'a}zquez} P.,  {Vazdekis} A.,
  {Ricciardelli} E.,  {Cardiel} N.,  {Cenarro} A.~J.,  {Gorgas} J.,
  {Peletier} R.~F.,  2011, \mn@doi [\aap] {10.1051/0004-6361/201116842}, \href
  {https://ui.adsabs.harvard.edu/abs/2011A&A...532A..95F} {532, A95}

\bibitem[\protect\citeauthoryear{{Fujita}}{{Fujita}}{2004}]{Fujita_2004_PASJ}
{Fujita} Y.,  2004, \mn@doi [\pasj] {10.1093/pasj/56.1.29}, \href
  {https://ui.adsabs.harvard.edu/abs/2004PASJ...56...29F} {56, 29}

\bibitem[\protect\citeauthoryear{{Gallazzi}, {Charlot}, {Brinchmann}, {White}
  \& {Tremonti}}{{Gallazzi} et~al.}{2005}]{Gallazzi_2005_MNRAS}
{Gallazzi} A.,  {Charlot} S.,  {Brinchmann} J.,  {White} S. D.~M.,   {Tremonti}
  C.~A.,  2005, \mn@doi [\mnras] {10.1111/j.1365-2966.2005.09321.x}, \href
  {https://ui.adsabs.harvard.edu/abs/2005MNRAS.362...41G} {362, 41}

\bibitem[\protect\citeauthoryear{{Gao} et~al.,}{{Gao}
  et~al.}{2020}]{Gao_2020_A&A}
{Gao} F.,  et~al., 2020, \mn@doi [\aap] {10.1051/0004-6361/201937178}, \href
  {https://ui.adsabs.harvard.edu/abs/2020A&A...637A..94G} {637, A94}

\bibitem[\protect\citeauthoryear{{Geller}, {Diaferio}  \& {Kurtz}}{{Geller}
  et~al.}{1999}]{Geller_1999_ApJL}
{Geller} M.~J.,  {Diaferio} A.,   {Kurtz} M.~J.,  1999, \mn@doi [\apjl]
  {10.1086/312024}, \href
  {https://ui.adsabs.harvard.edu/abs/1999ApJ...517L..23G} {517, L23}

\bibitem[\protect\citeauthoryear{{Gerhard}}{{Gerhard}}{1981}]{Gerhard_1981_MNRAS}
{Gerhard} O.~E.,  1981, \mn@doi [\mnras] {10.1093/mnras/197.1.179}, \href
  {https://ui.adsabs.harvard.edu/abs/1981MNRAS.197..179G} {197, 179}

\bibitem[\protect\citeauthoryear{{Gill}, {Knebe}  \& {Gibson}}{{Gill}
  et~al.}{2005}]{Gill_2005_MNRAS}
{Gill} S. P.~D.,  {Knebe} A.,   {Gibson} B.~K.,  2005, \mn@doi [\mnras]
  {10.1111/j.1365-2966.2004.08562.x}, \href
  {https://ui.adsabs.harvard.edu/abs/2005MNRAS.356.1327G} {356, 1327}

\bibitem[\protect\citeauthoryear{{Gomes} et~al.,}{{Gomes}
  et~al.}{2016}]{Gomes_2016_A&A}
{Gomes} J.~M.,  et~al., 2016, \mn@doi [\aap] {10.1051/0004-6361/201525976},
  \href {https://ui.adsabs.harvard.edu/abs/2016A&A...588A..68G} {588, A68}

\bibitem[\protect\citeauthoryear{{Gunn} \& {Gott}}{{Gunn} \&
  {Gott}}{1972}]{Gunn_1972_ApJ}
{Gunn} J.~E.,  {Gott} III J.~R.,  1972, \mn@doi [\apj] {10.1086/151605}, \href
  {https://ui.adsabs.harvard.edu/abs/1972ApJ...176....1G} {176, 1}

\bibitem[\protect\citeauthoryear{{Haines} et~al.,}{{Haines}
  et~al.}{2015}]{Haines_2015_ApJ}
{Haines} C.~P.,  et~al., 2015, \mn@doi [\apj] {10.1088/0004-637X/806/1/101},
  \href {https://ui.adsabs.harvard.edu/abs/2015ApJ...806..101H} {806, 101}

\bibitem[\protect\citeauthoryear{{Hernquist} \& {Mihos}}{{Hernquist} \&
  {Mihos}}{1995}]{Hernquist_1995_ApJ}
{Hernquist} L.,  {Mihos} J.~C.,  1995, \mn@doi [\apj] {10.1086/175940}, \href
  {https://ui.adsabs.harvard.edu/abs/1995ApJ...448...41H} {448, 41}

\bibitem[\protect\citeauthoryear{{Herpich}, {Stasi{\'n}ska}, {Mateus}, {Vale
  Asari}  \& {Cid Fernandes}}{{Herpich} et~al.}{2018}]{Herpich_2018_MNRAS}
{Herpich} F.,  {Stasi{\'n}ska} G.,  {Mateus} A.,  {Vale Asari} N.,   {Cid
  Fernandes} R.,  2018, \mn@doi [\mnras] {10.1093/mnras/sty2391}, \href
  {https://ui.adsabs.harvard.edu/abs/2018MNRAS.481.1774H} {481, 1774}

\bibitem[\protect\citeauthoryear{{Houghton}}{{Houghton}}{2015}]{Houghton_2015_MNRAS}
{Houghton} R.~C.~W.,  2015, \mn@doi [\mnras] {10.1093/mnras/stv1113}, \href
  {https://ui.adsabs.harvard.edu/abs/2015MNRAS.451.3427H} {451, 3427}

\bibitem[\protect\citeauthoryear{{Huertas-Company}, {Foex}, {Soucail}  \&
  {Pell{\'o}}}{{Huertas-Company} et~al.}{2009}]{Huertas-Company_2009_A&A}
{Huertas-Company} M.,  {Foex} G.,  {Soucail} G.,   {Pell{\'o}} R.,  2009,
  \mn@doi [\aap] {10.1051/0004-6361/200912621}, \href
  {https://ui.adsabs.harvard.edu/abs/2009A&A...505...83H} {505, 83}

\bibitem[\protect\citeauthoryear{{Huertas-Company}, {Aguerri}, {Bernardi},
  {Mei}  \& {S{\'a}nchez Almeida}}{{Huertas-Company}
  et~al.}{2011}]{Huertas-Company_2011_A&A}
{Huertas-Company} M.,  {Aguerri} J.~A.~L.,  {Bernardi} M.,  {Mei} S.,
  {S{\'a}nchez Almeida} J.,  2011, \mn@doi [\aap]
  {10.1051/0004-6361/201015735}, \href
  {https://ui.adsabs.harvard.edu/abs/2011A&A...525A.157H} {525, A157}

\bibitem[\protect\citeauthoryear{{Iannuzzi} \& {Dolag}}{{Iannuzzi} \&
  {Dolag}}{2012}]{Iannuzzi_2012_MNRAS}
{Iannuzzi} F.,  {Dolag} K.,  2012, \mn@doi [\mnras]
  {10.1111/j.1365-2966.2012.22017.x}, \href
  {https://ui.adsabs.harvard.edu/abs/2012MNRAS.427.1024I} {427, 1024}

\bibitem[\protect\citeauthoryear{{Jaff{\'e}}, {Smith}, {Candlish}, {Poggianti},
  {Sheen}  \& {Verheijen}}{{Jaff{\'e}} et~al.}{2015}]{Jaffe_2015_MNRAS}
{Jaff{\'e}} Y.~L.,  {Smith} R.,  {Candlish} G.~N.,  {Poggianti} B.~M.,  {Sheen}
  Y.-K.,   {Verheijen} M. A.~W.,  2015, \mn@doi [\mnras]
  {10.1093/mnras/stv100}, \href
  {https://ui.adsabs.harvard.edu/abs/2015MNRAS.448.1715J} {448, 1715}

\bibitem[\protect\citeauthoryear{{Jaff{\'e}} et~al.,}{{Jaff{\'e}}
  et~al.}{2018}]{Jaffe_2018_MNRAS}
{Jaff{\'e}} Y.~L.,  et~al., 2018, \mn@doi [\mnras] {10.1093/mnras/sty500},
  \href {https://ui.adsabs.harvard.edu/abs/2018MNRAS.476.4753J} {476, 4753}

\bibitem[\protect\citeauthoryear{{Katgert}, {Biviano}  \& {Mazure}}{{Katgert}
  et~al.}{2004}]{Katgert_2004_ApJ}
{Katgert} P.,  {Biviano} A.,   {Mazure} A.,  2004, \mn@doi [\apj]
  {10.1086/380118}, \href
  {https://ui.adsabs.harvard.edu/abs/2004ApJ...600..657K} {600, 657}

\bibitem[\protect\citeauthoryear{{Kauffmann} et~al.,}{{Kauffmann}
  et~al.}{2003a}]{Kauffmann_2003_MNRAS_Stellar}
{Kauffmann} G.,  et~al., 2003a, \mn@doi [\mnras]
  {10.1046/j.1365-8711.2003.06291.x}, \href
  {https://ui.adsabs.harvard.edu/abs/2003MNRAS.341...33K} {341, 33}

\bibitem[\protect\citeauthoryear{{Kauffmann} et~al.,}{{Kauffmann}
  et~al.}{2003b}]{Kauffmann_2003_MNRAS_The}
{Kauffmann} G.,  et~al., 2003b, \mn@doi [\mnras]
  {10.1046/j.1365-8711.2003.06292.x}, \href
  {https://ui.adsabs.harvard.edu/abs/2003MNRAS.341...54K} {341, 54}

\bibitem[\protect\citeauthoryear{{Kauffmann}, {White}, {Heckman}, {M{\'e}nard},
  {Brinchmann}, {Charlot}, {Tremonti}  \& {Brinkmann}}{{Kauffmann}
  et~al.}{2004}]{Kauffmann_2004_MNRAS}
{Kauffmann} G.,  {White} S. D.~M.,  {Heckman} T.~M.,  {M{\'e}nard} B.,
  {Brinchmann} J.,  {Charlot} S.,  {Tremonti} C.,   {Brinkmann} J.,  2004,
  \mn@doi [\mnras] {10.1111/j.1365-2966.2004.08117.x}, \href
  {https://ui.adsabs.harvard.edu/abs/2004MNRAS.353..713K} {353, 713}

\bibitem[\protect\citeauthoryear{{Kroupa}}{{Kroupa}}{2001}]{Kroupa_2001_MNRAS}
{Kroupa} P.,  2001, \mn@doi [\mnras] {10.1046/j.1365-8711.2001.04022.x}, \href
  {https://ui.adsabs.harvard.edu/abs/2001MNRAS.322..231K} {322, 231}

\bibitem[\protect\citeauthoryear{{Lapi} \& {Cavaliere}}{{Lapi} \&
  {Cavaliere}}{2011}]{Lapi_2011_ApJ}
{Lapi} A.,  {Cavaliere} A.,  2011, \mn@doi [\apj]
  {10.1088/0004-637X/743/2/127}, \href
  {https://ui.adsabs.harvard.edu/abs/2011ApJ...743..127L} {743, 127}

\bibitem[\protect\citeauthoryear{{Larson}, {Tinsley}  \& {Caldwell}}{{Larson}
  et~al.}{1980}]{Larson_1980_ApJ}
{Larson} R.~B.,  {Tinsley} B.~M.,   {Caldwell} C.~N.,  1980, \mn@doi [\apj]
  {10.1086/157917}, \href
  {https://ui.adsabs.harvard.edu/abs/1980ApJ...237..692L} {237, 692}

\bibitem[\protect\citeauthoryear{{Lemze} et~al.,}{{Lemze}
  et~al.}{2012}]{Lemze_2012_ApJ}
{Lemze} D.,  et~al., 2012, \mn@doi [\apj] {10.1088/0004-637X/752/2/141}, \href
  {https://ui.adsabs.harvard.edu/abs/2012ApJ...752..141L} {752, 141}

\bibitem[\protect\citeauthoryear{{Lopes}, {Ribeiro}  \& {Brambila}}{{Lopes}
  et~al.}{2024}]{Lopes_2024_MNRAS}
{Lopes} P. A.~A.,  {Ribeiro} A. L.~B.,   {Brambila} D.,  2024, \mn@doi [\mnras]
  {10.1093/mnrasl/slad134}, \href
  {https://ui.adsabs.harvard.edu/abs/2024MNRAS.527L..19L} {527, L19}

\bibitem[\protect\citeauthoryear{{Lotz}, {Remus}, {Dolag}, {Biviano}  \&
  {Burkert}}{{Lotz} et~al.}{2019}]{Lotz_2019_MNRAS}
{Lotz} M.,  {Remus} R.-S.,  {Dolag} K.,  {Biviano} A.,   {Burkert} A.,  2019,
  \mn@doi [\mnras] {10.1093/mnras/stz2070}, \href
  {https://ui.adsabs.harvard.edu/abs/2019MNRAS.488.5370L} {488, 5370}

\bibitem[\protect\citeauthoryear{{Macci{\`o}}, {Dutton}  \& {van den
  Bosch}}{{Macci{\`o}} et~al.}{2008}]{Maccio_2008_MNRAS}
{Macci{\`o}} A.~V.,  {Dutton} A.~A.,   {van den Bosch} F.~C.,  2008, \mn@doi
  [\mnras] {10.1111/j.1365-2966.2008.14029.x}, \href
  {https://ui.adsabs.harvard.edu/abs/2008MNRAS.391.1940M} {391, 1940}

\bibitem[\protect\citeauthoryear{{Mahajan}, {Mamon}  \&
  {Raychaudhury}}{{Mahajan} et~al.}{2011}]{Mahajan_2011_MNRAS}
{Mahajan} S.,  {Mamon} G.~A.,   {Raychaudhury} S.,  2011, \mn@doi [\mnras]
  {10.1111/j.1365-2966.2011.19236.x}, \href
  {https://ui.adsabs.harvard.edu/abs/2011MNRAS.416.2882M} {416, 2882}

\bibitem[\protect\citeauthoryear{{Mahdavi}, {Geller}, {B{\"o}hringer}, {Kurtz}
  \& {Ramella}}{{Mahdavi} et~al.}{1999}]{Mahdavi_1999_ApJ}
{Mahdavi} A.,  {Geller} M.~J.,  {B{\"o}hringer} H.,  {Kurtz} M.~J.,   {Ramella}
  M.,  1999, \mn@doi [\apj] {10.1086/307280}, \href
  {https://ui.adsabs.harvard.edu/abs/1999ApJ...518...69M} {518, 69}

\bibitem[\protect\citeauthoryear{{Mamon} \& {{\L}okas}}{{Mamon} \&
  {{\L}okas}}{2005}]{Mamon_2005_MNRAS}
{Mamon} G.~A.,  {{\L}okas} E.~L.,  2005, \mn@doi [\mnras]
  {10.1111/j.1365-2966.2005.09400.x}, \href
  {https://ui.adsabs.harvard.edu/abs/2005MNRAS.363..705M} {363, 705}

\bibitem[\protect\citeauthoryear{{Mamon}, {Biviano}  \& {Bou{\'e}}}{{Mamon}
  et~al.}{2013}]{Mamon_2013_MNRAS}
{Mamon} G.~A.,  {Biviano} A.,   {Bou{\'e}} G.,  2013, \mn@doi [\mnras]
  {10.1093/mnras/sts565}, \href
  {https://ui.adsabs.harvard.edu/abs/2013MNRAS.429.3079M} {429, 3079}

\bibitem[\protect\citeauthoryear{{Mamon}, {Cava}, {Biviano}, {Moretti},
  {Poggianti}  \& {Bettoni}}{{Mamon} et~al.}{2019}]{Mamon_2019_A&A}
{Mamon} G.~A.,  {Cava} A.,  {Biviano} A.,  {Moretti} A.,  {Poggianti} B.,
  {Bettoni} D.,  2019, \mn@doi [\aap] {10.1051/0004-6361/201935081}, \href
  {https://ui.adsabs.harvard.edu/abs/2019A&A...631A.131M} {631, A131}

\bibitem[\protect\citeauthoryear{{Maraston} \& {Str{\"o}mb{\"a}ck}}{{Maraston}
  \& {Str{\"o}mb{\"a}ck}}{2011}]{Maraston_2011_MNRAS}
{Maraston} C.,  {Str{\"o}mb{\"a}ck} G.,  2011, \mn@doi [\mnras]
  {10.1111/j.1365-2966.2011.19738.x}, \href
  {https://ui.adsabs.harvard.edu/abs/2011MNRAS.418.2785M} {418, 2785}

\bibitem[\protect\citeauthoryear{{Maraston}, {Str{\"o}mb{\"a}ck}, {Thomas},
  {Wake}  \& {Nichol}}{{Maraston} et~al.}{2009}]{Maraston_2009_MNRAS}
{Maraston} C.,  {Str{\"o}mb{\"a}ck} G.,  {Thomas} D.,  {Wake} D.~A.,   {Nichol}
  R.~C.,  2009, \mn@doi [\mnras] {10.1111/j.1745-3933.2009.00621.x}, \href
  {https://ui.adsabs.harvard.edu/abs/2009MNRAS.394L.107M} {394, L107}

\bibitem[\protect\citeauthoryear{{Maraston} et~al.,}{{Maraston}
  et~al.}{2013}]{Maraston_2013_MNRAS}
{Maraston} C.,  et~al., 2013, \mn@doi [\mnras] {10.1093/mnras/stt1424}, \href
  {https://ui.adsabs.harvard.edu/abs/2013MNRAS.435.2764M} {435, 2764}

\bibitem[\protect\citeauthoryear{{Martig}, {Bournaud}, {Teyssier}  \&
  {Dekel}}{{Martig} et~al.}{2009}]{Martig_2009_ApJ}
{Martig} M.,  {Bournaud} F.,  {Teyssier} R.,   {Dekel} A.,  2009, \mn@doi
  [\apj] {10.1088/0004-637X/707/1/250}, \href
  {https://ui.adsabs.harvard.edu/abs/2009ApJ...707..250M} {707, 250}

\bibitem[\protect\citeauthoryear{{Martig} et~al.,}{{Martig}
  et~al.}{2013}]{Martig_2013_MNRAS}
{Martig} M.,  et~al., 2013, \mn@doi [\mnras] {10.1093/mnras/sts594}, \href
  {https://ui.adsabs.harvard.edu/abs/2013MNRAS.432.1914M} {432, 1914}

\bibitem[\protect\citeauthoryear{{Mart{\'\i}nez}, {Muriel}  \&
  {Coenda}}{{Mart{\'\i}nez} et~al.}{2016}]{Martinez_2016_MNRAS}
{Mart{\'\i}nez} H.~J.,  {Muriel} H.,   {Coenda} V.,  2016, \mn@doi [\mnras]
  {10.1093/mnras/stv2295}, \href
  {https://ui.adsabs.harvard.edu/abs/2016MNRAS.455..127M} {455, 127}

\bibitem[\protect\citeauthoryear{{Mercurio} et~al.,}{{Mercurio}
  et~al.}{2021}]{Mercurio_2021_A&A}
{Mercurio} A.,  et~al., 2021, \mn@doi [\aap] {10.1051/0004-6361/202142168},
  \href {https://ui.adsabs.harvard.edu/abs/2021A&A...656A.147M} {656, A147}

\bibitem[\protect\citeauthoryear{{Mihos} \& {Hernquist}}{{Mihos} \&
  {Hernquist}}{1996}]{Mihos_1996_ApJ}
{Mihos} J.~C.,  {Hernquist} L.,  1996, \mn@doi [\apj] {10.1086/177353}, \href
  {https://ui.adsabs.harvard.edu/abs/1996ApJ...464..641M} {464, 641}

\bibitem[\protect\citeauthoryear{{Moore}, {Katz}, {Lake}, {Dressler}  \&
  {Oemler}}{{Moore} et~al.}{1996}]{Moore_1996_Natur}
{Moore} B.,  {Katz} N.,  {Lake} G.,  {Dressler} A.,   {Oemler} A.,  1996,
  \mn@doi [\nat] {10.1038/379613a0}, \href
  {https://ui.adsabs.harvard.edu/abs/1996Natur.379..613M} {379, 613}

\bibitem[\protect\citeauthoryear{{Moore}, {Lake}  \& {Katz}}{{Moore}
  et~al.}{1998}]{Moore_1998_ApJ}
{Moore} B.,  {Lake} G.,   {Katz} N.,  1998, \mn@doi [\apj] {10.1086/305264},
  \href {https://ui.adsabs.harvard.edu/abs/1998ApJ...495..139M} {495, 139}

\bibitem[\protect\citeauthoryear{{Munari}, {Biviano}, {Borgani}, {Murante}  \&
  {Fabjan}}{{Munari} et~al.}{2013}]{Munari_2013_MNRAS}
{Munari} E.,  {Biviano} A.,  {Borgani} S.,  {Murante} G.,   {Fabjan} D.,  2013,
  \mn@doi [\mnras] {10.1093/mnras/stt049}, \href
  {https://ui.adsabs.harvard.edu/abs/2013MNRAS.430.2638M} {430, 2638}

\bibitem[\protect\citeauthoryear{{Munari}, {Biviano}  \& {Mamon}}{{Munari}
  et~al.}{2014}]{Munari_2014_A&A}
{Munari} E.,  {Biviano} A.,   {Mamon} G.~A.,  2014, \mn@doi [\aap]
  {10.1051/0004-6361/201322450}, \href
  {https://ui.adsabs.harvard.edu/abs/2014A&A...566A..68M} {566, A68}

\bibitem[\protect\citeauthoryear{{Navarro}, {Frenk}  \& {White}}{{Navarro}
  et~al.}{1996}]{Navarro_1996_ApJ}
{Navarro} J.~F.,  {Frenk} C.~S.,   {White} S. D.~M.,  1996, \mn@doi [\apj]
  {10.1086/177173}, \href
  {https://ui.adsabs.harvard.edu/abs/1996ApJ...462..563N} {462, 563}

\bibitem[\protect\citeauthoryear{{Noguchi}}{{Noguchi}}{1988}]{Noguchi_1988_A&A}
{Noguchi} M.,  1988, \aap, \href
  {https://ui.adsabs.harvard.edu/abs/1988A&A...203..259N} {203, 259}

\bibitem[\protect\citeauthoryear{{Oguri}, {Bayliss}, {Dahle}, {Sharon},
  {Gladders}, {Natarajan}, {Hennawi}  \& {Koester}}{{Oguri}
  et~al.}{2012}]{Oguri_2012_MNRAS}
{Oguri} M.,  {Bayliss} M.~B.,  {Dahle} H.,  {Sharon} K.,  {Gladders} M.~D.,
  {Natarajan} P.,  {Hennawi} J.~F.,   {Koester} B.~P.,  2012, \mn@doi [\mnras]
  {10.1111/j.1365-2966.2011.20248.x}, \href
  {https://ui.adsabs.harvard.edu/abs/2012MNRAS.420.3213O} {420, 3213}

\bibitem[\protect\citeauthoryear{{Okabe}, {Smith}, {Umetsu}, {Takada}  \&
  {Futamase}}{{Okabe} et~al.}{2013}]{Okabe_2013_ApJL}
{Okabe} N.,  {Smith} G.~P.,  {Umetsu} K.,  {Takada} M.,   {Futamase} T.,  2013,
  \mn@doi [\apjl] {10.1088/2041-8205/769/2/L35}, \href
  {https://ui.adsabs.harvard.edu/abs/2013ApJ...769L..35O} {769, L35}

\bibitem[\protect\citeauthoryear{{Oman}, {Hudson}  \& {Behroozi}}{{Oman}
  et~al.}{2013}]{Oman_2013_MNRAS}
{Oman} K.~A.,  {Hudson} M.~J.,   {Behroozi} P.~S.,  2013, \mn@doi [\mnras]
  {10.1093/mnras/stt328}, \href
  {https://ui.adsabs.harvard.edu/abs/2013MNRAS.431.2307O} {431, 2307}

\bibitem[\protect\citeauthoryear{{Pallero}, {G{\'o}mez}, {Padilla}, {Bah{\'e}},
  {Vega-Mart{\'\i}nez}  \& {Torres-Flores}}{{Pallero}
  et~al.}{2022}]{Pallero_2022_MNRAS}
{Pallero} D.,  {G{\'o}mez} F.~A.,  {Padilla} N.~D.,  {Bah{\'e}} Y.~M.,
  {Vega-Mart{\'\i}nez} C.~A.,   {Torres-Flores} S.,  2022, \mn@doi [\mnras]
  {10.1093/mnras/stab3318}, \href
  {https://ui.adsabs.harvard.edu/abs/2022MNRAS.511.3210P} {511, 3210}

\bibitem[\protect\citeauthoryear{{Pasquali}, {Gallazzi}, {Fontanot}, {van den
  Bosch}, {De Lucia}, {Mo}  \& {Yang}}{{Pasquali}
  et~al.}{2010}]{Pasquali_2010_MNRAS}
{Pasquali} A.,  {Gallazzi} A.,  {Fontanot} F.,  {van den Bosch} F.~C.,  {De
  Lucia} G.,  {Mo} H.~J.,   {Yang} X.,  2010, \mn@doi [\mnras]
  {10.1111/j.1365-2966.2010.17074.x}, \href
  {https://ui.adsabs.harvard.edu/abs/2010MNRAS.407..937P} {407, 937}

\bibitem[\protect\citeauthoryear{{Pasquali}, {Smith}, {Gallazzi}, {De Lucia},
  {Zibetti}, {Hirschmann}  \& {Yi}}{{Pasquali}
  et~al.}{2019}]{Pasquali_2019_MNRAS}
{Pasquali} A.,  {Smith} R.,  {Gallazzi} A.,  {De Lucia} G.,  {Zibetti} S.,
  {Hirschmann} M.,   {Yi} S.~K.,  2019, \mn@doi [\mnras]
  {10.1093/mnras/sty3530}, \href
  {https://ui.adsabs.harvard.edu/abs/2019MNRAS.484.1702P} {484, 1702}

\bibitem[\protect\citeauthoryear{{Peng} et~al.,}{{Peng}
  et~al.}{2010}]{Peng_2010_ApJ}
{Peng} Y.-j.,  et~al., 2010, \mn@doi [\apj] {10.1088/0004-637X/721/1/193},
  \href {https://ui.adsabs.harvard.edu/abs/2010ApJ...721..193P} {721, 193}

\bibitem[\protect\citeauthoryear{{Peng}, {Maiolino}  \& {Cochrane}}{{Peng}
  et~al.}{2015}]{Peng_2015_Natur}
{Peng} Y.,  {Maiolino} R.,   {Cochrane} R.,  2015, \mn@doi [\nat]
  {10.1038/nature14439}, \href
  {https://ui.adsabs.harvard.edu/abs/2015Natur.521..192P} {521, 192}

\bibitem[\protect\citeauthoryear{{Planck Collaboration} et~al.,}{{Planck
  Collaboration} et~al.}{2016}]{PlanckCollaboration_2016_A&A}
{Planck Collaboration} et~al., 2016, \mn@doi [\aap]
  {10.1051/0004-6361/201525830}, \href
  {https://ui.adsabs.harvard.edu/abs/2016A&A...594A..13P} {594, A13}

\bibitem[\protect\citeauthoryear{{Poggianti} \& {Barbaro}}{{Poggianti} \&
  {Barbaro}}{1997}]{Poggianti_1997_A&A}
{Poggianti} B.~M.,  {Barbaro} G.,  1997, \mn@doi [\aap]
  {10.48550/arXiv.astro-ph/9703067}, \href
  {https://ui.adsabs.harvard.edu/abs/1997A&A...325.1025P} {325, 1025}

\bibitem[\protect\citeauthoryear{{Postman} \& {Geller}}{{Postman} \&
  {Geller}}{1984}]{Postman_1984_ApJ}
{Postman} M.,  {Geller} M.~J.,  1984, \mn@doi [\apj] {10.1086/162078}, \href
  {https://ui.adsabs.harvard.edu/abs/1984ApJ...281...95P} {281, 95}

\bibitem[\protect\citeauthoryear{{Rhee}, {Smith}, {Choi}, {Yi}, {Jaff{\'e}},
  {Candlish}  \& {S{\'a}nchez-J{\'a}nssen}}{{Rhee}
  et~al.}{2017}]{Rhee_2017_ApJ}
{Rhee} J.,  {Smith} R.,  {Choi} H.,  {Yi} S.~K.,  {Jaff{\'e}} Y.,  {Candlish}
  G.,   {S{\'a}nchez-J{\'a}nssen} R.,  2017, \mn@doi [\apj]
  {10.3847/1538-4357/aa6d6c}, \href
  {https://ui.adsabs.harvard.edu/abs/2017ApJ...843..128R} {843, 128}

\bibitem[\protect\citeauthoryear{{Rhee}, {Smith}, {Choi}, {Contini}, {Jung},
  {Han}  \& {Yi}}{{Rhee} et~al.}{2020}]{Rhee_2020_ApJS}
{Rhee} J.,  {Smith} R.,  {Choi} H.,  {Contini} E.,  {Jung} S.~L.,  {Han} S.,
  {Yi} S.~K.,  2020, \mn@doi [\apjs] {10.3847/1538-4365/ab7377}, \href
  {https://ui.adsabs.harvard.edu/abs/2020ApJS..247...45R} {247, 45}

\bibitem[\protect\citeauthoryear{{Roberts} \& {Parker}}{{Roberts} \&
  {Parker}}{2020}]{Roberts_2020_MNRAS}
{Roberts} I.~D.,  {Parker} L.~C.,  2020, \mn@doi [\mnras]
  {10.1093/mnras/staa1213}, \href
  {https://ui.adsabs.harvard.edu/abs/2020MNRAS.495..554R} {495, 554}

\bibitem[\protect\citeauthoryear{{Ryzhov} et~al.,}{{Ryzhov}
  et~al.}{2025}]{Ryzhov_2025_ApJS}
{Ryzhov} O.,  et~al., 2025, \mn@doi [\apjs] {10.3847/1538-4365/ad93cd}, \href
  {https://ui.adsabs.harvard.edu/abs/2025ApJS..276...55R} {276, 55}

\bibitem[\protect\citeauthoryear{{Salerno}, {Mart{\'\i}nez}  \&
  {Muriel}}{{Salerno} et~al.}{2019}]{Salerno_2019_MNRAS}
{Salerno} J.~M.,  {Mart{\'\i}nez} H.~J.,   {Muriel} H.,  2019, \mn@doi [\mnras]
  {10.1093/mnras/sty3456}, \href
  {https://ui.adsabs.harvard.edu/abs/2019MNRAS.484....2S} {484, 2}

\bibitem[\protect\citeauthoryear{{Sampaio}, {de Carvalho}, {Ferreras},
  {Lagan{\'a}}, {Ribeiro}  \& {Rembold}}{{Sampaio}
  et~al.}{2021}]{Sampaio_2021_MNRAS}
{Sampaio} V.~M.,  {de Carvalho} R.~R.,  {Ferreras} I.,  {Lagan{\'a}} T.~F.,
  {Ribeiro} A.~L.~B.,   {Rembold} S.~B.,  2021, \mn@doi [\mnras]
  {10.1093/mnras/stab673}, \href
  {https://ui.adsabs.harvard.edu/abs/2021MNRAS.503.3065S} {503, 3065}

\bibitem[\protect\citeauthoryear{{Sampaio}, {de Carvalho},
  {Arag{\'o}n-Salamanca}, {Merrifield}, {Ferreras}  \& {Cornwell}}{{Sampaio}
  et~al.}{2024}]{Sampaio_2024_MNRAS}
{Sampaio} V.~M.,  {de Carvalho} R.~R.,  {Arag{\'o}n-Salamanca} A.,
  {Merrifield} M.~R.,  {Ferreras} I.,   {Cornwell} D.~J.,  2024, \mn@doi
  [\mnras] {10.1093/mnras/stae1533}, \href
  {https://ui.adsabs.harvard.edu/abs/2024MNRAS.532..982S} {532, 982}

\bibitem[\protect\citeauthoryear{{Sarzi} et~al.,}{{Sarzi}
  et~al.}{2006}]{Sarzi_2006_MNRAS}
{Sarzi} M.,  et~al., 2006, \mn@doi [\mnras] {10.1111/j.1365-2966.2005.09839.x},
  \href {https://ui.adsabs.harvard.edu/abs/2006MNRAS.366.1151S} {366, 1151}

\bibitem[\protect\citeauthoryear{{Shen}, {Mo}, {White}, {Blanton}, {Kauffmann},
  {Voges}, {Brinkmann}  \& {Csabai}}{{Shen} et~al.}{2003}]{Shen_2003_MNRAS}
{Shen} S.,  {Mo} H.~J.,  {White} S. D.~M.,  {Blanton} M.~R.,  {Kauffmann} G.,
  {Voges} W.,  {Brinkmann} J.,   {Csabai} I.,  2003, \mn@doi [\mnras]
  {10.1046/j.1365-8711.2003.06740.x}, \href
  {https://ui.adsabs.harvard.edu/abs/2003MNRAS.343..978S} {343, 978}

\bibitem[\protect\citeauthoryear{{Smith}, {Choi}, {Lee}, {Rhee},
  {Sanchez-Janssen}  \& {Yi}}{{Smith} et~al.}{2016}]{Smith_2016_ApJ}
{Smith} R.,  {Choi} H.,  {Lee} J.,  {Rhee} J.,  {Sanchez-Janssen} R.,   {Yi}
  S.~K.,  2016, \mn@doi [\apj] {10.3847/1538-4357/833/1/109}, \href
  {https://ui.adsabs.harvard.edu/abs/2016ApJ...833..109S} {833, 109}

\bibitem[\protect\citeauthoryear{{Solanes} \& {Salvador-Sole}}{{Solanes} \&
  {Salvador-Sole}}{1990}]{Solanes_1990_A&A}
{Solanes} J.~M.,  {Salvador-Sole} E.,  1990, \aap, \href
  {https://ui.adsabs.harvard.edu/abs/1990A&A...234...93S} {234, 93}

\bibitem[\protect\citeauthoryear{{Spinoso}, {Bonoli}, {Dotti}, {Mayer}, {Madau}
   \& {Bellovary}}{{Spinoso} et~al.}{2017}]{Spinoso_2017_MNRAS}
{Spinoso} D.,  {Bonoli} S.,  {Dotti} M.,  {Mayer} L.,  {Madau} P.,
  {Bellovary} J.,  2017, \mn@doi [\mnras] {10.1093/mnras/stw2934}, \href
  {https://ui.adsabs.harvard.edu/abs/2017MNRAS.465.3729S} {465, 3729}

\bibitem[\protect\citeauthoryear{{Springel} \& {Hernquist}}{{Springel} \&
  {Hernquist}}{2003}]{Springel_2003_MNRAS}
{Springel} V.,  {Hernquist} L.,  2003, \mn@doi [\mnras]
  {10.1046/j.1365-8711.2003.06206.x}, \href
  {https://ui.adsabs.harvard.edu/abs/2003MNRAS.339..289S} {339, 289}

\bibitem[\protect\citeauthoryear{{Stasi{\'n}ska} et~al.,}{{Stasi{\'n}ska}
  et~al.}{2008}]{Stasinska_2008_MNRAS}
{Stasi{\'n}ska} G.,  et~al., 2008, \mn@doi [\mnras]
  {10.1111/j.1745-3933.2008.00550.x}, \href
  {https://ui.adsabs.harvard.edu/abs/2008MNRAS.391L..29S} {391, L29}

\bibitem[\protect\citeauthoryear{{Stasi{\'n}ska}, {Costa-Duarte}, {Vale Asari},
  {Cid Fernandes}  \& {Sodr{\'e}}}{{Stasi{\'n}ska}
  et~al.}{2015}]{Stasinska_2015_MNRAS}
{Stasi{\'n}ska} G.,  {Costa-Duarte} M.~V.,  {Vale Asari} N.,  {Cid Fernandes}
  R.,   {Sodr{\'e}} L.,  2015, \mn@doi [\mnras] {10.1093/mnras/stv078}, \href
  {https://ui.adsabs.harvard.edu/abs/2015MNRAS.449..559S} {449, 559}

\bibitem[\protect\citeauthoryear{{Stasi{\'n}ska}, {Trevisan}  \& {Vale
  Asari}}{{Stasi{\'n}ska} et~al.}{2022}]{Stasinska_2022_FrASS}
{Stasi{\'n}ska} G.,  {Trevisan} M.,   {Vale Asari} N.,  2022, \mn@doi
  [Frontiers in Astronomy and Space Sciences] {10.3389/fspas.2022.913485},
  \href {https://ui.adsabs.harvard.edu/abs/2022FrASS...9.3485S} {9, 913485}

\bibitem[\protect\citeauthoryear{{Tempel}, {Tuvikene}, {Kipper}  \&
  {Libeskind}}{{Tempel} et~al.}{2017}]{Tempel_2017_A&A}
{Tempel} E.,  {Tuvikene} T.,  {Kipper} R.,   {Libeskind} N.~I.,  2017, \mn@doi
  [\aap] {10.1051/0004-6361/201730499}, \href
  {https://ui.adsabs.harvard.edu/abs/2017A&A...602A.100T} {602, A100}

\bibitem[\protect\citeauthoryear{{Thomas}, {Maraston}  \& {Johansson}}{{Thomas}
  et~al.}{2011}]{Thomas_2011_MNRAS}
{Thomas} D.,  {Maraston} C.,   {Johansson} J.,  2011, \mn@doi [\mnras]
  {10.1111/j.1365-2966.2010.18049.x}, \href
  {https://ui.adsabs.harvard.edu/abs/2011MNRAS.412.2183T} {412, 2183}

\bibitem[\protect\citeauthoryear{{Thomas} et~al.,}{{Thomas}
  et~al.}{2013}]{Thomas_2013_MNRAS}
{Thomas} D.,  et~al., 2013, \mn@doi [\mnras] {10.1093/mnras/stt261}, \href
  {https://ui.adsabs.harvard.edu/abs/2013MNRAS.431.1383T} {431, 1383}

\bibitem[\protect\citeauthoryear{{Treu}, {Ellis}, {Kneib}, {Dressler}, {Smail},
  {Czoske}, {Oemler}  \& {Natarajan}}{{Treu} et~al.}{2003}]{Treu_2003_ApJ}
{Treu} T.,  {Ellis} R.~S.,  {Kneib} J.-P.,  {Dressler} A.,  {Smail} I.,
  {Czoske} O.,  {Oemler} A.,   {Natarajan} P.,  2003, \mn@doi [\apj]
  {10.1086/375314}, \href
  {https://ui.adsabs.harvard.edu/abs/2003ApJ...591...53T} {591, 53}

\bibitem[\protect\citeauthoryear{{Trussler}, {Maiolino}, {Maraston}, {Peng},
  {Thomas}, {Goddard}  \& {Lian}}{{Trussler}
  et~al.}{2020}]{Trussler_2020_MNRAS}
{Trussler} J.,  {Maiolino} R.,  {Maraston} C.,  {Peng} Y.,  {Thomas} D.,
  {Goddard} D.,   {Lian} J.,  2020, \mn@doi [\mnras] {10.1093/mnras/stz3286},
  \href {https://ui.adsabs.harvard.edu/abs/2020MNRAS.491.5406T} {491, 5406}

\bibitem[\protect\citeauthoryear{{Turner} \& {Gott}}{{Turner} \&
  {Gott}}{1976}]{Turner_1976_ApJS}
{Turner} E.~L.,  {Gott} III J.~R.,  1976, \mn@doi [\apjs] {10.1086/190403},
  \href {https://ui.adsabs.harvard.edu/abs/1976ApJS...32..409T} {32, 409}

\bibitem[\protect\citeauthoryear{{Valk} \& {Rembold}}{{Valk} \&
  {Rembold}}{2025}]{Valk_2025_MNRAS}
{Valk} G.~A.,  {Rembold} S.~B.,  2025, \mn@doi [\mnras]
  {10.1093/mnras/stae2779}, \href
  {https://ui.adsabs.harvard.edu/abs/2025MNRAS.536.2730V} {536, 2730}

\bibitem[\protect\citeauthoryear{{Vollmer}, {Cayatte}, {Balkowski}  \&
  {Duschl}}{{Vollmer} et~al.}{2001}]{Vollmer_2001_ApJ}
{Vollmer} B.,  {Cayatte} V.,  {Balkowski} C.,   {Duschl} W.~J.,  2001, \mn@doi
  [\apj] {10.1086/323368}, \href
  {https://ui.adsabs.harvard.edu/abs/2001ApJ...561..708V} {561, 708}

\bibitem[\protect\citeauthoryear{{Weinmann}, {van den Bosch}, {Yang}  \&
  {Mo}}{{Weinmann} et~al.}{2006}]{Weinmann_2006_MNRAS}
{Weinmann} S.~M.,  {van den Bosch} F.~C.,  {Yang} X.,   {Mo} H.~J.,  2006,
  \mn@doi [\mnras] {10.1111/j.1365-2966.2005.09865.x}, \href
  {https://ui.adsabs.harvard.edu/abs/2006MNRAS.366....2W} {366, 2}

\bibitem[\protect\citeauthoryear{{Weinmann}, {Kauffmann}, {van den Bosch},
  {Pasquali}, {McIntosh}, {Mo}, {Yang}  \& {Guo}}{{Weinmann}
  et~al.}{2009}]{Weinmann_2009_MNRAS}
{Weinmann} S.~M.,  {Kauffmann} G.,  {van den Bosch} F.~C.,  {Pasquali} A.,
  {McIntosh} D.~H.,  {Mo} H.,  {Yang} X.,   {Guo} Y.,  2009, \mn@doi [\mnras]
  {10.1111/j.1365-2966.2009.14412.x}, \href
  {https://ui.adsabs.harvard.edu/abs/2009MNRAS.394.1213W} {394, 1213}

\bibitem[\protect\citeauthoryear{{Wetzel}}{{Wetzel}}{2011}]{Wetzel_2011_MNRAS}
{Wetzel} A.~R.,  2011, \mn@doi [\mnras] {10.1111/j.1365-2966.2010.17877.x},
  \href {https://ui.adsabs.harvard.edu/abs/2011MNRAS.412...49W} {412, 49}

\bibitem[\protect\citeauthoryear{{Wetzel}, {Tinker}, {Conroy}  \& {van den
  Bosch}}{{Wetzel} et~al.}{2013}]{Wetzel_2013_MNRAS}
{Wetzel} A.~R.,  {Tinker} J.~L.,  {Conroy} C.,   {van den Bosch} F.~C.,  2013,
  \mn@doi [\mnras] {10.1093/mnras/stt469}, \href
  {https://ui.adsabs.harvard.edu/abs/2013MNRAS.432..336W} {432, 336}

\bibitem[\protect\citeauthoryear{{Wright}, {Lagos}, {Power}, {Stevens},
  {Cortese}  \& {Poulton}}{{Wright} et~al.}{2022}]{Wright_2022_MNRAS}
{Wright} R.~J.,  {Lagos} C. d.~P.,  {Power} C.,  {Stevens} A. R.~H.,  {Cortese}
  L.,   {Poulton} R. J.~J.,  2022, \mn@doi [\mnras] {10.1093/mnras/stac2042},
  \href {https://ui.adsabs.harvard.edu/abs/2022MNRAS.516.2891W} {516, 2891}

\bibitem[\protect\citeauthoryear{{York} et~al.,}{{York}
  et~al.}{2000}]{York_2000_AJ}
{York} D.~G.,  et~al., 2000, \mn@doi [\aj] {10.1086/301513}, \href
  {https://ui.adsabs.harvard.edu/abs/2000AJ....120.1579Y} {120, 1579}

\bibitem[\protect\citeauthoryear{{Zabludoff} \& {Mulchaey}}{{Zabludoff} \&
  {Mulchaey}}{1998}]{Zabludoff_1998_ApJ}
{Zabludoff} A.~I.,  {Mulchaey} J.~S.,  1998, \mn@doi [\apj] {10.1086/305355},
  \href {https://ui.adsabs.harvard.edu/abs/1998ApJ...496...39Z} {496, 39}

\bibitem[\protect\citeauthoryear{{Zeldovich}, {Einasto}  \&
  {Shandarin}}{{Zeldovich} et~al.}{1982}]{Zeldovich_1982_Natur}
{Zeldovich} I.~B.,  {Einasto} J.,   {Shandarin} S.~F.,  1982, \mn@doi [\nat]
  {10.1038/300407a0}, \href
  {https://ui.adsabs.harvard.edu/abs/1982Natur.300..407Z} {300, 407}

\bibitem[\protect\citeauthoryear{{de Carvalho}, {Costa}, {Moura}  \&
  {Ribeiro}}{{de Carvalho} et~al.}{2019}]{deCarvalho_2019_MNRAS}
{de Carvalho} R.~R.,  {Costa} A.~P.,  {Moura} T.~C.,   {Ribeiro} A.~L.~B.,
  2019, \mn@doi [\mnras] {10.1093/mnrasl/slz084}, \href
  {https://ui.adsabs.harvard.edu/abs/2019MNRAS.487L..86D} {487, L86}

\makeatother
\end{thebibliography}

%%%%%%%%%%%%%%%%%%%%%%%%%%%%%%%%%%%%%%%%%%%%%%%%%%

%%%%%%%%%%%%%%%%% APPENDICES %%%%%%%%%%%%%%%%%%%%%

\appendix

\section{Additional Figures}

In order to complement the discussion based on the PPS diagram presented throughout this work, we present here additional figures showing the distribution of P and R galaxies across the PPS using the zones defined by \citet{Pasquali_2019_MNRAS}, rather than those of \citetalias{Rhee_2017_ApJ}. \citet{Pasquali_2019_MNRAS} investigated the distribution of time since infall of satellite galaxies from the YZiCS simulations in the PPS and defined eight distinct zones, which delineate the shape of the contours of the distribution of infall times, thus effectively segmenting the PPS into regions that constrain galaxies to relatively narrow time-since-infall ranges. The mean infall time of galaxies decreases from the innermost zone 1 ($T_{\rm{inf}} \sim 5.4 \,$Gyr) to the outermost zone 8 ($T_{\rm{inf}} \sim 1.4\,$Gyr). In contrast to the \citetalias{Rhee_2017_ApJ} zones, those defined by \citet{Pasquali_2019_MNRAS} are restrict to cluster-centric distances within $r_{200}$. It is important to note that the figures presented here are intended solely to complement the main discussion, as the \citet{Pasquali_2019_MNRAS} zones provides narrower infall time intervals than those of \citetalias{Rhee_2017_ApJ}, making it more informative in certain parts of the analysis. Nevertheless, the results remain essentially unchanged regardless of the chosen PPS segmentation scheme. 

\begin{figure*}
    \centering
    \includegraphics[width = 0.9\linewidth]{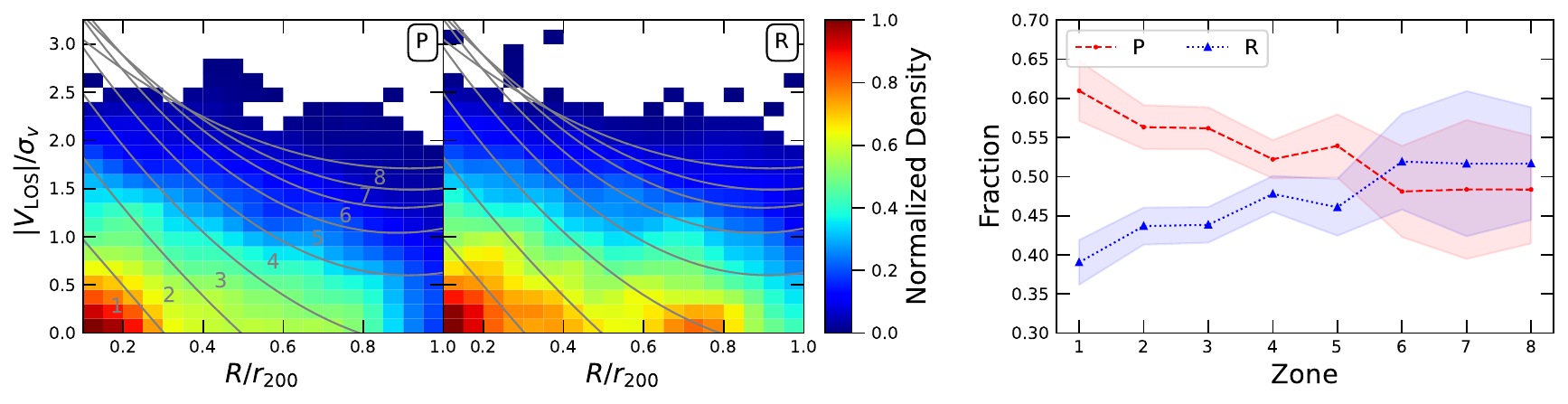}
    \caption{Same as Fig.~\ref{fig:PPS_Dist_All_Gal}, but using the PPS zones defined by \citet{Pasquali_2019_MNRAS}, instead of those defined by \citetalias{Rhee_2017_ApJ}.} 
    \label{fig:PPS_Dist_All_Gal_Pasquali}
\end{figure*}

\begin{figure*}
    \centering
    \includegraphics[width = 0.9\linewidth]{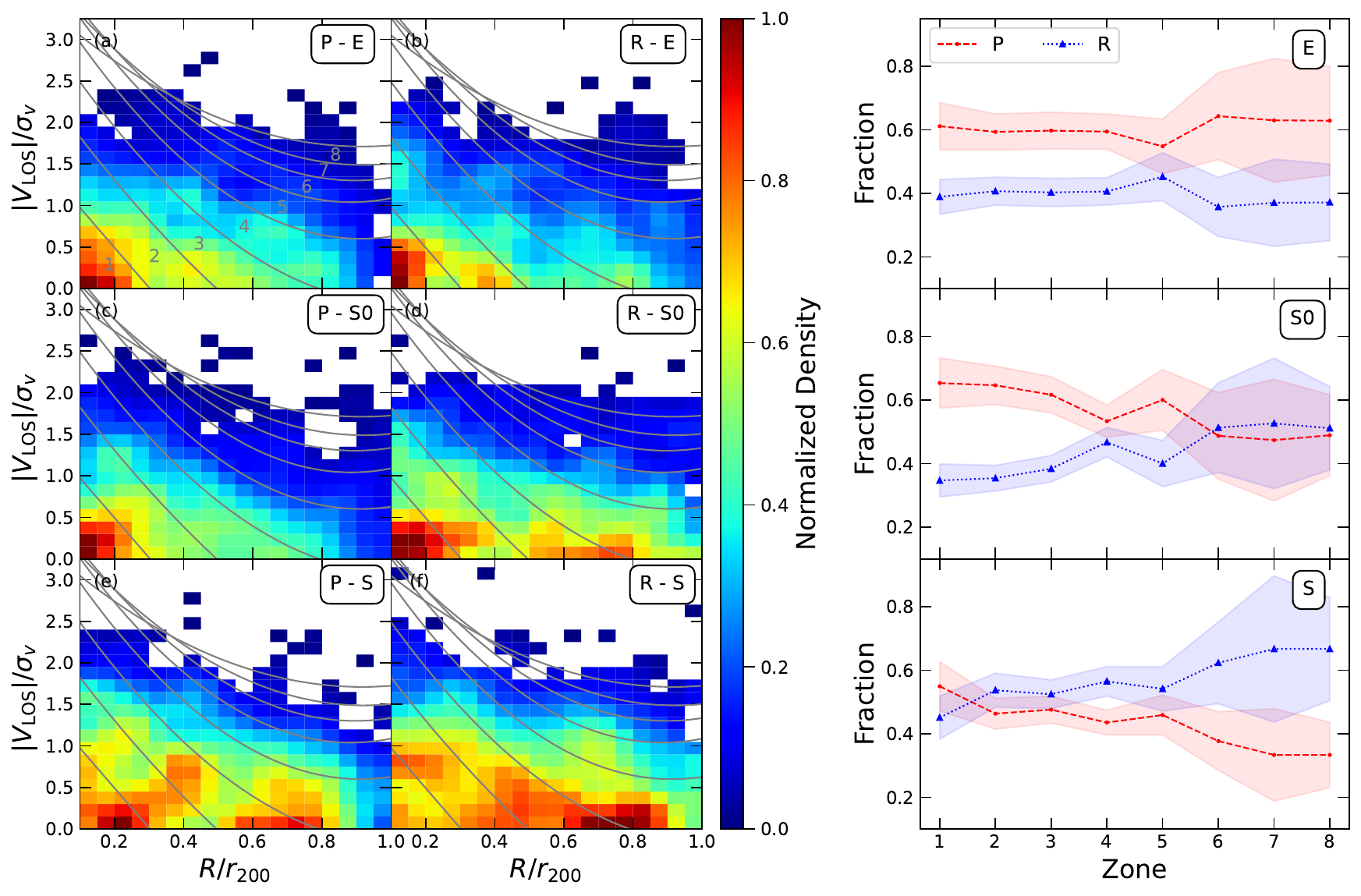}
    \caption{Same as Fig.~\ref{fig:PPS_Dist_Morpho}, but using the PPS zones defined by \citet{Pasquali_2019_MNRAS}, instead of those defined by \citetalias{Rhee_2017_ApJ}.} 
    \label{fig:PPS_Dist_Morpho_Pasquali}
\end{figure*}

Fig.~\ref{fig:PPS_Dist_All_Gal_Pasquali} displays the distributions of the full P and R populations across the \citet{Pasquali_2019_MNRAS} zones, while the results for the morphological classes are presented in Fig.~\ref{fig:PPS_Dist_Morpho_Pasquali}. 

% Don't change these lines
\bsp	% typesetting comment
\label{lastpage}
\end{document}